\documentclass[preprint,aps,prx,amsmath,amssymb,nobibnotes,citeautoscript,floatfix,longbibliography,raggedbottom,balancelastpage]{revtex4-2} 
\usepackage[usenames,dvipsnames]{color}
\usepackage{graphicx,xfrac,microtype}
\usepackage[bookmarks=false,colorlinks]{hyperref}
\hypersetup{hypertexnames=false,linkcolor=RubineRed,citecolor=Plum,filecolor=Mulberry,urlcolor=RoyalBlue}
\usepackage{tabularx}
\usepackage{booktabs}
\usepackage{comment}
 \usepackage{float}
\usepackage[version=4]{mhchem}
\usepackage{fancyhdr}
\usepackage{xr}
\usepackage[capitalise, noabbrev]{cleveref}

\newcommand{\lw}[1]{\textcolor{ForestGreen}{#1}}

\newcommand{\RN}[1]{%
  \textup{\uppercase\expandafter{\romannumeral#1}}
}

\begin{document}

\title{Synthetic accessibility and sodium ion conductivity of the Na$_{8-x}$A$^{x}$P$_2$O$_9$ (NAP) high-temperature sodium superionic conductor framework}
\author{Lauren N.\ Walters} 
\affiliation{Division of Materials Science, Lawrence Berkeley National Laboratory, California 94720, USA}

\author{Yuxing Fei} 
\affiliation{Division of Materials Science, Lawrence Berkeley National Laboratory, California 94720, USA}

\author{Bernardus Rendy} 
\affiliation{Division of Materials Science, Lawrence Berkeley National Laboratory, California 94720, USA}

\author{Xiaochen Yang} 
\affiliation{Division of Materials Science, Lawrence Berkeley National Laboratory, California 94720, USA}

\author{Mouhamad Diallo} 
\affiliation{Division of Materials Science, Lawrence Berkeley National Laboratory, California 94720, USA}

\author{KyuJung Jun} 
\affiliation{Division of Materials Science, Lawrence Berkeley National Laboratory, California 94720, USA}

\author{Grace Wei} 
\affiliation{Division of Materials Science, Lawrence Berkeley National Laboratory, California 94720, USA}

\author{Matthew J. McDermott} 
\affiliation{Division of Materials Science, Lawrence Berkeley National Laboratory, California 94720, USA}

\author{Andrea Giunto} 
\affiliation{Division of Materials Science, Lawrence Berkeley National Laboratory, California 94720, USA}

\author{Tara Mishra} 
\affiliation{Division of Materials Science, Lawrence Berkeley National Laboratory, California 94720, USA}

\author{Fengyu Shen} 
\affiliation{Energy Storage and Distributed Resources Division, Lawrence Berkeley National Laboratory, California 94720, USA}

\author{David Milsted} 
\affiliation{Division of Materials Science, Lawrence Berkeley National Laboratory, California 94720, USA}

\author{May Sabai Oo} 
\affiliation{Division of Materials Science, Lawrence Berkeley National Laboratory, California 94720, USA}

\author{Haegyeom Kim}
\affiliation{Division of Materials Science, Lawrence Berkeley National Laboratory, California 94720, USA}

\author{Michael C.\ Tucker}
\affiliation{Energy Storage and Distributed Resources Division, Lawrence Berkeley National Laboratory, California 94720, USA}

\author{Gerbrand Ceder}
\affiliation{Division of Materials Science, Lawrence Berkeley National Laboratory, California 94720, USA}

\begin{abstract}
\centering




%
Advancement of solid state electrolytes (SSEs) for all solid state batteries typically focuses on modification of a parent structural framework for improved conductivity, \textit{e.g.} cation substitution for an immobile ion or varying the concentration of the mobile ion. Therefore, novel frameworks can be disruptive by enabling fast ion conduction aided by different structure and diffusion mechanisms, and unlocking optimal conductors with different properties (\textit{e.g.} mechanical properties, sintering needs, electrochemical stability) than previously published. Herein, we perform a high throughput survey of an understudied structural framework for sodium ion conduction, Na$_{8-x}$A$^x$P$_2$O$_9$ (NAP), to understand the family's thermodynamic stability, synthesizability, and ionic conduction. We first show that the parent phase Na$_4$TiP$_2$O$_9$ (NTP) undergoes a structural distortion (with accompanying conductivity transition) due to unstable phonons from a pseduo-Jahn Teller mode in the 1D titanium chains. Then, screening of cation-substituted structural candidates with \textit{ab initio} and machine-learned potential calculations reveal a number of candidates that are thermodynamically stable, likely synthesizable, and have high predicted ionic conductivities. High throughput experimental trials and subsequent methodology optimization of one Na$_4$SnP$_2$O$_9$ (NSP) highlight collective challenges to the synthesis pathways for sodium phosphate materials via solid state synthesis. Our results demonstrate that NAP is a highly tunable conduction framework whose high temperature conductivity transition has heretofore eliminated it from significant research interest. By expanding the structural toolkit for SSE design, we increase the number of useful sodium ion electrolytes for integration into safe and accessible solid state batteries.

\end{abstract}

\maketitle

\section{Introduction}

First principles design of new solid state electrolytes (SSEs) often incorporates structural framework that are engineered to allow for fast ion conductivity.
When new frameworks are reported, they are dissected, categorized, and integrated into high-throughput studies to understand the conductivity pathway and diffusion mechanisms, and how these might be optimized.
For lithium SSEs, the significant number of frameworks have allowed studies to distill key features and design principles utilizing high-throughput materials screening and machine learning-based approaches \cite{intro_LiConduct_Physics1, intro_LiConduct_Physics2, intro_LiConduct_Physics3, intro_LiConduct_Physics4, intro_LiConduct_Physics5, intro_LiConduct_Physics6, intro_LiConduct_Physics7, intro_LiConduct_Review1, intro_LiConduct_ML2, intro_LiConduct_ML3}.
However, rational development of sodium and other non-lithium ionic conductors is impeded by the relatively smaller pool of reported examples to draw from, and the only handful of studies that have attempted to featurize the frameworks \cite{Design_Na_SSE_1, Design_Na_SSE_2}.
Therefore, uncovering a single new framework provides a key additional data point for the computational modeling and machine learning studies necessary for data driven sodium SSE development.

The NTP superionic conductor class was first introduced by T. Takahashi, \textit{et al.} at the International Conference on Solid State Electrolytes in 1981 \cite{NTP_original} (visualized in \autoref{fig:intro-figure}).
Subsequent studies were released in the 1980s and 1990s to clarify the structure of the low and high-temperature polymorphs, and verify a one-order-of-magnitude change in conductivity upon phase transition near 250\,$^\circ$C \cite{NTP_1, NTP_2, NTP_3_and_NFP,NTP_4, NTP_5, NTP_6}.
The framework presented in NTP falls into a general formula Na$_{8-x}$A$^{x}$P$_2$O$_9$ (NAP) and is composed of large, highly-coordinated sodium cages pinned by 1D cation chains.
Despite the initial success and ease of synthesis of the original NTP structure, 
there are only handful of other reported cation substituted NAP phases \cite{NAP_1, NAP_2, NTP_3_and_NFP}.
However, no studies known to the authors describe detailed the mechanism behind the structural or conductivity transitions, underscoring the limited knowledge and research activity on NTP, despite its high ionic mobility.
Moreover, within the original Google set, only a handful of materials were found that had low activation energies for sodium ion diffusion, and NTP had one of the lowest.
The lack of reported structural and conductivity investigations, particularly with modern first principles calculations, in conjunction a narrow understanding of the chemical flexibility of the structure, demands an immediate and broad survey of the NAP framework.
%

Historically, investigation of new materials represents a bottleneck towards accelerating new phase discovery; trial-and-error-based experimental synthesis development is limited by time, resources, and expertise, with no guarantee of success.
Meanwhile, as new materials are predicted, critical \textit{a priori} questions for emerging materials families include synthesizability and properties of prospective phases.
However, output from first principles calculations and machine learning models can fail to reflect the structures and properties of real-world samples.
This gap in knowledge between computationally predicted and experimentally realized results limits the scope that ambitious computational research can reasonably tackle.
To address this challenge, the integrated use of computational and experimental tools such as open-access material databases, machine-learned potentials, synthesis science models, and self-driving labs is essential. 
Feedback between these different strategies fosters more efficient sampling of materials spaces.
Moreover, the culmination of positive and negative results can paint a more complete picture of the materials discovery process, introducing new insights towards the synthesis development and phase characterization processes.

We here conduct a survey of the NAP framework to report key components necessary for development, including thermodynamic stability, synthesizability, conductivity predictions and measurements, and the sodium diffusion pathway.
Experimental synthesis attempts of 3+, 4+, and 5+ cation substitutions yields successful synthesis of one new material, Na$_4$SnP$_2$O$_9$ (NSP).
A closer look into the physics behind previously reported NTP gives crucial insight into why NAPs have a sudden drop in ionic conductivity upon distortion to lower temperature phase, and what modifications might be necessary to suppress this transition.

\section{Methods \label{sec:methods}}

\subsection{Experimental Methods}
\subsubsection{Solid state synthesis}
Reactions were completed with traditional, powdered solid state (ceramic) synthesis. Reagents were purchased from Sigma-Aldrich (Millipore Sigma) and BeanTown Chemical at generally 98\% or greater purity. Due to the significant number of reagents used, the full list of precursors used, CAS numbers, manufacturer, and purity are included in the Supplemental Materials. 
As sodium pyrophosphate is purchased as hydrated \ce{Na4P2O7}$\cdot$10\ce{H2O}, it is heated for at least 24 hours at 70\,$^\circ$C under vacuum to remove crystal water before use.

For all syntheses, stoichiometric amounts of reagent are weighed within a 0.1\% threshold and then mixed in air. Mixing is done via either wet (ball) milling, pestle and mortar grinding, or dry ball milling.
%
Wet (ball) milling is carried out as described by Z. Lun, et al. \cite{Wet_Milling} First, a slurry is made by spin mixing ethanol and precursors together in plastic vial with 10 zirconia balls for 10 minutes within a Hauschild Speedmixer machine (DAC 150.1 FVZ-K) at 2,000\,rpm. The slurry is then immediately pipetted into a crucible. The crucible is heated at 80\,$^\circ$C in air to fully remove the ethanol.
%
Mortar and pestle mixing involves placing powdered precursor in a pestle and mortar, and then grinding them by hand for at least 10 minutes until fully ground.
%
For dry ball-milling, the powder is first carefully ground using a pestle and mortar for 15 minutes in the air. This mixture is then placed in a 50\,mL zirconia jar with 5mm -zirconia balls and milled (SPEX 8000M Mixer/Mill high energy ball mill) for 50 min. The ball milling followed by hand grinding with mortar and pestle are repeated twice, with an additional final hand grinding for 15 minutes. The pestle and mortar steps are found necessary to ensure adequate mixing due to the stickiness of many phosphate precursors.
After ball-milling, samples are pelletized into 13\,mm pellets with a Carver Manual Bench Top Pellet Press (model 3851-0). Pelletization is completed by applying 4\,tons of pressure for 5 minutes.
Samples are then heated at a default of 2\,$^\circ$C/min to 150\,$^\circ$C, held for 12 hours (calcination), and then ramped at 5\,$^\circ$C/min to the dwell temperature for 12 hours. Other heating profiles are used where indicated. Natural (uncontrolled) cooling is completed in the furnace without opening the door.
After cooling, samples are scraped from crucibles and collected in vials.

\subsubsection{A-Lab Synthesis}
Exploratory solid state synthesis is performed using the A-Lab, an autonomous and automated solid state synthesis laboratory where full details of the infrastructure are described by N. Szymanski, \textit{et al.} \cite{Alab}. Briefly, the A-Lab performs solid state synthesis in three stages reflective of human (manual) synthesis: (\textit{i}) a precursor preparation stage, (\textit{ii}) a heating stage, and (\textit{iii}) a sample recovery and characterization stage. At each stage, there is a 6-axis robot arm that is programmed with high-precision waypoints to move samples through a series of tasks carried out by devices as implemented within AlabOS by Y. Fei, \textit{et al.} \cite{Alab_os}
To prepare samples, powders are dispensed and weighed simultaneously by using an automatic dispenser balance (Quantos, Mettler Toledo) into a plastic vial containing 10 zirconia balls.  Ethanol is added and the sample is mixed at 2000\,rpm for 10 minutes to create a slurry. The slurry is pipetted from the vial and dispensed into a crucible. After drying at 80$^\circ$C, the sample is left mixed and densified at the bottom of a crucible. The samples are then moved to furnaces and heated according to the programmed profile. The default heating profile is 2\,$^\circ$C/min to 300\,$^\circ$C, followed by 5\,$^\circ$C/min up to the maximum dwell temperature, followed by a 12-hour hold, and then uncontrolled cooling in the furnace to room temperature. Finally, the samples are recovered autonomously by vertically shaking a crucible covered with a plastic cap with a 10-mm alumina ball inside to dislodge (mill) the sample from the crucible wall and generate powder. Finally, the sample is shaken through a steel mesh and pressed onto a disc for automated powder XRD characterization, and the remaining samples are stored inside a uniquely-labeled plastic vial for further characterization.

\subsubsection{Characterization}
Sample characterization is completed primarily with powdered x-ray diffraction (XRD) on an Aeris Minerals diffractometer (Malvern Panalytical) utilizing Cu K$\alpha$ radiation.
Scans are performed between 10\,$^\circ$ and 100\,$^\circ$ $2\theta$ range for 8 minutes, with a step size of $\approx$0.01\,$^\circ$ and an active length of 5.542\,$^\circ$ unless otherwise noted.
\textit{In situ} x-ray diffraction targeting \ce{Na4SnP2O9} samples were carried out on a Bruker \textit{in situ} XRD (model: D8 ADVANCE) with Anton Paar HTK 1200N non-ambient heating chamber. Stoichiometric powders are prepared and loaded onto a 0.8 mm depth alumina sample holder (Anton Paar) and inserted into the flow cell. Scans are performed every 10\,$^\circ$C during heating within the range of 10$^\circ$ and 60$^\circ$ 2$\theta$, and at a step length of 0.02\,$^\circ$ with 0.04 s per scan step for 2 minutes (low fidelity scan). Samples are held for 4 hours at a 900\,$^\circ$C dwell. Scans are taken every 10 minutes at a step size of 0.02 $^\circ$ with 0.2 s per scan step during a total time of nearly 10 minutes (high fidelity scan). During uncontrolled cooling, temperatures are held every 100\,$^\circ$C for a low fidelity scan.

Analysis of XRD patterns is completed through Rietveld refinement delivered by Dara and Profex, utilizing the refinement kernel BGMN \cite{Dara, Profex}.
Dara is an in-house automated phase identification and Rietveld refinement algorithm for powder XRD pattern analysis which ranks possible phase sets based on a quality of fit figure of merit.
Initial structure searches for phase identification are performed by Dara on known crystal structures from the Inorganic Crystal Structure Database (ICSD) \cite{ICSD}. Ultimate phase identification is performed by human analysis with the aid of best case matches from Dara.
%
Final refinements are completed within Profex. The background is fit by Lagrangian polynomials. Crystalline phase site occupancies are assumed to be nominally stoichiometric. Lattice parameters are permitted to vary by 1\% when permitted by symmetry. A preferred orientation model with fourth-order spherical harmonics (gewicht=SPHAR4) is used during refinement. The crystalline size parameters k\textsubscript{1} and B\textsubscript{1} are constrained isotropically in the range 0.0 to 0.1 with initial value of 0. The microstrain effect k\textsubscript{2} is constrained between 0 and 0.1.
Isotropic thermal displacement parameters are set to vary by up to 0.01\,\AA$^2$ for oxygen sites.

Electrochemical impedance spectroscopy (EIS) was completed at low and elevated temperatures to assess sodium conductivity. The annealed sample is first ball milled using a high-energy ball mill (SPEX 8000$\cdot$M Mixer/Mill) with zirconia balls and jar to reduce the particle size. Then the sample is pressed into a 6\,mm pellet with 2\,tons of weight for 3 minutes and sintered at 900\,$^\circ$C for 16 hours before cooling naturally in the furnace. For room temperature measurements, pellets are sanded finely and attached to two indium foils as ion-blocking electrodes. EIS measurements were performed using an EC-Lab Electrochemistry SP300 system. The measurements were conducted at the initial open-circuit voltage in the frequency range of 7 MHz–10 mHz with the application of a 10-mV signal amplitude. For high temperature measurements, the pellets were Au-coated by 2 minutes sputtering with a gold source. Silver paste (AG-I) purchased from fuelcellmaterials is then applied and platinum electrodes are attached with a platinum mesh during a 15 minute heat at 900\,$^\circ$C. High temperature EIS is performed with a BioLogic potentiostat (VSP-300) within a Barnstead International tube furnace (21100). Measurements are taken every 50\,$^\circ$C up to 800\,$^\circ$C, and every 50\,$^\circ$C during controlled cooling.
%

Scanning electron microscopy (SEM) energy-dispersive X-ray spectroscopy (EDS) measurements are taken using a Thermofisher Phenom XL G2 system.
SEM-EDS data are collected and quantified in automated fashion with Auto-SEMEDS, a software developed in-house. 
Details on the employed method are reported elsewhere \cite{EDS_Fitting_Andrea_Method}.
For each analyzed sample, 200 point spectra are randomly collected from tens of particles. The electron beam energy is fixed at 15\,kV and a total of 50,000 counts per spectrum is collected. The spectra are then quantified with the peak-to-background method \cite{PeakToBackground_1, PeakToBackground_2}, using the pure precursors as standards. Only compositions with analytical errors less than 10\% are considered for this study.
From each measured composition -- \textit{i.e.} vector of elemental atomic fractions -- precursor molar fractions are obtained via a non-negative least squares optimization algorithm, employing the precursors elemental atomic fractions as fixed vector bases. The obtained coefficients are normalized to a total of 1.


Transmission electron microscopy (TEM) EDS measurements were taken using the ThemIS microscope using a superXG2 detector at 300\,kV, housed at the National Center for Electron Microscopy.
The beam current is recorded to be $\approx$0.5\,nA using a spot size 3. During EDS collection, the electron counts varied between 2,000\,cps to 4300\,cps. The dwell time was fixed at 25\,$\mu$s.
Elemental quantification was done using the Thermofisher Velox software (version: 3.7.0.872).
Na, Sn, O, and P compositional quantification used K, L, K, and K edges of the elements, respectively.
We note that the samples were first imaged at room temperature; however, noticeable beam damage was observed in the HAADF-STEM micrograph of the particle before and after EDS mapping (shown in Supplemental).
To prevent beam damage the EDS mapping was obtained under cryo-conditions using a Gatan 915 holder.
As EDS is measured by suspending the particles on a foamvar stabilized lacey carbon support Cu grid, both C and Cu EDS peaks were used for deconvolution.
The total summed up EDS spectrum is shown in the Supplemental. 

\subsection{Computational Methods}

\subsubsection{Density Functional Theory Calculations}

Density functional theory calculation were carried out with Vienna Ab initio Simulation Package (VASP) leveraging the generalized gradient approximation functional Perdew-Burke-Ernzerhof \cite{VASP_1, VASP_2, VASP_3, VASP_4, PAW_1, PAW_2, PBE}. DFT settings, including K-point density, plane wave cutoff, pseudopotential, magnetic moments, and smearing were those recommended by pymatgen's \textit{MPRelaxSet} \cite{pymatgen, pymatgen_MPSets}. However force convergence during relaxation was tightened to 10$^{-2}$\,eV/\AA.
Energetic corrections to static calculations on optimized structures were completed with the MaterialsProject2020Compatibility class \cite{MPCorrections_1, MPCorrections_2}. 
Phonon calculations at 0\,K utilizing the quasiharmonic approximation were executed with the Phonopy package on a 2$\times$2$\times$2 supercell with a minimum lattice parameter length of 10\,\AA\,\cite{QHA, Phonopy}. Phonon band structures sample the Brillion zone with reciprocal space paths recommended by SeeK-path \cite{seekkpath1, seekkpath2}.
Sodium-vacancy ordering distributions are found with structures enumerated by pymatgen's OrderDisorderedStructureTransformation class, wherein symmetrically equivalent structures are removed.

\subsubsection{Diffusivity Calculations}
Diffusivity calculations for all relevant materials were completed with molecular dynamics calculations with machine-learned potentials created with fine-tuned CHGNet \cite{CHGNet}.
CHGNet was fine-tuned separately for each structure. Each dataset for fine tuning was generated by collecting DFT structures from on-the-fly MD with FLARE \cite{Flare_1} at 5 applied strains (-1, 0, 1, 2, 3\%) and 4 temperatures (400, 600, 800, 1000\,K) initiated from the DFT relaxed structure. Mean absolute errors for the validation energy, force, and stress are provided in the Supplemental. 
Molecular dynamics (MD) simulations were run for 50\,ps in N-P-T, followed by a minimum of 2\,ns under N-V-T conditions. Simulations were performed with the Berendsen thermostat \cite{Berendsen_thermostat}.

For each phase, MD simulations were performed at 8 temperatures: 300, 400, 500, 600, 700, 800, 900 and 1,000\,K. Na-ion self-diffusion coefficients were computed from the mean squared displacements (MSD) of the MD simulation trajectories, where the squared displacements were averaged over time intervals of duration $\Delta$t. Only the portion of the MSD curve with $\Delta$t $>$\,1\,ps and $\Delta$t\,$<$\,50\% of the total simulation time was considered in the analysis. Activation energies and room temperature diffusivities were extrapolated by fitting the diffusivities to an Arrhenius relation. Ionic conductivities were calculated from Na-ion self diffusivities assuming the Nernst-Einstein relation. Error bars for the diffusivities were computed using the empirical error estimation scheme proposed by X. He \textit{et al.} \cite{Diffusivity_Errors}

\subsubsection{Precursor Prediction Optimization}
Precursor prediction models were utilized to suggest and build a list of precursors for NSP. Precursor sets were filtered safety, accessibility, product separability (if a solid byproduct is made), and the model-dependent confidence ranking.
%
%
Reaction network \cite{McDermott_reactionNetwork_model, Matt_Selectivity_Model} was performed with the NSP phase set as the target, and with a hyperdimensional chemical system of ``B-Br-C-Cl-F-H-K-Li-N-Na-O-P-S-Sn", and a reaction temperature of 600\,$^\circ$C. 34,880 pairwise reactions were found to have negative or close to negative reaction energy to form the target, and were ranked by additional phase competition metrics.
s4 \cite{TextMinedSynthesisRecipes} was run with the target composition as input. 10 precursor sets were output, which are modifications to previously reported syntheses for chemically similar targets.
The synthesis planning algorithm \cite{Wenhao_Autunomous_Synthesis} was executed with the respective target composition as input. 3 precursor sets were found which had negative reaction energies and fulfilled other thermodynamic metrics of synthesizability.
A full list of precursor sets chosen, their reaction energy from Materials Project, total DFT energies, and their relative model outputs is shown in the Supplemental.

\section{Results}

\subsection{NAP Structural Model \label{sec:ntp-structure-model}}

\begin{figure}
    \centering
    \includegraphics[width=1.0\linewidth]{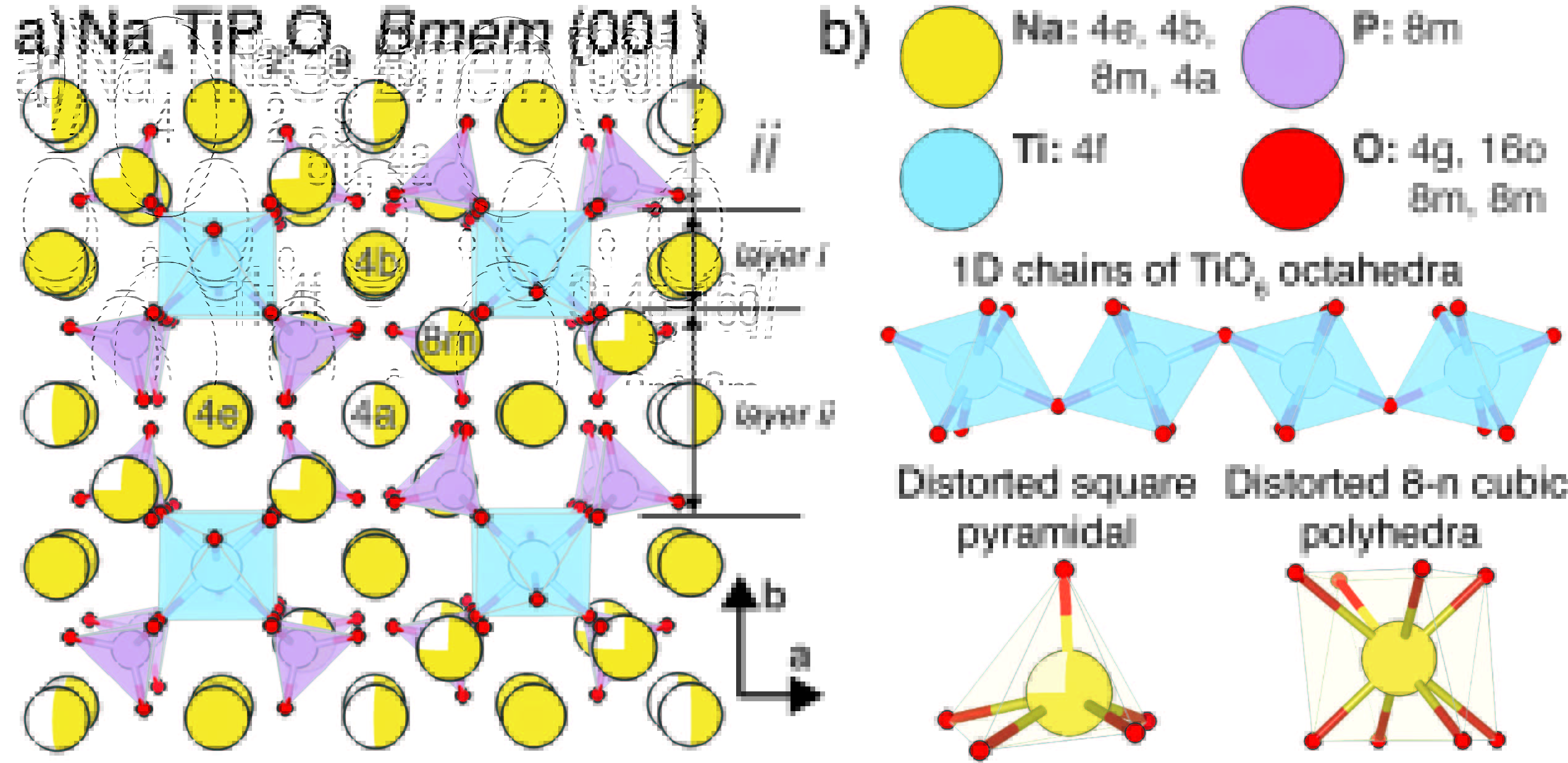}
    \caption{Structural visualization of \ce{Na4TiP2O9}, showing a) multiple formula units with patterned \textit{i} and \textit{ii} layers in (001) b) components of the structure, including atoms and their respective Wyckoff sites, titanium octahedra chains, and the two types of sodium polyhedra, 5-n square pyramidal and 8-cn cubic polyhedra.}
    \label{fig:intro-figure}
\end{figure}

The generalized Na$_{8-x}$A$^x$P$_2$O$_9$ structural framework is derived from the high temperature, orthorhombic \textit{Bmem} (67) \ce{Na4TiP2O9} parent phase (HT-NTP, ICSD 39702), visualized in \autoref{fig:intro-figure}. HT-NTP is a layered structure composed of five 1D chains of polyhedra stacked into two distinct layers in the a-c plane. Na$^+$ ions are primarily located in large 8 coordination number (cn) distorted cubic polyhedra cages (\autoref{fig:intro-figure}b). One-dimensional, corner-connected TiO$_6$ octahedra chains (\autoref{fig:intro-figure}b), surrounded by chains of phosphate tetrahedra that alternate with sodium square pyramidal 5-cn polyhedra, scaffold the structure.
Layer \textit{i} holds the titanium octahedra and the Na2 \textit{4b} sodium site, while the thicker layer \textit{ii} contains the 8-cn cubic Na1 \textit{4e} and Na4 \textit{4a} cages framed by the Na3 \textit{8m}-PO$_4$ columns.
The low temperature, monoclinic \textit{P2/c} (13) polymorph (LT-NTP, ICSD 39901) is topologically similar to HT-NTP, but displays compression and elongation of the TiO$_6$ octahedra, and a significant distortion of sodium polyhedra resulting in changes to Na site occupations \cite{NTP_1}.
For HT-NTP, high sodium mobility 
($2.6\times10^{-2}\Omega^{-1}\mathrm{cm}^{-1}$ at 300\,$^\circ$C \cite{NTP_1})
was previously reported to be promoted by easy transport between and through the large 2D sodium channels in the a-c plane, and is likely supported by partial sodium disorder on the Na2 4\textit{b}, Na4 4\textit{a}, and Na3 8\textit{m} sites \cite{NAP_2}.
%

We note that coupled temperature-dependent conductivity-structure transitions define ionic conductivity for all known NAP phases.
The conductivity of NTP decreases upon cooling when the orthorhombic phase transforms to the monoclinic phase at approximately 300\,$^\circ$C \cite{NTP_1}.
Similar conductivity transitions are measured near 200-300\,$^\circ$C for the other two reported NAP materials \ce{Na_{4.6}FeP2O_{8.6}F_{0.4}} and \ce{Na{(Al,Cr)}P_2O_{9-x}F_{2x}} \cite{NAP_1}.

\begin{figure}
    \centering
    \includegraphics[width=0.7\linewidth]{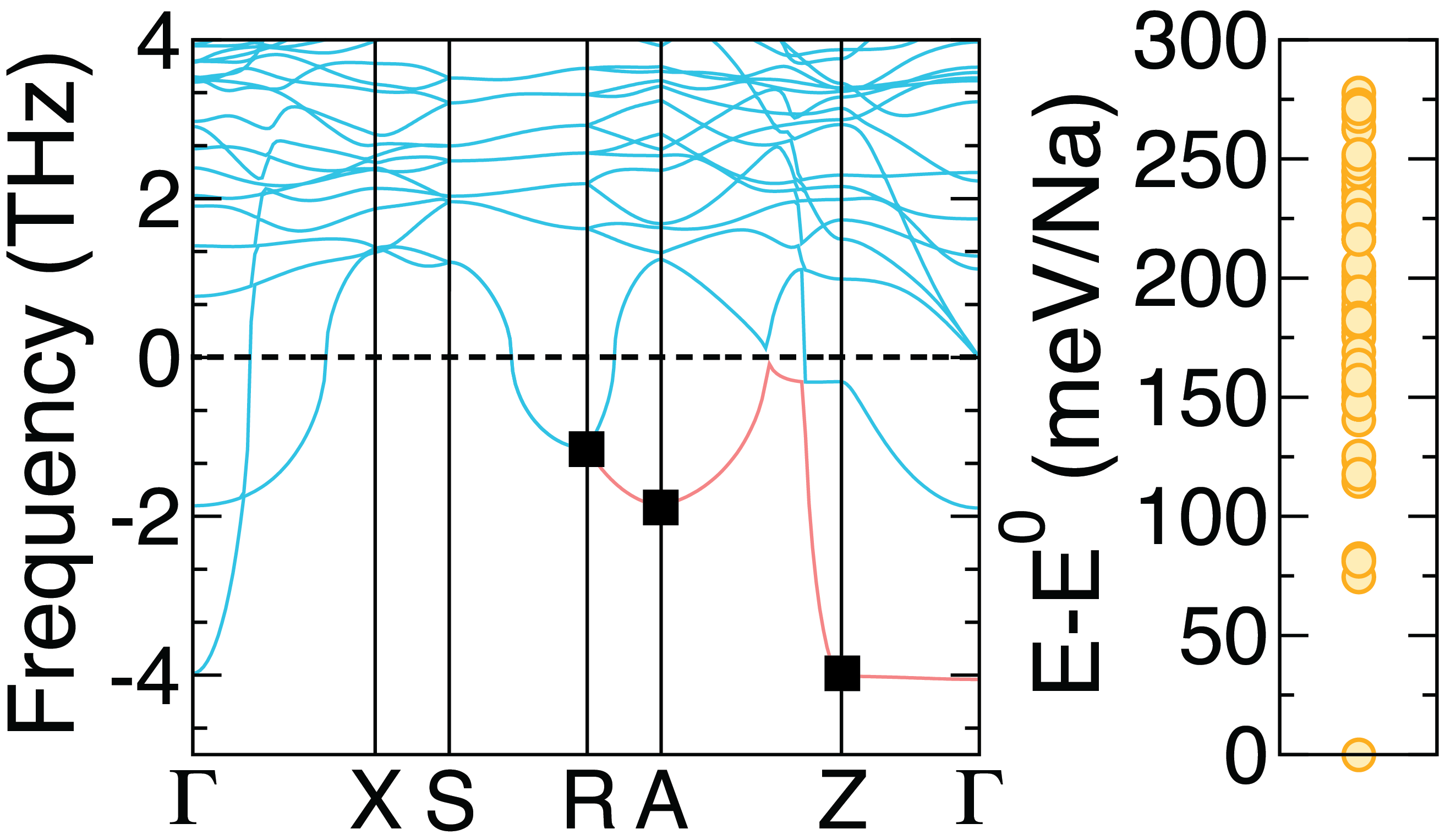}
    \caption{Figure showing (left) HT-NTP lattice dynamics calculations show negative modes at \textit{R}, \textit{A}, and \textit{Z} visualized by the green phonon band (right) energetic variation with sodium occupational disorder in HT-NTP.}
    \label{fig:NTP-dynamic-stability}
\end{figure}

To better understand the conductivity transition in NTP, we use \textit{ab initio} calculations to explore specific structural features in NTP that might impact conductivity: the stability and sodium ordering energies of the high T symmetric structure. \autoref{fig:NTP-dynamic-stability} (left) shows the phonon band structure and (right) variation of the energy with sodium occupational disorder in HT-NTP. We note that the high temperature phase is dynamically unstable due to a number of negative phonon frequencies at the $\Gamma$, $R$, $A$, and $Z$ points (depiction of the full Brillouin zone path is shown in the Supplemental.
Visual inspection of the negative modes at \textit{q}-points in the Brillion zone elucidates key instabilities:
($\Gamma$: 0, 0, 0) and ($Z$: 0, 0, 0.5) pseudo-Jahn Teller like TiO$_6$ elongation/compression,
($A$: 0, 0.5, 0.5), Ti-octahedral rotation with dimerization of Na-Na in layer \textit{ii}, and
($R$: 0.3006, 0.3006, 0.5) collective wave-like wobbling of the oxygen atoms. Modulation of individual unstable phonon mode followed by DFT structural relaxation is not able to recover the monoclinic ground state (see Supplemental). 
Therefore, the structural transition to the low temperature monoclinic phase is likely driven by the combination and interaction of multiple unstable phonon modes.

We also explore the energetics of sodium ordering. The right panel of \autoref{fig:NTP-dynamic-stability} depicts the energy per sodium for 300 symmetrically inequivalent sodium occupations in HT-NTP. The lowest energy configuration, found to be a fully vacant Na4 site configuration, is set to zero. Higher energy configurations range from 74.4\,meV/Na up to 277.6\,meV/Na. 
In addition, we perform DFT calculations on the LT-NTP phase, where we explore the filling of inequivalent Na sites.
In LT-NTP, the Na4 site is distorted such that there are now four inequivalent cation centers in the same 1D chain along the c-axis.
Only one is reported to be filled \cite{NTP_1}.
Cell volume and shape are fixed to the experimental values, while atomic position are allowed to move during relaxation.
We find that there is an increase ranging from 447\,meV/Na and 969\,meV/Na as sodium is moved to other sites previously reported to be vacant see Supplemental. 
This large energy change with site occupation variation represents a possible barrier for Na mobility in the LT-NTP. Together, we find that the sodium site energy landscape is not flat, and structural distortions to sodium polyhedra impact this landscape substantially.

\begin{figure}
    \centering
    \includegraphics[width=1.0\linewidth]{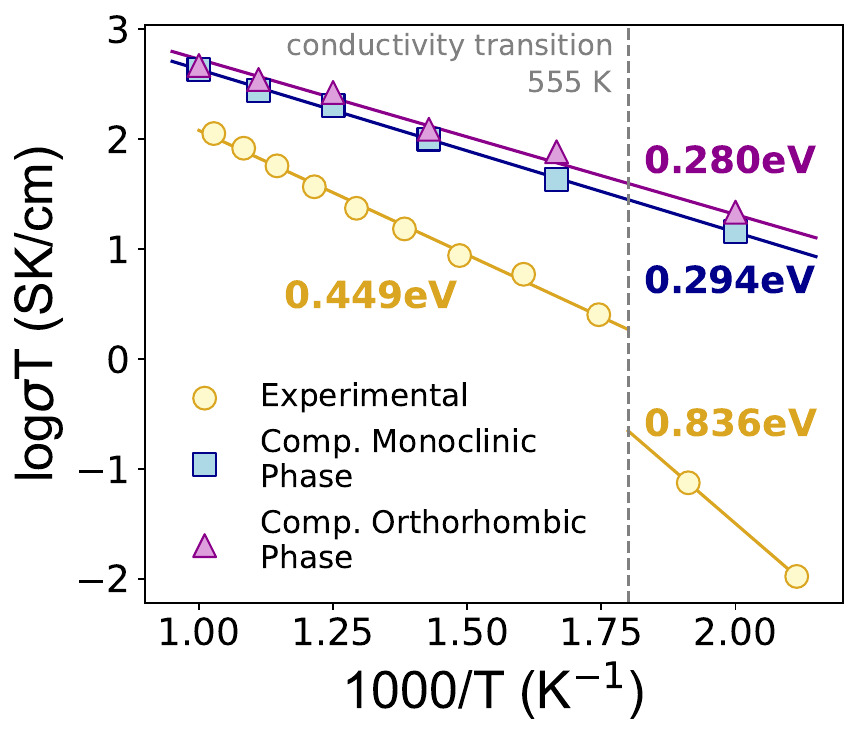}
    \caption{
    Computationally predicted and experimentally measured (golden line) temperature-dependent conductivity of NTP.
    Pink triangles give the computed conductivity in the DFT-optimized single crystal XRD structures for the orthorhombic high temperature phase, and blue squares give computed conductivity in the low temperature monoclinic phase. The dashed gray line indicates the measured conductivity transition, occurring at approximately 555\,K.}
    \label{fig:ntp-conducivity}
\end{figure}

We verify the synthesis, characterization, and impedance of NTP. NTP samples are made by wet ball milling \ce{Na2CO3}, \ce{NH4H2PO4}, and \ce{TiO2} precursors, followed by heating at 950$^{\circ}$C. PXRD confirms the presence of monoclinic \ce{Na4TiP2O9} with some \ce{Na4P2O7} (1.05w\%) impurity found during refinement  (see Supplemental)
Temperature-dependent EIS is then performed on a conventionally-sintered pellet of $\approx$99\% phase pure NTP.
For comparison to experimental conductivities, machine-learning molecular dynamics (MLMD) diffusivity calculations were completed on the experimentally reported monoclinic and orthorhombic phases.
\autoref{fig:ntp-conducivity} plots the experimentally measured and MLMD calculated conductivities as log($\sigma T$) versus 1000/$T$ for the temperature range of 150 to 850\,$^\circ$C.
EIS measures a conductivity transition near 200-300\,$^{\circ}$C, in agreement with the literature \cite{NTP_1}.
The low and high temperature conductivity regimes exhibit activation energies of 0.836\,eV and 0.449\,eV, respectively.
MLMD diffusivity calculations predict much lower activation energies of 0.294\,eV and 0.280\,eV for LT-NTP and HT-NTP, respectively. Furthermore, no conductivity transition is identified the MLMD calculations.
Interestingly, MLMD diffusivity calculations demonstrates much higher conductivity and much lower diffusion barriers for both phases than that of the experimental NTP sample \autoref{fig:ntp-conducivity}. This supports the hypothesis that the experimentally observed conductivity transition might have structural origins that occur from the temperature-dependent opening of a percolation pathway.

Our work with NTP shows us that the high temperature NTP structure becomes unstable upon cooling, leading to distortions in Na polyhedra and cation octahedra \cite{Design_Na_SSE_1, Design_Na_SSE_2}. This structural instability increases the activation energy for Na motion, limiting the room tempterature Na-ion conductivity.
For NAP to be a useful solid state electrolyte, it is critical to understand if cation substitution might be able to lower activation energies, increase sodium conductivity, and either stabilize the orthorhombic phase or significantly lower the phase transition temperature.
Therefore, we complete a high-throughput exploration of possible A site substitutions within Na$_{8-x}$A$^{x}$P$_2$O$_9$.

\subsection{High Throughput Computation Investigation of Cation Substitution \label{sec:high-throughput-screen}}
We evaluate the possibility of cation substitution on the A (Ti) site into orthorhombic NAP through high-throughput density functional theory computations.
A total of 24 cation candidates with oxidation states ranging from 3+ to 5+ are chosen, with sodium content adjusted to balance charge: Al, Sc, Ti, V, Cr, Mn, Fe, Co, Ga, Ge, Y, Zr, Nb, Mo, Ru, In, Sn, Sb, La, Hf, Ce, Ta, W, and Bi. 
When the A cation has a 3+ charge, all Na sites are filled.
For a 4+ cation, we arrange the Na ions according to the ground state in NTP (vacant Na4 Wyckoff site).
Similarly, sodium occupation for A\,5+ was set by keeping Na4 and half of Na3 sites vacant, in line with common occupancies in NTP \cite{NTP_4, NTP_5}. 
To our knowledge, all substituted \ce{Na_{8-x}A^xP2O9} are unreported in literature.
The substituted NAP candidates are structurally optimized with DFT and manually inspected after relaxation to ensure that symmetry, structure topology, and sodium site occupancies does not change.

\begin{figure}
    \centering
    \vspace*{-16.5mm}
    \includegraphics[width=0.7\linewidth]{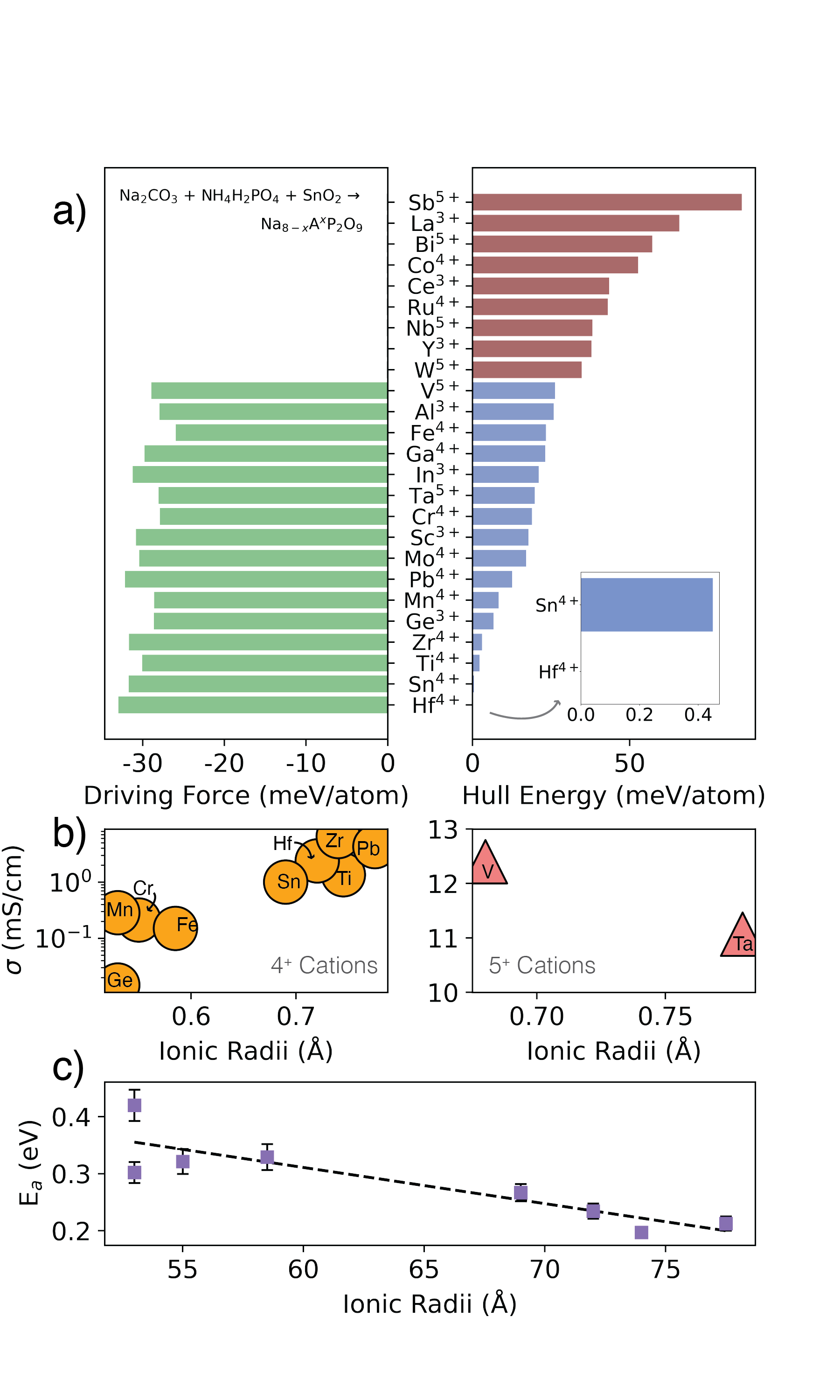}
    \vspace*{-12mm}
    \caption{Results from high throughput computational screening for candidate structures created through substitution of the A site cation. a) left panel: the reaction energy to create the NAP compound from the most stable binary A oxide, sodium carbonate, and ammonium phosphate for (meta)stable phases; right panel: energy above the convex hull.
    b) Room temperature extrapolated conductivities calculated with MLMD plotted versus ionic radii for (left) 4+ and (right) 5+ cations substituted into the NAP structure.
    c) Activation energies calculated with MLMD plotted versus ionic radii for 4+ cations. 
    }
    \label{fig:candidate-structure-thermo}
\end{figure}

To evaluate thermodynamic stability, the composition-energy convex hull is generated with pymatgen's Phase Diagram class \cite{pymatgen, MaterialsProject_PhaseDiagram} utilizing DFT total energies as approximations for 0\,K enthalpies.
A phase diagram for each substituted candidate is created by querying all entries within the corresponding chemical space from the Materials Project \cite{MaterialsProject}.
In the right graph of \autoref{fig:candidate-structure-thermo}a, we show the energy above the convex hull ($E_{hull}$) for each candidate \cite{MaterialsProject, MaterialsProject_PhaseDiagram}. Fifteen phases shown in blue (inclusive of NTP), are within 30\,meV of the convex hull. While there is no precise energy criterion to separate synthesizable and nonsynthesizable phases, previous work has indicated that many structures with $E_{hull}<30$ can be synthesized \cite{metastability1, metastability2}. One composition, \ce{Na4HfP2O9}, is on the hull. The remaining nine out of twenty-five phases with $E_{hull}>30$\,meV/atom (shown in red) are excluded from future screening as it is less likely they are synthesizable.

Na-ion conductivity for each phase is calculated from molecular dynamics simulations with machine-learned potentials created with fine-tuned CHGNet \cite{CHGNet} between $T\,=\,300$\,K to 1,000\,K. Room temperature extrapolated (RT) conductivities for (meta)stable phases with 4+ and 5+ cation substitutions are shown in the left and right panel, respectively, of \autoref{fig:candidate-structure-thermo}b. 
Calculations associated with 3+ cations, where all sodium Wyckoff sites are filled, exhibit no Na-ion hopping during the time of the simulation. For example, the probability density and number of hops for \ce{Na5AlP2O9} is shown in the Supplemental 
at 1000\,K. No hoping is observed for up to 2\,ns, indicating that it is likely an ionic insulator.
We find that 5+ cations have high predicted room temperature conductivities: 11.1\,mS/cm (\ce{Na3TaP2O9}) and 12.4\,mS/cm (\ce{Na3VP2O9}). The RT conductivites are slightly lower for the 4+ cations, between 0.01 and 6.5\,mS/cm.
%
We plot the activation energy at 600\,K versus ionic radius for 4+ cations in \autoref{fig:candidate-structure-thermo}c. The increase of activation energy with ionic radius is supportive of a possible pillaring effect, wherein structural barriers to conduction are forced open with bigger cations. Smaller ions (Fe, Ga, Mn) are calculated to have activation barriers approximately 0.1-0.2\,eV greater than the larger ions.

Finally, we apply a filter for potential synthesizability by calculating the Gibbs reaction free energies ($\Delta G_{rxn}$) to each candidate NAP phase from select precursor sets.
A negative reaction energy infers a favorable thermodynamic driving force for that precursor set to create the target composition.
We obtain suggested synthesis protocols from precursor prediction models, including Reaction Network and s4\cite{McDermott_reactionNetwork_model, TextMinedSynthesisRecipes}.
Reaction network precursor prediction model that ranks candidate precursor sets on thermodynamic driving force and thermodynamic selectivity metrics \cite{McDermott_reactionNetwork_model}.
s4 is natural language processing model that suggests synthesis recipes for a target inspired by previously reported methods found successful for chemically similar products.
From these algorithms, we find that highly ranked precursor sets always contain \ce{Na2CO3} and \ce{NH4H2PO4}, or \ce{Na4P2O7}, and a binary oxide for A as precursors (see Supplemental 
for all s4 outputs).
Therefore, we consider two general precursor sets. Precursor set (\textit{i}) consists of \ce{Na2CO3}, \ce{NH4H2PO4}, and \ce{A_{y}O_{z}}, and precursor set (\textit{ii}) contains \ce{Na4P2O7} and \ce{A_{y}O_{z}}. 
The left panel of \autoref{fig:candidate-structure-thermo}a depicts that there is a negative reaction energy for the formation of all candidate phases when \ce{Na2CO3}, \ce{NH4H2PO4}, and \ce{A_{y}O_{z}} precursors are used, supporting synthesizability. We also calculate the energy for an additional reaction of sodium pyrophosphate and binary A oxides shown in the Supplemental. 
Here, we see that only a limited number of NAP candidates have a negative reaction energy: \ce{Na4HfP2O9}($\Delta G_{rxn}$=-31\,meV/atom), \ce{Na4SnP2O9} ($\Delta G_{rxn}$=-18\,meV/atom), \ce{Na4ZrP2O9} ($\Delta G_{rxn}$=-18\,meV/atom), \ce{Na4MoP2O9} ($\Delta G_{rxn}$=-2\,meV/atom), and \ce{Na3TaP2O9} ($\Delta G_{rxn}$=-15\,meV/atom).
Therefore, we choose the first precursor set containing \ce{Na2CO3}, \ce{NH4H2PO4}, and \ce{A_{y}O_{z}} to attempt to screen potential precursors.

\subsection{Automated Synthesis Screening}

We perform experimental solid state synthesis with the self-driving, automated A-Lab \cite{Alab, Alab_os}. Reactions are attempted for all candidates with $E_{hull}<30$\,meV/atom and $\Delta G_{rxn}<$0.
Dwell temperatures in the range 600\,$^{\circ}$C through 1,100\,$^{\circ}$C are considered. Samples are characterized by powder XRD. A summary of the synthesis trial results can be found in the Supplemental. 
Syntheses are only considered successful if the candidate NAP is detected by Rietveld refinement analysis of measured XRD patterns, regardless of ultimate phase purity.

Synthesis attempts were unsuccessful for a majority of NAP phases suggested in \autoref{fig:candidate-structure-thermo}a, despite relatively low $E_{hull}$s and a negative driving force for formation of the target. 
Resulting XRD patterns for unsuccessful synthesis attempts provide clues about the reaction pathway, and possible reasons why the target was not created. 
In the cases of \ce{Na4ZrP2O9}, \ce{Na4HfP2O9}, \ce{Na4CrP2O9}, \ce{Na4MoP2O9} and \ce{Na4GeP2O9} there is no reaction between the intermediate \ce{Na4P2O7} and the binary oxide, until the maximum furnace temperature was reached or a hard, solid bulk forms. In these cases, there does not appear to be enough driving force to form the target, even when the sample melts.
For \ce{Na4FeP2O9} and \ce{Na3VP2O9} phase competition to NaAO$_x$ oxides as the final product.
In these cases, the \ce{NaFeO2} and \ce{NaVO3} impurities are both on the convex hull.
In limited cases, the A cation reacts with phosphate, as in the case of the attempted synthesis of \ce{Na4GaP2O9} and formed \ce{GaPO4}.
\ce{Na3TaP2O9} was made, but in a topologically inequivalent structure previously reported (see Supplemental) \cite{Na3TaP2O9}.
%
In particular, one interesting case was the unsuccessful synthesis attempts of thermodynamically stable \ce{Na4HfP2O9}. At maximum dwell temperature for the furnace (1,100\,$^\circ$C), only \ce{Na4P2O7} and \ce{HfO2} are formed (see Supplemental).
SEM/EDS confirms the presence of micron-sized separate phases of the precursor. This indicates that either the integration of Hf into the NAP is not favorable, or there might be a high diffusion barrier at maximum laboratory dwell temperatures within solid state conditions.
If the latter is true, some methodology changes to improve sluggish kinetics could be dwell temperature increases, precursor changes, or general synthesis methodology changes (such as use of sol-gel methods).
%
%

We successfully created one new NAP phase, \ce{Na4SnP2O9}. Refinement of XRD patterns in the supplemental indicates a significant amount of \ce{SnO2} and \ce{Na4P2O7} ($>30$ wt\%) impurities. In the subsequent sections, we optimize the synthesis of this material and characterize the structure and ionic conductivity of the phase pure sample.

\subsection{Rational Synthesis Optimization \label{sec:NSP-synthesis-optimization}}

\begin{figure}
\includegraphics[width=0.7\linewidth]{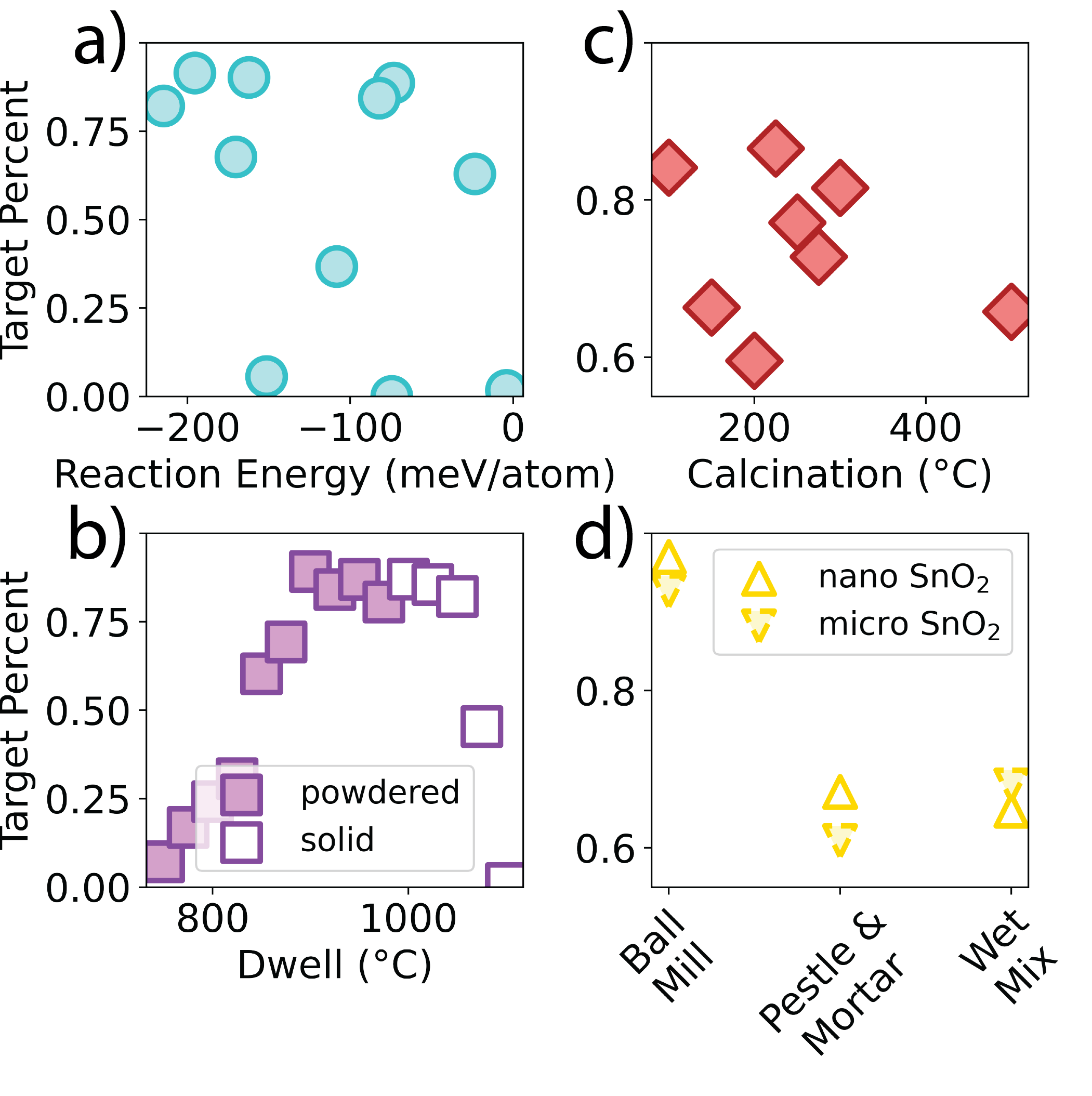}
\caption{Figure showing experimental synthesis conditions that impact \ce{Na4SnP2O9} phase purity.
a) Reaction energies of precursor sets are plotted compared to the highest phase purity recovered for that set.
b) The dwell temperature is plotted with respect to recovered phase purity for the reaction of \ce{Na2CO3},\ce{NH4H2PO4}, and \ce{SnO2}. A calcination step composed of a hold at 100\,$^\circ$C for 12 hours is completed prior to a 12 hour hold at the respective dwell temperature. Sample recovered as a powder is shown with filled squares, and hard, solid sample with low yield is visualized as empty squares.
c) The calcination temperature is plotted with respect to recovered phase purity for the reaction of \ce{Na2CO3},\ce{NH4H2PO4}, and \ce{SnO2}. Calcination is held for 12 hours before heating to the dwell temperature of 925\,$^\circ$C, which is held for 12 hours.
d) Mixing methods for \ce{Na4P2O7} and \ce{SnO2} with ball milling, pestle and mortar, and wet mixing are plotted with respect to recovered phase purity when nanometer-sized \ce{SnO2} (filled, up triangle) and micrometer-sized \ce{SnO2} (dashed, down triangle) reagent are used.
}
    \label{fig:synthesis-optimization}
\end{figure}

%
In this section, we evaluate modifications to key parts of the NSP solid state synthesis recipe in order to produce a phase pure sample of \ce{Na4SnP2O9}: (\textit{i}) precursor identity, (\textit{ii}) dwell temperature, (\textit{iii}) calcination temperature, and (\textit{iv}) mixing method in \autoref{fig:synthesis-optimization}.

\subsubsection{Precursors}
Previous literature investigating synthesis reactions has found very different reaction products and purity by varying precursor identity \cite{Matt_Selectivity_Model, piro, Tyrel_RationalSynthesis, TextMinedSynthesisRecipes}. Therefore, we first explore yield for a number of different precursor sets. To gather possible precursors, we utilize a number of different avenues, including past literature and solid state synthesis precursor prediction algorithms \cite{NTP_1, McDermott_reactionNetwork_model, Wenhao_Autunomous_Synthesis}. The full list of attempted precursor sets, their source, reaction energy, and their highest yield synthesis with wet spin mixing can be found in the Supplemental.
Due to the high number of possible precursor sets found by reaction network and s4 (greater than 30,000), we filter precursors by safety, cost, commercial availability, number of competing phases, and thermodynamic driving force. Furthermore, we did not consider routes that necessitate an additional processing step after heating, such as a washing/separation step. Therefore, it is likely that additional precursor sets exist for which NSP could be synthesized, but that have not been attempted within this study.

\autoref{fig:synthesis-optimization}a shows the phase purity, found through Rietveld Refinement, for considered routes plotted against reaction energies.
The highest yield synthesis were found to be (\textit{R1})\ce{Na2CO3}, \ce{NH4H2PO4}, \ce{SnO2} ($\Delta G_{rxn}$=-154.9/,meV/atom); (\textit{R8}) \ce{Na2HPO4}, \ce{Sn} ($\Delta G_{rxn}$=-204.5\,meV/atom); (\textit{R9}) \ce{Na4P2O7}, \ce{SnO2} ($\Delta G_{rxn}$=-28.4\,meV/atom). Reactions found to have little or no yield include (\textit{R2}) \ce{Na2CO3}, \ce{NH4H2PO4}, \ce{Sn} ($\Delta G_{rxn}$=-239.8\,meV/atom); (\textit{R11}) \ce{Na2SnO3}, \ce{(NaPO3)6} ($\Delta G_{rxn}$=-112.8\,meV/atom). In particular, it was surprising to find such low yield for \textit{R11}, given that it was predicted to a successful synthesis route in the Synthesis Planning Algorithm based on its inverse hull metric \cite{Wenhao_Autunomous_Synthesis}.
We observe little correspondence between reaction energy and yield (\autoref{fig:synthesis-optimization}a). This is important to note, as reaction energies often serve as the basis for previously published thermodynamic models for solid state synthesis.
Of all reactions, those that best combine reaction phase purity and precursor accessibility is the $R1$ set: the common precursors \ce{Na2CO3}, \ce{NH4H2PO4}, \ce{SnO2}. We therefore next explore if temperature profile, including calcination and dwell temperatures, can be modified to increase phase purity.

\subsubsection{Maximum Dwell Temperature \label{sec:dwell-temp}}

%
In \autoref{fig:synthesis-optimization}b, we show the outcome of phase purity (target percent) with varying dwell temperatures and for a 12-hour hold time for the reaction of \ce{Na2CO3}, \ce{NH4H2PO4}, \ce{SnO2}, prepared via wet spin mixing. Powdered final product is indicated by filled-in purple squares. Samples with low weight recovery due to mechanically tough bulk products are indicated by empty purple squares. We found that purity increases more or less linearly with dwell temperatures between 750\,$^\circ$C to 900\,$^\circ$C. No product is found below 750\,$^\circ$C. Between 900\,$^\circ$C to 1000\,$^\circ$C, the target percent is stabilized near 85\%. The target percentage recovered stays the same between 1,000\,$^\circ$C to 1,050, however, very little mass is recovered due to the formation of hard, bulk solid sample. Finally, at a dwell temperature of 1,100\,$^\circ$C, only impurity (\ce{Na4P2O7} and \ce{SnO2}) is recovered from the hard sample. The low weight recovery and phase impurity at high temperatures correspond well with the 1,086\,$^\circ$C sample melting point measured by thermogravimetric analysis.
We therefore argue that reaction temperatures between 900,$^\circ$C to just below 1,000,$^\circ$C are expected to lead to maximum phase purity for the $R1$ precursor set. However, optimization of the dwell temperature alone is not enough to generate acceptable phase purity for property testing.

\subsubsection{Calcination Temperature}

We explore impact of a low temperature calcination on target recovery for NSP. Low temperature calcination is often reported within solid state synthesis methods, and is integrated into the default temperature profiles for previously reported self-driving labs \cite{Alab}. We subject precursor set $R1$ (\ce{Na2CO3}, \ce{NH4H2PO4}, \ce{SnO2}) to a range of low temperature holds for 12 hours, before increasing the temperature to 925\,$^\circ$C for an additional 12 hour dwell step.
\autoref{fig:synthesis-optimization}c shows the recovered target percentage when calcination temperatures between 80-500\,$^\circ$C are applied. Significant variation in phase purity is observed. Calcination between 80-150\,$^\circ$C leads to phase purity $>85\%$. When the calcination temperature is increased to 200-275\,$^\circ$C, the phase purity decreases to $<$70\%. The phase purity goes back up to above 80\% near 300\,$^\circ$C, and then at temperatures above 500\,$^\circ$ decreases again. These results suggest a strong impact of calcination temperature on target phase purity. \textit{In situ} XRD and \autoref{sec:dwell-temp} indicate that the target should not form before approximately 750\,$^\circ$C. Therefore the calcination temperature is likely changing the reaction pathway or sample conditions before the target forming step.

To better understand how the calcination temperature influences target fraction, we explore the reaction pathway for R1, and the likely changes to it from heating profile variance.
We use the pairwise reaction scheme recently evaluated by Szymanski et al. \cite{Pairwise}, which breaks a synthesis pathway down to pairwise reactions between precursors or intermediates (see Supplemental). By evaluating the compositionally unconstrained reaction energy \cite{MaterialsProject_ReactionInterface1, MaterialsProject_ReactionInterface2} between all pairs of precursors, we find that \ce{Na2CO3} and \ce{NH4H2PO4} have the highest driving force to react, leading to the assumption that the first phase to form in the reaction pathway should be a sodium phosphate. Previously, V.A. Nicholas, et al. \cite{Sodium_Phosphate_Experimental} found that the product of pairwise reactions of ammonium phosphate and sodium carbonate is temperature-dependent. We therefore completed a number of pairwise reactions between \ce{Na2CO3} and \ce{NH4H2PO4} (see Supplemental)
and find that between 100-500\,$^\circ$C, a complex set of sodium phosphate phases may form, including \ce{Na3PO4} at low temperatures, and higher linked phosphates such as sodium pyrophosphate (linked \ce{PO4} pairs), and the triply corner-connected \ce{PO4} containing sodium tripolyphosphate polymer \ce{Na5P3O10} at T\,$\geq$\,300\,$^\circ$C.

We hypothesize that lower yield may occur for reaction pathways containing phases with a higher degree of phosphate polymerization due to strong P-O-P bonds.
The NAP structure contains only isolated orthophosphate groups, meaning that condensed polyphosphates must fully decompose to make the final target.
Polyphosphates may be introduced through precursor (\textit{e.g} \ce{Na5P3O10}) or possibly through tuning of the reaction pathway.
Variation in heating profile, such as the introduction of a calcination step, could present one way to tune the reaction pathway.
The Supplemental 
provides evidence that low temperature holds can modify the outcome of sodium and phosphate precursor reactions.
We completed reactions at 80\,$^\circ$C, 150\,$^\circ$C, 300\,$^\circ$C, and 500\,$^\circ$C with different sodium and phosphate sources also used in this study. We found that while \ce{Na4P2O7} was the most common phase to form at a large temperature range, higher hold temperatures at 500\,$^\circ$C were able to synthesize sodium tripolyphosphate.
While high yeild NSP can be created through the direct reaction of \ce{Na4P2O7} (cornersharing PO4 tetrahedra) or \ce{Na2HPO4} (isolated PO4 tetrahedra) with \ce{SnO2}, utilization of \ce{Na5P3O10} (mass balanced with additional \ce{Na2CO3}), target yield is nearly undetectable $>$1\% (see Supplemental).
More work is necessary to understand the impact on solid state synthesis of molecular units within crystalline solids used as precursors and found to be intermediate phases.

\subsubsection{Precursor Preparation}

\begin{figure}
    \centering
    \includegraphics[width=1.0\linewidth]{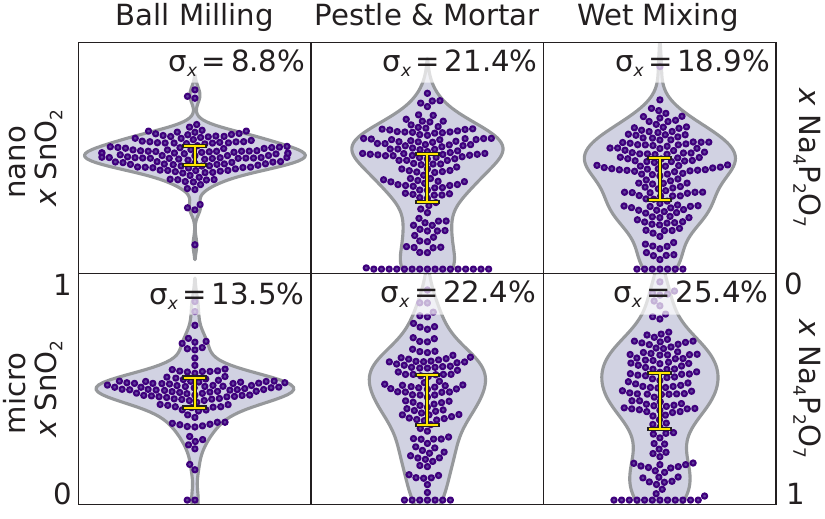}
    \caption{Violin plots illustrate the distributions of SEM EDS composition measurements on as-prepared precursor mixes, as a function of mixing method and \ce{SnO2} precursor size \cite{seaborn}
    Each purple data point represents the molar fractions ($x$, reported on the y-axes) of the \ce{Na4P2O7} and \ce{SnO2} precursors obtained from a point spectrum randomly collected in the sample.
    The molar fractions are normalized, such that $x_{\ce{Na4P2O7}} + x_{\ce{SnO2}}= 1$.
    The spread of data points along the x-axis, combined with the shape of the violin plot, provides a visual representation of the density of points for a given $x$ value.
    In each panel, we report the standard deviation of molar fractions ($\sigma_x$), also represented with a yellow bar.
    }
    \label{fig:eds-mixing-analysis}
\end{figure}

We investigate the effect of utilizing ball milling, pestle and mortar, or wet (spin) mixing to prepare precursors on final phase purity of NSP. Additionally, we explore the effect of $<$100 nanometers and 325 mesh (micrometer diameter) \ce{SnO2} particle size on target purity. To avoid excessive stickiness and ammonia gas release during ball milling, R1 set (\ce{Na2CO3}, \ce{NH4H2PO4}, \ce{SnO2}) is substituted for the similar phase purity ($>$70\%) R9 set (\ce{Na4P2O7} and \ce{SnO2}).

\autoref{fig:synthesis-optimization}d shows the recovered target phase purity with different precursor preparation methods. Regardless of the use of micrometer or nanometer tin oxide, wet mixing and grinding by pestle and mortar both exhibit phase purities below 70 wt\% by Rietveld refinement of XRD. However, when precursors are ball milled together, we are able to consistently achieve phase purity over 90\%. When utilizing the ball milled precursors and $<$100\,nm nanoparticle \ce{SnO2}, XRD impurity peaks are small or almost undetectable.

To better understand the difference between the precursor preparation methods, SEM EDS characterization is performed on the as-prepared precursor mixes. Visual analysis, automated particle size analysis, and particle composition  analysis is employed for each set of precursors and mixing methods.
\autoref{fig:eds-mixing-analysis} shows a set of violin plots illustrating the distributions of composition measurements obtained from individual SEM EDS spot spectra randomly collected across tens of different particles in each sample \cite{seaborn}.
The y-axes quantify the molar fractions ($x$) of \ce{Na4P2O7} and \ce{SnO2} precursors obtained from each spot spectrum.
We note that the measured molar fractions are normalized such that $x_{\ce{Na4P2O7}} + x_{\ce{SnO2}}= 1$.
The spread of data points along the x-axes, together with the shapes of the violin profiles, give a visual representation of the density of points measured for a given $x$.
Within each panel, the standard deviation of the precursor molar fractions ($\sigma_x$) is reported and visually represented by the yellow bars.
We emphasize that the composition measured via EDS spectrum quantification reflects the average composition of the material within the electron-beam interaction volume, with an approximate diameter of a few micrometers.
Hence, analyzing the distribution of measured compositions provides insight into the homogeneity of intermixing between the two precursors.
Specifically, in the case of ideal precursor intermixing at the nanometer level, the measured compositions would always be $x_{\ce{Na4P2O7}} = x_{\ce{SnO2}}= 0.5$.
In contrast, with no intermixing and large precursor particles, the values of $x_{\ce{SnO2}}$ would be either 0 or 1 (and vice versa for $x_{\ce{Na4P2O7}}$).
In other words, as the degree of intermixing decreases, the dispersion of measured compositions will increase.
Therefore, the values of $\sigma_x$, which quantify the spread in measured molar fractions, can be used to quantitatively compare the intermixing homogeneity of the precursors across different samples.
%

\autoref{fig:eds-mixing-analysis} shows that ball milling yields the smallest values of $\sigma_x$, with 8.8\% for nano-\ce{SnO2} and 13.5\% for micron-\ce{SnO2}, indicating improved precursor intermixing with respect to the two other methods employed, whose $\sigma_x\geq$18.9\%.
In addition, the measured values of $\sigma_x$ also suggest that smaller tin oxide particles lead to improved precursor mixing and homogeneity. For instance, when wet mixing is performed, the sample with micrometer \ce{SnO2} has a $\sigma_x$ of 25.4\%, while $\sigma_x$ decreases to 18.9\% when utilizing nanometer \ce{SnO2}.
We find wet spin mixing to produce an ethanol slurry returns minimally lower particle homogeneity than a traditional pestle and mortar grind when utilizing nano-\ce{SnO2} (versus pestle and mortar mixing with micro-
\ce{SnO2}).
Finally, we highlight that these trends of precursor intermixing are visually confirmed by collecting EDS maps across the different samples (see Suppplemental). Ball-milled samples qualitatively show homogeneous composition, while single \ce{SnO2} particles could be clearly observed with the other two methods.

We also explore particle size distributions in the samples. Particles sizes were measured automatically across pin stubs within the SEM with in-house software. Due to the large range of particle sizes and the limited SEM imaging resolution, particle sizes were binned into size categories of small (0.8-15\,$\mu$m), medium (15-100\,$\mu$m), and large (100-700\,$\mu$m). Up to 400 particles were counted for the small particle size, and then 100 particles each were measured for the medium and large sizes. 
The Supplemental shows particle sizes documented for each scale category. At the small size scale, we see little difference in particle size distribution for samples created with different mixing methods. Medium and larger size particles are predominantly observed in the ball-milled sample with nanoparticle tin oxide, and is corroborated by the SEM observation shown in the Supplemental. 
Given that significant particle mixing is found in \autoref{fig:eds-mixing-analysis}, we hypothesize that precursor size is initially reduced with all mixing methods, including ball milling. Any large particles might be made via assembly of small particles into larger aggregates, such as those seen in the Supplemental. 

Taken together, the panels in \autoref{fig:synthesis-optimization} demonstrate that multiple factors play a role in ultimate sample phase purity. Careful selection of precursors, their mixing, and optimization of temperature profile are required to yield the phase pure NSP sample.
Precursor selection was found to vary ultimate phase purity from approximately 0 to 80\%.
Maximum dwell temperature demonstrated a window for ultimate phase purity of ranging from 875 to 975\,$^\circ$C.
Investigation of precursor preparation and starting particle size reveal a strong correlation between mixing methodology and phase purity, with only ball milling and nano-\ce{SnO2} yielding a phase pure sample.
SEM EDS analysis suggests that ball milling is most effective at creating a homogeneous precursor mix. Reduction of particle size, as observed with SEM imaging, is not directly found with ball milling. In fact, particles with $\gg$5\,$\mu$m radius are primarily found in ball milled samples, possibly suggesting particle reduction, mixing, and subsequent aggregation (see Supplemental).
%

\subsection{NSP Discovery and Property Characterization \label{sec:nsp-characterization}}

\begin{figure*}
    \centering
    \includegraphics[width=1.0\linewidth]{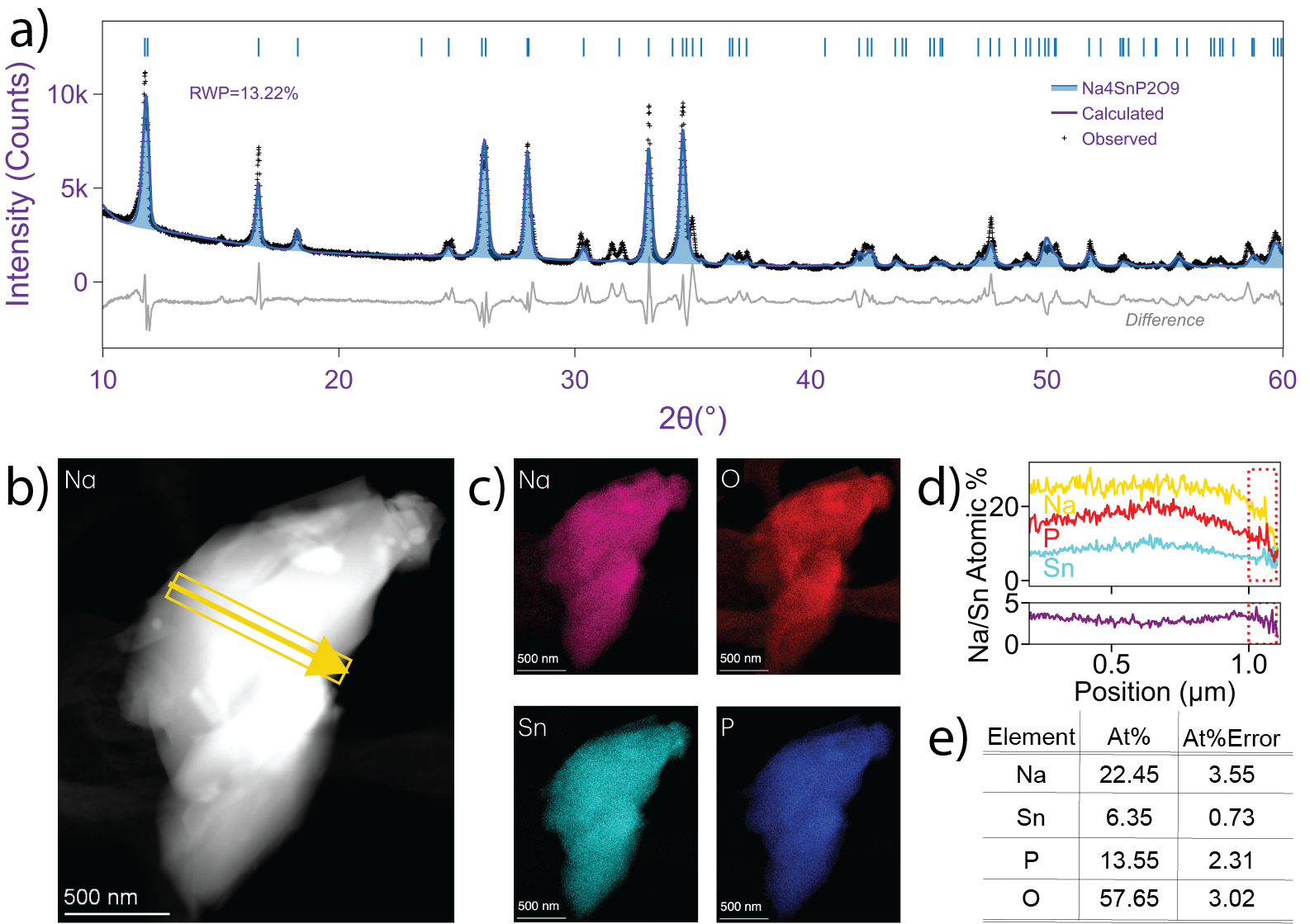}
    \caption{Characterization of NSP.
    (a) powder XRD. Curves show the observed pattern, refinement for \ce{Na4SnP2O9}, and the difference. Peak positions and RWP are included for reference.
    (b) HAADF-STEM image of a representative NAP particle from which EDS data is collected.
    (c) Cryo-EDS Sn, Na, O, and P intensity maps.
    (d) A compositional line profile of Na, P, and Sn composition along the trajectory in (b). The Sn composition remains nearly uniform across the bulk, while the Na and P composition decreases close to the particle surface.
    (e) The elemental ratios from line scan, where quantification procedure is described in \autoref{sec:methods}.}
    \label{fig:NSP-struture-microscopy-characterization}
\end{figure*}

By optimizing the synthesis condition of NSP, the maximum purity \ce{Na4SnP2O9} was obtained by solid state synthesis. Starting from high energy balled milled nano \ce{SnO2} and dried \ce{Na4P2O7}, the mix was pelletized and heated to 950$^\circ$C for 12 hours. \autoref{fig:NSP-struture-microscopy-characterization} shows characterization of the sample with multiple techniques.

%
\autoref{fig:NSP-struture-microscopy-characterization}a shows the powder XRD obtained from 2 hour scan on the Aeris Minerals diffractometer (Malvern Panalytical) utilizing Cu K$\alpha$ radiation for relatively phase pure NSP.
Peak matching is first pursued to find a structure solution.
We leverage the data-driven search algorithm Dara \cite{Dara} to complete phase search with and without the NSP target.
Without using the target phase for the refinement, the DARA phase search results in two phases which are synthesized at high-pressure or reducing conditions, and are thus unlikely to be stabilized at the synthesis conditions. Using those phases, the Rwp is 24\% and more than 75\% of the peaks are unmatched.
Leveraging DARA with the target available yields the NSP Orthorhombic phase with Rwp\,=\,13.22\%.
The majority of peaks are matched to this reference structure with some small unidentifiable peaks.

A more detailed structural refinement search was completed manually for the long scan pattern in \autoref{fig:NSP-struture-microscopy-characterization}a. Various structural solutions were attempted, including tin replacement for titanium into the experimental NTP monoclinic structure \cite{NTP_1}, distinct sodium orderings and occupations of Sn-substituted HT-NTP, and phonon modulated structures of Sn-substituted HT-NTP (see Supplemental). 
Relatively high Rwp figures of merit were recovered for all. The best structure solution comes from the tin substitution for titanium in the orthorhombic HT-NTP prototype.
However, some mismatch between the observed pattern and calculated pattern are found at unmatched peaks at about 15$^\circ$, 31.5$^\circ$, and at $2\theta>50^\circ$. Additionally, differences in intensity are still found for major peaks, as clearly seen in the gray difference curve.
However, most peaks are shown to be fit well, including major peaks, with minimal lattice and particle microstrain.
Therefore, the diffraction data, while not definitive, suggest the assumption that the target orthorhombic \textit{Bmem} NSP phase was made, with possible distortions, is valid.

To evaluate the possibility of a temperature-driven structural transition, powder XRD was performed between 100 to 800\,$^\circ$C (shown in the Supplemental).
Based on this set of scans, we conclude that two HT-NSP phases are likely possible. We find from changes in peak location, shape, and splitting that a high temperature \ce{Na4SnP2O9} phase might be stable from approximately 400 to 650\,$^\circ$C, which would transform again at $T \approx 650\,^\circ$C.
Attempts to find structural models for both high temperature phases with Dara, and with modifications to the orthorhombic NSP and experimental NTP structures were unsuccessful (see Supplemental for more details). 
Furthermore, we find that the unfitted peaks at room temperature, located at 15$^\circ$ and 31$^\circ$\,$2\theta$, 
disappear and then reappear upon cooling, indicating that they are likely associated with a symmetry-breaking structural distortion in the current NSP structure solution rather than with impurities.
As such, further structure refinement for NSP, both for ambient and high temperature polymorphs, is necessary to better resolve the phases presented in this study.

Microscopy on the sample was completed on the micrometer scale (SEM-EDS, visualized in Supplemental)
and nanometer scale (HAADF-STEM, \autoref{fig:NSP-struture-microscopy-characterization}b). The EDS profile was measured across the depicted arrow in \autoref{fig:NSP-struture-microscopy-characterization}b and is presented in \autoref{fig:NSP-struture-microscopy-characterization}d.
The EDS map and line scan reveal a homogenous sample with a Na/Sn ratio of approximately 4.5\,Na:1\,Sn within the bulk sample.
However, within approximately 100\,nm of the surface, the Na and P content decreases considerably, whereas the Sn concentration remains nearly uniform. Therefore, there is some indication that as-synthesized NSP particles could have a tin rich surface.
Overall, the characterization results from STEM support the synthesis of a new phase with bulk composition, approximately \ce{Na4SnP2O9}, with some Na deficiency of the particles' surface.

\begin{figure}
    \centering
    \includegraphics[width=1.0\linewidth]{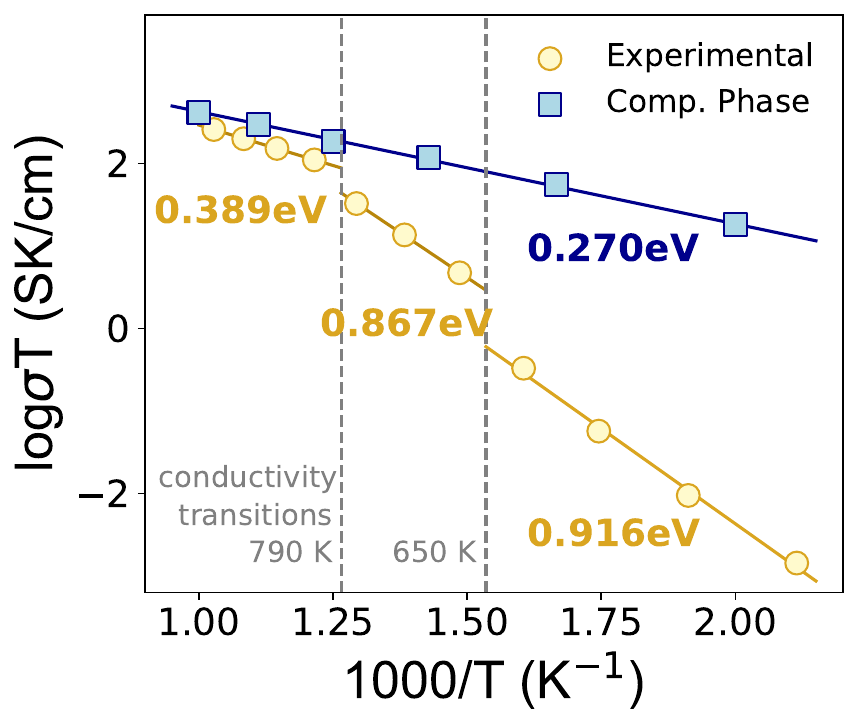}
    \caption{Experimentally measured and computationally predicted conductivity for \ce{Na4SnP2O9}.}
    \label{fig:nsp-conductivity}
\end{figure}

Pellets of the nearly phase pure NSP were conventionally sintered, achieving greater than 90\% density with no indication of sodium loss.
Ambient EIS demonstrates near insulator-like behavior at room temperature ($\sigma \approx 10^{-7}$\,S/cm). We note that likely water absorption on the pellet surface, leading to possible proton conduction, necessitates an air free environment below 100\,$^\circ$C for accurate measurements (see Supplemental).
\autoref{fig:nsp-conductivity} shows the experimentally measured and computationally calculated activation energies for NSP over the temperature range of 150\,$^\circ$C to 850\,$^\circ$C.
The computational activation energy is relatively low (0.27\,eV) (blue), consistent with the high predicted room temperature conductivity (1\,mS/cm) from AIMD calculations.
The experimentally measured activation energy is much higher with a value of 0.916\,eV between room temperature and 650\,$^\circ$C.
Moreover, two conductivity transitions are apparent in experimental EIS. Between 650\,$^\circ$C and 800\,$^\circ$C, $E_a$\,=\,0.867\,eV, and above 800\,$^\circ$C the activation energy drops considerably to 0.389\,eV.
Both the XRD and the disagreement between the computational and experimentally activation energies hint at the likelihood that the low temperature structure model for NSP is not sufficiently accurate.

\subsection{Conductivity Model of NAP}

\begin{figure}
    \centering
    \includegraphics[width=0.3\linewidth]{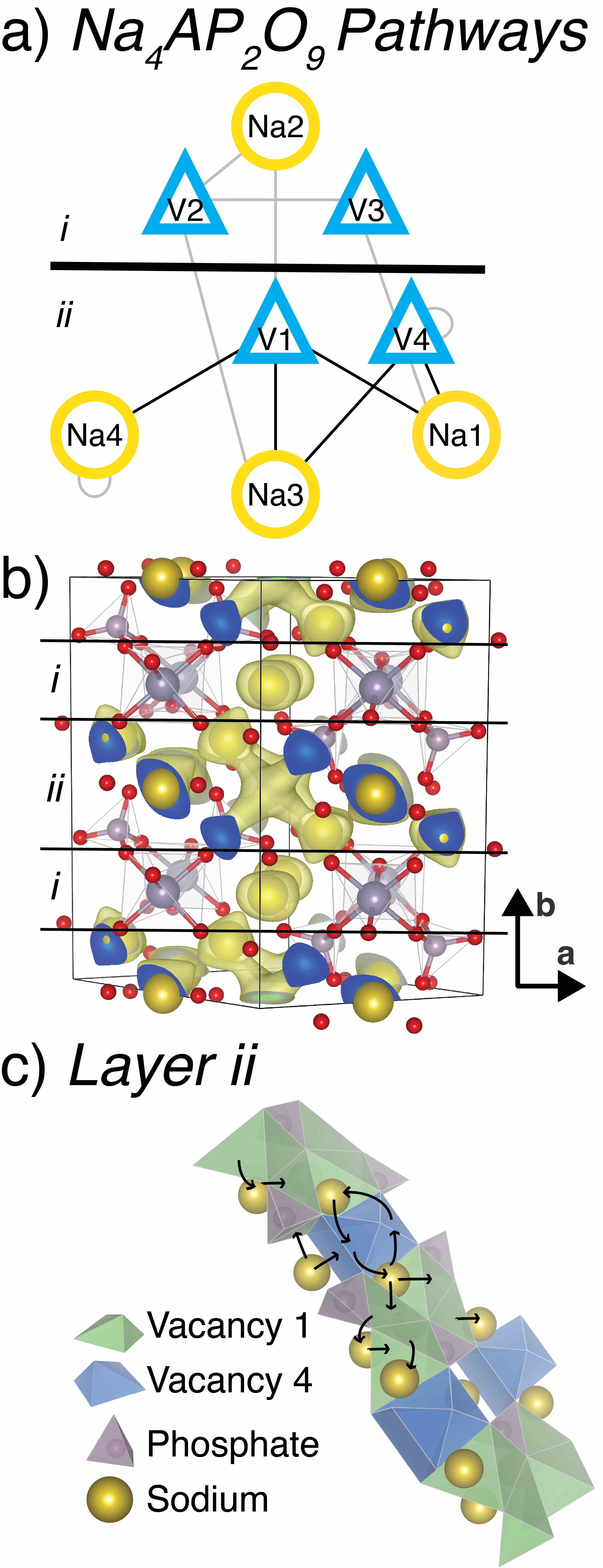}
    \caption{Ionic conductivity behavior of \ce{Na4AP2O9} phases. (a) Schematic of possible symmetrically inequivalent conduction pathways sodium could taken through a parent NAP structure. Lines connect vacancies and filled sodium sites  with face-sharing polyhedra. Solid black lines represent paths found likely to be taken based on MD calculations on NTP, while grey lines show unused routes. (b) Visualized probability density of \ce{Na4TiP2O9} during a 1\,ps molecular dynamics run at 600\,K. Structural layers \textit{i} and \textit{ii} are annotated.  (c) Model of structural layer \textit{ii} showing the location of symmetrically inequivalent vacant polyhedra, vacancy 1 (green) and vacancy 4 (blue). The path sodium is likely to use based on face sharing polyhedra and paths taken in molecular dynamics calculations are shown with black arrows.}
    \label{fig:nap-conduction}
\end{figure}

Experimentally observed NAP phases (\ce{Na4TiP2O9}, \ce{Na4SnP2O9},  \ce{Na_{4.6}FeP2O_{8.6}F_{0.4}}, and \ce{Na{(Al,Cr)}P_2O_{9-x}F_{2x}}) are superionic conductors at elevated temperatures. The coupled structural-conductivity transition for all but NSP manifests between 245-270\,$^\circ$C. Below the transition temperature, all NAP materials have measured conductivities of $\approx$10$^{-7}$\,S/cm \cite{NAP_1, NTP_3_and_NFP, NTP_1}. Measurements from NTP and NSP at room temperature show activation energies greater than 800\,meV.

To better understand how sodium ions move through an NAP crystal, we enumerate possible conductivity pathways.
A pathway is defined as a shared polyhdra face between sodium or vacant sites.
\autoref{fig:nsp-conductivity}a shows all possible sodium paths through orthorhombic NTP (see Supplemental 
for name and visualization of all vacancy locations).
It should be noted the size and connectivity of vacancies may vary slightly for higher temperature phases due to bond length, sodium content, and for polyhedra distortion for other NAP structures.
For all four sodium sites, there exist multiple possible conductivity pathways through the structure.
All sodium sites are connected to at least two different vacant interstitial sites via polyhedra face-sharing pathways except the Na4 site, which only shares a polyhedra face with one vacancy (V1).
In particular, the V1 vacancy in layer \textit{ii} is connected to all four sodium sites.
Movement between layers includes pathways through either the V1, V2, or V3 vacancies.

\autoref{fig:nsp-conductivity}b visualizes the probability density of sodium during a 1\,ps molecular dynamics run at 600\,K for NTP. Layers \textit{i} and \textit{ii} in the a-c plane are marked with black lines.
From the probability map, we see that there is limited probability for sodium to exist in between the \textit{i} and \textit{ii} layers in the structure.
The only sodium site in layer \textit{i}, Na2, are also isolated.
Based upon MD location probabilities, the majority of sodium conduction is expected to occur in layer \textit{ii} between Na1, Na3, and Na4.
The hops between sites in layer \textit{ii}, as read from the probability diagram, are visualized in \autoref{fig:nsp-conductivity}c as black arrows. Here, we see that hopping occurs between sodium sites and through largely triangular and rectangular polyhedral faces. As sodium hops between sites, it will move between either a V1 or V4 vacancy.
Mean squared displacements (MSD) 
over 2\,ns for AIMD calculations of four NAP candidates further shows directional hopping: MSD ranges from 0 to 50\,\AA for [010] for all candidates, and 75-200\,\AA in [100] and [001] (see Supplemental).
Therefore, based on these calculations, the NAP framework is hypothesized to be primarily a two dimensional conductor on the a-c plane.

We introduce the possibility of vacancy mediated conduction within NAP \cite{Vacancy_Mediated}. \autoref{sec:high-throughput-screen} points out that with 3+ cation substitution, sodium content is at 5 and every sodium site is full. For all compositions tested, no hopping was found at elevated temperatures. Alternatively, cation substitution with 5+ oxidation states and a sodium content of 3 recovered the highest predicted conductivities. Moreover, conductivity pathways were found to change slightly for 4+ substitution. For instance, calculations on \ce{Na3VP2O9} at 700\,K find significant conductivity in [010]. Very low barrier and MSD is found in [001] for \ce{Na3TaP2O9} at 700\,K at the expense of conduction in [100].


Interestingly, AIMD calculations simulated using the high-temperature NAP structure does not exhibit any phase transition nor non-Arrhenius behavior, in contrast to our experimental measurements. Calculations performed on NTP show more or less the same conductivity for the low temperature monoclinic phase as the high temperature orthorhombic phase.
Experimentally measured ionic conductivities of the higher temperature phases are approximately two orders of magnitude higher than the lower temperature phases.
Assuming pure phase NSP sample was synthesized, there are two possible outcomes: (\textit{i}) the structures investigated are not physical, and better structure resolution is necessary. In this case, once a better structure solution is found, AIMD calculations would recover a much lower conductivity for sodium. On the other hand (\textit{ii}) the mechanism preventing the conductivity is not captured by AIMD calculations, and some other mechanism, such as grain boundary resistance or anti-collaborative movement of the sodium ions, is inhibiting conductivity. Sintering technique and the observed tin rich surface on NSP could also play a role.
Due to the numerous possible sources, future work to uncover the origin of the discrepancies is necessary.

\section{Discussion}
In this article, we presented a survey of the NAP superionic conductor framework, including the key structural components of the parent NTP, the thermodynamic behavior and synthesizability of candidate cation substitutions, and characterization of the novel \ce{Na4SnP2O9}. The extent of methods used here is meant to ensure a baseline of knowledge for any researcher interested in the NAP framework.
Computational work in \autoref{sec:high-throughput-screen} demonstrates NAP's flexibility, which is predicted accommodate cations of different sizes and charges through structural distortions and varying sodium concentration.
Experimental work underscores challenges with synthesis when employing traditional solid state methods for both phase discovery through cation substitution, and when attempting to recover phase pure materials.
Our work highlights the NAP framework as one that has remained under-published since it was reported on decades ago. The NTP framework itself is promising: it consists of commercially available and cheap materials, including common transition metal and post transition metal elements, sodium, and phosphate. Clear three dimensional percolation pathways through the structure foreshadow a general promise of high ionic conductivity.
Thermodynamic analysis suggests that many other materials within this family could be discovered. Many rest in relatively untouched areas of their convex hulls, and still others have large driving forces despite no successful synthesis reported here. We attribute some of these dependencies to experimental limitations. Other labs with the capability to utilize methods with more intimate precursor mixing, such as hydrothermal or sol-gel methods, may encounter these exciting new phases.
Through this survey, state-of-the art tools highlight the pervasive barriers to new materials development ranging from synthesizability and property prediction to experimental characterization.

The focal point of our work is the discovery of new structural frameworks for sodium ion conductors.
Despite decades of efforts to develop all solid state batteries, very few are commercially available in part due to the challenge of developing effective solid state electrolytes. While high ionic conductivity is absolutely necessary for a solid state electrolyte, there are other engineering constraints: densification, electrochemical stability, air stability, and materials and processing costs are primary considerations \cite{SSE_practical_considerations_1, SSE_practical_considerations_2}.
Identifying materials that fulfill all SSE requirements for solid state batteries necessitates high throughput materials screening and targeted synthesis efforts that eclipse a single research project.

There are two primary approaches for discovery of solid state ion conductors: the discovery of new structural frameworks, and the discovery of new phases within a known framework type.
Rational discovery of the former often relies on executing upon design rules for high conductivity \cite{intro_LiConduct_Physics5, intro_LiConduct_Physics6, intro_LiConduct_Physics8}, while the latter can be iterated with computational or experimental high throughput screening.
Thus, while more materials can be found through chemical modification of known frameworks, the discovery of new frameworks is a likely prerequisite to disruptive large scale discovery of new conductive phases.
Furthermore, new frameworks provide another physical reference for structures and chemistries that can support fast ion mobility for continued AI models in the battery field.
In lithium ion electrolytes, design rules are better established, including bottleneck size and rational introduction of lithium vacancies \cite{intro_LiConduct_Physics1, intro_LiConduct_Physics2, intro_LiConduct_Physics3, intro_LiConduct_Physics4, intro_LiConduct_ML2, intro_LiConduct_ML3}. Fundamental key structural and compositional descriptors are not as well understood for sodium ion electrolytes due to limited number of studies that focus on a much smaller number of materials \cite{Na_SSE_1, Na_SSE_2, Na_SSE_3, Na_SSE_4, Design_Na_SSE_1, Design_Na_SSE_2}.

Our computational work on NAP materials highlights many descriptors that overlap with known sodium SSEs. For instance, many of the main structural features in NASICON attributed to conductivity include a high, flexible Na$^+$/V$_{Na+}$ ratio, sodium ordering, polyhedra corner-sharing known to be favorable for Li-ion conductors, and large, triangular bottlenecks that allow for an efficient 3D percolation pathway \cite{Design_NASICON_SSE_1, Design_NASICON_SSE_2, intro_LiConduct_Physics5}.
Studies by S. Wang \textit{et al.} \cite{Design_Na_SSE_1} and D. Park, \textit{et al.} \cite{Design_Na_SSE_2} assert common features of Na SSEs include high coordination number, weak Na bonding, minimal electronegativity differences between ions, distinct Na$^+$ channels, and face sharing sodium sites, as these generally allow for larger bottlenecks within a conductivity channel.
Our NAP framework shares many of these features. All four sodium sites have coordination numbers $\geq$5. Visualization of diffusion pathways in MD calculations reveal spacious and distinct Na-ion channels in structural layer \textit{ii}. The flexible sodium content and layer size, shown from the large number of cations that can be substituted into the A site, reveal tunability in the system. While calculations with 4+ cations suggest that NAP is a 2D conductor, some structural and sodium content modifications support 3D conduction through the Na2 site. Within the classes of NAP highlighted here, 5+ cation substituted phases (\ce{Na3VP2O9} and \ce{Na3TaP2O9}) show particular promise; both are three-dimensional conductors at higher temperatures and have room temperature conductivities on the order of 10\,mS/cm. The high predicted conductivities for compositions with lower sodium content point to vacancy-mediated ion mobility and vacancy-dependent percolation pathways \cite{Vacancy_Mediated}, which also suggest the possibility for partial substitution of known 4+ phases (NTP/NSP) for sodium content tuning.

Further challenges highlighted within the work were that of new synthesis method discovery and optimization. Specifically, there are two primary challenges that must be addressed by synthesis science to disruptively accelerate new materials discovery.
Computational models must be able to better filter which materials are synthesizable. This means high accuracy for predictions on whether materials can be made and when materials cannot be made to minimize false positives. Significant experimental resources are often dedicated to make predicted materials, particularly those that are metastable, which ultimately may not be able to be synthesized.
Additionally, synthesis science should be better at understanding the effect of synthesis conditions, and what synthesis conditions are optimal for phase pure sample creation. When synthesizing materials for application-driven research, high purity may be necessary for property characterization.
Innovative approaches to synthesis methodology optimization include the integration of active learning models, particularly with self-driving laboratories \cite{Alab, Wenhao_Autunomous_Synthesis, ActiveLearning_SDL}, which are uniquely positioned to be able to quickly, reproducibly, and precisely optimize synthesis method and sample purity for a wide range of experimental conditions.

Despite these calls to action, \autoref{sec:NSP-synthesis-optimization} underscores challenges in synthesis methodology optimization and prediction of successful experimental conditions \textit{a priori}.
For instance, the change in temperature profile during heating has a large effect on \ce{Na4SnP2O9} phase purity. Dwell temperatures are required to be within a limited range ($\approx$\,875-975\,$^\circ$C) for maximum target yield. Even more complicated is a low temperature hold (calcination) step. Within a 500\,$^\circ$C window, the phase purity ranged from 59\% to 89\% without demonstrating obvious trends. It is hypothesized target yield variation is due to the low temperature reaction of sodium carbonate and ammonium phosphate to form sodium polyphosphate phases. Previous research has shown \ce{Na}-\ce{PO4} reactions to be highly temperature-dependent \cite{Pyrophosphate_bonding}. The formation of different phases would lead to different local bonding environments, including with isolated orthophosphate polyanions, as in NAP, or polyphosphate, whose strong P-O-P bonds are projected to have an impact on efficient target phase formation. While the early reaction of sodium phosphate has been shown to modify NAP reaction pathways, other solid state syntheses may experience diverse behavior from other molecular units within solids.

We also spotlight the effect of processing method on powdered precursors within NSP synthesis optimization. In our work, the soft \ce{Na4P2O7} and \ce{SnO2} precursors are shown to mix more homogenously when ball milled compared to other mixing methods. Due to the lack of phase purity reached by simple re-heat and re-grind methods or increasing dwell temperature, and the relatively low driving force $\approx$18\,meV/atom, it is projected that this reaction is diffusion limited. Therefore, it is comprehensible that broken-down, well-mixed precursors assembled and pressed with milling into larger particles are likely to yield high phase purity. We hypothesize that for other systems with sluggish kinetics, simple decreases in measured particle size could be insufficient to ensure proper particle contact. However, mechanical properties of precursors, as well as reaction pathway, are sure to modify this rule of thumb.
%

Certainly, some unsettled questions in this study have their place in a discussion, and demonstrate the challenges of high throughput and multi-material phase investigations.
In particular, a lack of completed structural model to refine XRD patterns demonstrates the challenges associated with new materials characterization. Every attempt by the authors was made to find more physical solutions: manual refinement and direct structure/occupation modifications, search tree algorithms to identify reported phases, and modulation of unstable lattice modes. Non-ambient XRD shows disappearance of unmatched peaks at high temperatures, followed by recovery of the original XRD patterns upon cooling, supporting these peaks from the NSP phase rather than unmatched impurities. Additionally, microscopy and spectroscopy analyses support recovered samples as homogeneous, with no observed impurities. While our combined techniques supports the hypothesis that \ce{Na4SnP2O9} was made, more characterization is essential.
We also highlight the discrepancy between computationally reported and experimentally measured conductivities for both NTP and NSP systems. From experience, we believe the likely origins include inaccurate structural models in \autoref{sec:nsp-characterization} for NSP, insufficient densification from sintering \cite{sintering_densification_1, sintering_densification_2}, and possible tin rich particle surfaces as measured by TEM-EDS \cite{decreased_surface_conductivity_1, decreased_surface_conductivity_2, decreased_surface_conductivity_3}. However, while incomplete structure solutions may play a role in computational-experimental conductivity discrepancies for NSP, the NTP structure solution for HT and LT polymorphs is well understood \cite{NTP_1}.
Despite this, our high throughput investigation consistently yields high sodium mobility throughout 2D percolation pathways for 4+, and some 1D, 2D, and 3D pathways for 5+. It is possible that single particle conductivity measurements \cite{SingleParticleConductivity} and additional surface characterization from techniques such as X-ray photoelectron spectroscopy (XPS) could provide confidence in sodium conduction.

\section{Conclusion}

In this work, we performed an extensive survey of the little-reported high-temperature sodium superionic conductor class, NAP.
First, we examine the parent \ce{Na4TiP2O9} structural features such as weakly bonded, high coordination sodium sites and their role in Na-ion conduction.
We then explored the flexibility of cation substitution to Ti (A octahedral) site using computational thermodynamic hull models.
Experimental synthesis attempts within our self-driving lab, informed by state-of-the-art precursor prediction tools, led to the successful synthesis of one new NAP: \ce{Na4SnP2O9}. 
Synthesis optimization reveals this material to be an ionic conductor only at high temperature, with temperature-driven coupled structural-conductivity transitions present as in other known NAPs.
Impactful future work may include stabilization of high symmetry, conducting polymorphs at low temperatures, and expanding experimentally realized phases through alternative synthesis methods (\textit{e.g.} sol-gel) or partial A cation substitution.
For both topics, first principles calculations are likely to play a predictive key role, emphasizing the continued need for materials physics models to better understand conduction, battery, and synthesis systems.

\section{Authorship Contributions}
L.N.W. developed the theoretical formalism; conceived, planned, and executed experiments; performed computations; and wrote the manuscript.
Y.F. contributed the theoretical formalism; executed synthesis experiments; executed XRD characterization and analysis; and contributed to manuscript revision.
B.R. contributed to the theoretical formalism; executed synthesis experiments; and contributed to manuscript revision.
X.Y. executed synthesis experiments; performed EIS characterization; and contributed to manuscript revision.
M.D. executed synthesis experiments and EIS characterization.
K.J. contributed the theoretical formalism and performed computations.
G.W. contributed the theoretical formalism and performed computations.
M.J.M. contributed to the theoretical formalism; executed synthesis experiments; executed reaction network calculations.
A.G. contributed to the theoretical formalism; executed EDS characterization analysis; contributed to manuscript revision.
T.M. executed STEM-HAADF characterization and contributed to manuscript revision.
F.S. executed nonambient EIS characterization.
D.M. executed synthesis experiments.
M.S.O. executed synthesis experiments.
H.K. contributed to the theoretical formalism for synthesis and conductivity testing.
M.C.T. contributed to the theoretical formalism for conductivity testing.
G.C. contributed the theoretical formalism; supervised the project.

\begin{acknowledgments}
This work was funded and intellectually led by the D2S2 program within the U.S. Department of Energy, Office of Basic Energy Sciences, Materials Sciences and Engineering Division under Contract No. DE-AC02-05-CH11231.
The authors gratefully acknowledge research support from the HydroGEN Advanced Water Splitting Materials Consortium, established as part of the Energy Materials Network under the U.S. Department of Energy, Office of Energy Efficiency and Renewable Energy, Fuel Cell Technologies Office, under Contract Number DE-AC02-05CH11231. 
This work was funded in part by the U.S. Department of Energy under contract no. DE-AC02-05CH11231.
G.W. acknowledges support by the U.S. Department of Energy, Office of Science, Office of Advanced Scientific Computing Research, Department of Energy Computational Science Graduate Fellowship under Award Number DESC0023112.
Work at the Molecular Foundry was supported by the Office of Science, Office of Basic Energy Sciences, of the U.S. Department of Energy under Contract No. DE-AC02-05CH11231.
This research used resources of the National Energy Research Scientific Computing Center (NERSC), a Department of Energy Office of Science User Facility using NERSC award DDR-ERCAPm3844\_g.
A portion of the research was performed using computational resources sponsored by the Department of Energy's Office of Energy Efficiency and Renewable Energy and located at the National Renewable Energy Laboratory.
The authors thank Dr. Ekin Dogus Cubuk and Dr. Amil Merchant for their assistance and discussions on choosing structural frameworks to investigate.
The authors thank Dr. Carolin M. Sutter-Fella and Dr. Tim Kodalle for experimental help and discussion.
T.P.M. acknowledges the support of Dr. Karen Bustillo for help with setting up cryo measurements.
\end{acknowledgments}

\section{Disclaimer}

This report was prepared as an account of work sponsored by an agency of the United States Government. Neither the United States Government nor any agency thereof, nor any of their employees, makes any warranty, express or implied, or assumes any legal liability or responsibility for the accuracy, completeness, or usefulness of any information, apparatus, product, or process disclosed, or represents that its use would not infringe privately owned rights. Reference herein to any specific commercial product, process, or service by trade name, trademark, manufacturer, or otherwise does not necessarily constitute or imply its endorsement, recommendation, or favoring by the United States Government or any agency thereof. The views and opinions of authors expressed herein do not necessarily state or reflect those of the United States Government or any agency thereof.

\bibliography{AMain}
\end{document}


\vspace{-4.5em}
\begin{singlespace}
{\hypersetup{linkcolor=black}\tableofcontents}
\end{singlespace}

\clearpage

\section{NTP Thermodynamics and Stability}

$\bm{\mathrm{N}\mathrm{a}_\mathrm{8-x}A^\mathrm{x}\mathrm{P}_\mathrm{2}\mathrm{O}_\mathrm{9}}$ (NAP)

\subsection{Ordering of Sodium}

\begin{figure}[H]
    \centering
    \includegraphics[scale=0.5]{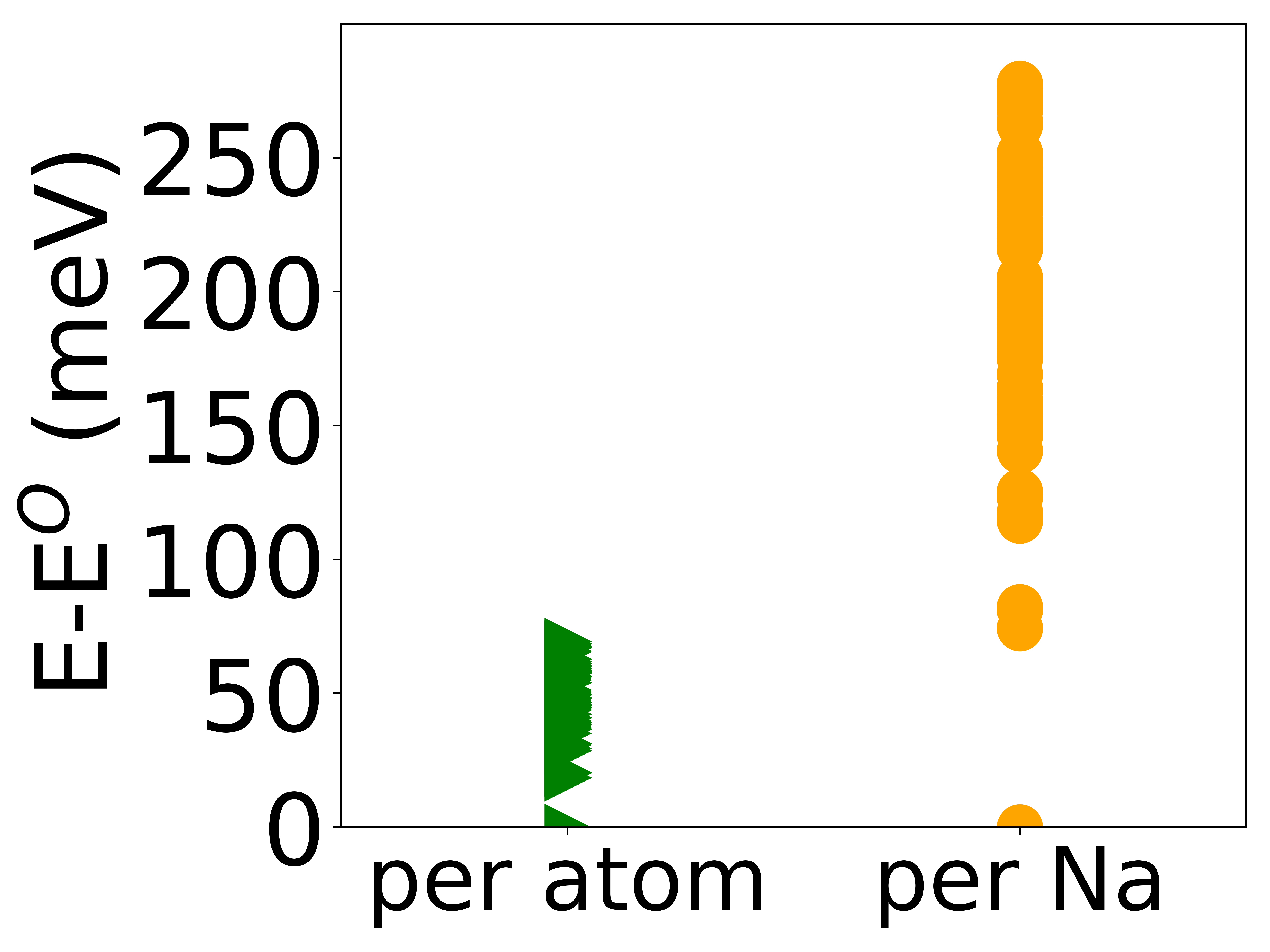}
    \caption{DFT total energies from NTP with different sodium orderings}
    \label{fig:NTP-Na-ordering}
\end{figure}

\subsection{Structural Labeling: Sodium}

\begin{figure}[H]
    \centering
    \includegraphics[scale=0.1]{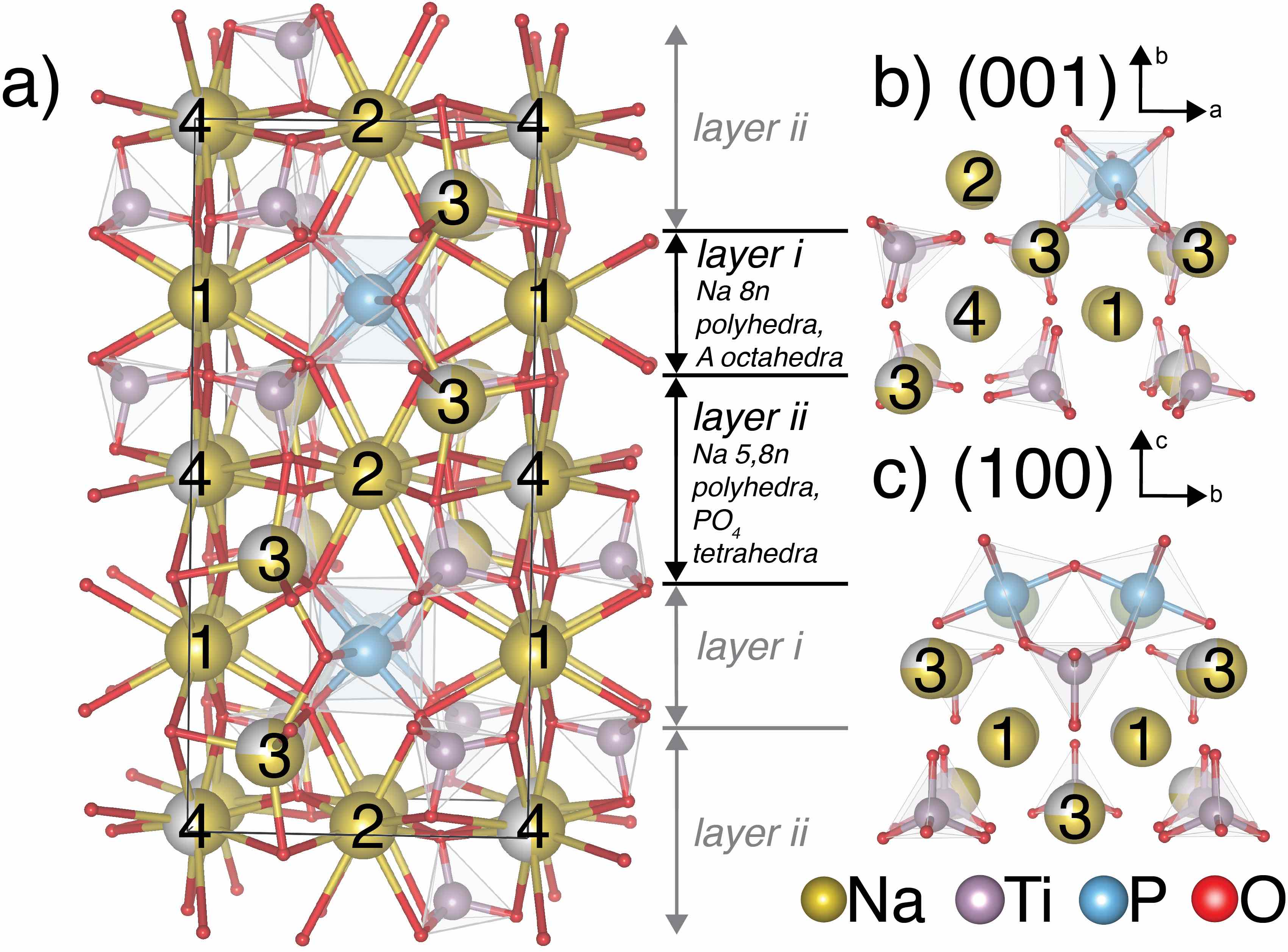}
    \caption{Reproduction of Figure 1, showing a structural visualization of NTP. In this reproduction, inequivalent sodium sites are labeled.}
    \label{fig:NTP-Na-Labels}
\end{figure}

\subsection{Structural Labeling: Vacancies for Conductivity Pathway Analysis}

\begin{figure}[H]
    \centering
    \includegraphics[width=0.9\linewidth]{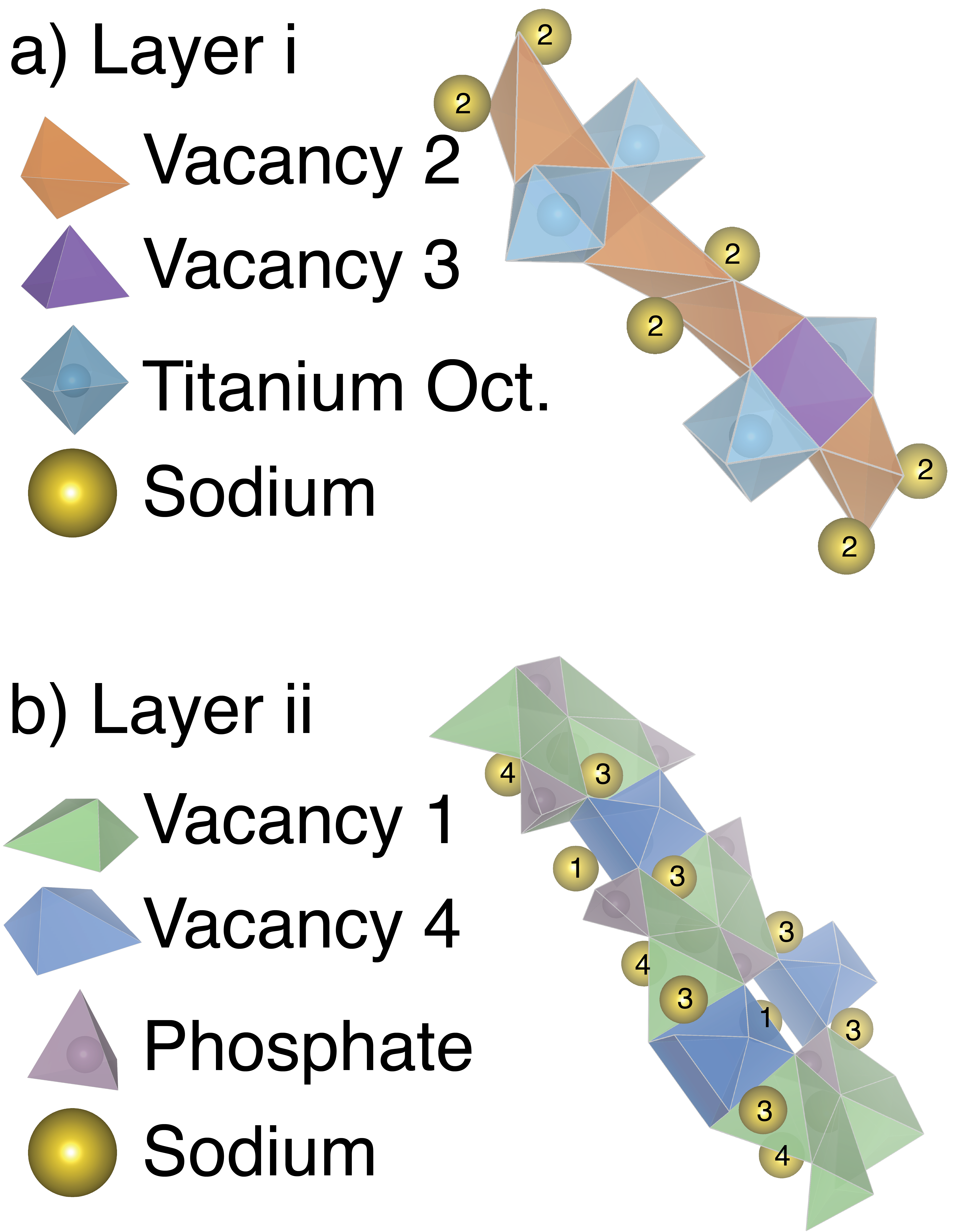}
    \caption{NTP Layers with vacancies visualized in their respective polyhedral sites.}
    \label{fig:NTP_vacancies_labeled}
\end{figure}

\section{High Throughput Cation Substitution with \ce{Na2CO3}, \ce{NH4H2PO4} Precursors}

Thermodynamic stability is assessed via hull energy and reaction energies. Reaction energies are found with a set of relatively common precursors, \ce{Na2CO3}, \ce{NH4H2PO4}, and \ce{A_yO_z}, where \ce{A_yO_z} is the most stable binary oxide on the A-O convex hull. Reactions for all targets are found to be thermodynamically favorable (negative reaction energy).

\begin{figure}[H]
    \centering
    \includegraphics[width=0.75\linewidth]{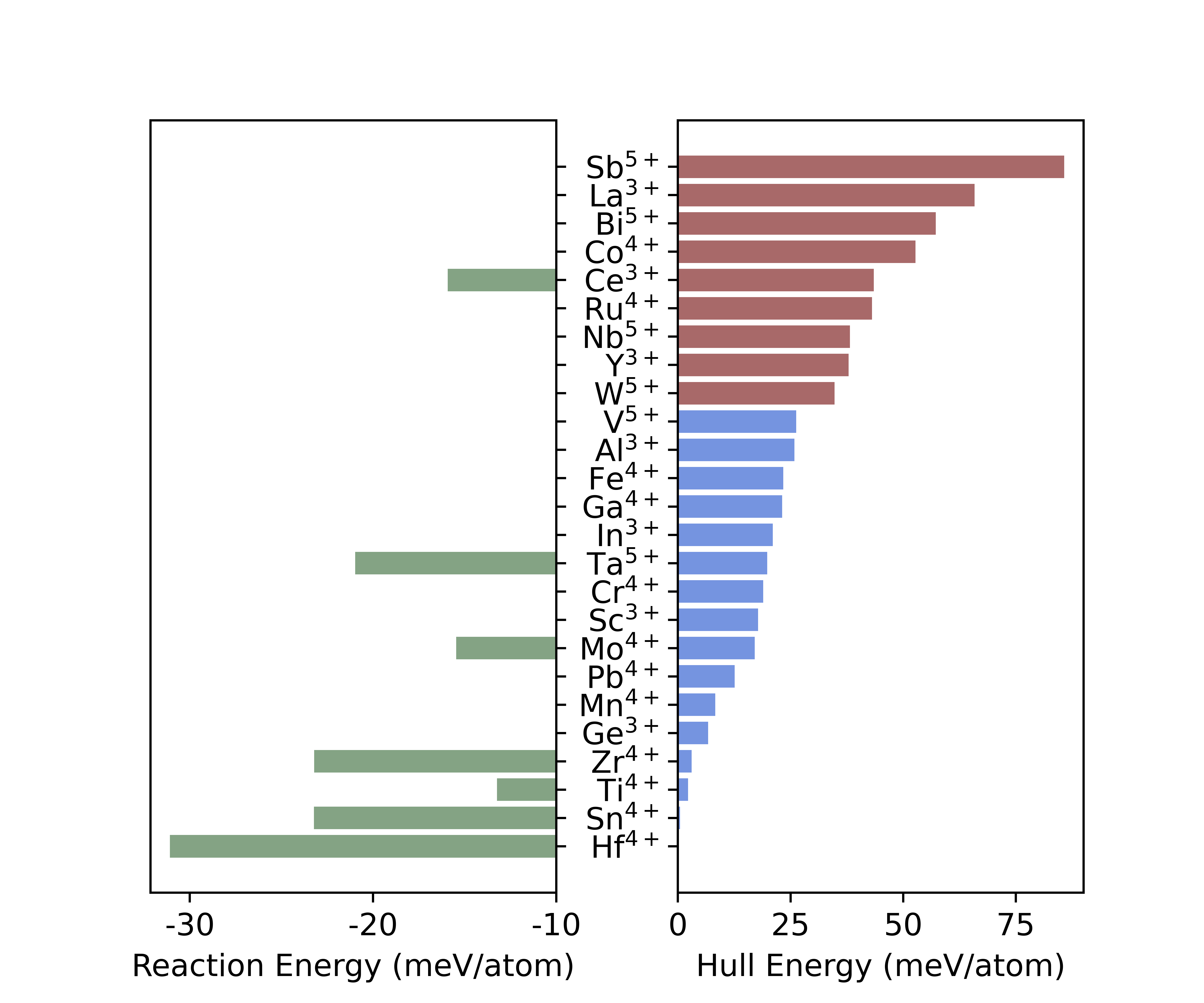}
    \caption{Hull energies and reaction energy for NAP candidate phases. Reactions are calculated with the precursor set of \ce{Na4P2O7} and \ce{A_yO_z}, where \ce{A_yO_z} is the most stable binary oxide on the A-O convex hull.}
    \label{fig:hull_driving_force_candidates_Na4P2O7}
\end{figure}

\section{Precursor Model Output}

\subsection{s4 Output via MatSynthesis}

\begin{table}[H]
\tiny
\caption{\label{tab:s4_Na3TaP2O9} Precursor recommendations for \ce{Na3TaP2O9} from the solid-state synthesis science analyzer (S4), as distributed by \href{https://matsynthesis.org}{matsynthesis.org} \cite{TextMinedSynthesisRecipes}.}
\begin{tabular}{lcc}
\hline \hline
\textbf{Ranking} & \textbf{Precursors} & \textbf{Reaction} \\
\hline
1&	NH4H2PO4, Na2CO3, Ta2O5 &	2 NH4H2PO4 + 1.5 Na2CO3 + 0.5 Ta2O5 == 1 Na3TaP2O9 + 1.5 CO2 + 3 H2O + 2 NH3 \\
2&	Na2CO3, NaPO3, Ta2O5	&0.5 Na2CO3 + 2 NaPO3 + 0.5 Ta2O5 == 1 Na3TaP2O9 + 0.5 CO2 \\
3&	NH4H2PO4, NaNO3, Ta2O5	&2.0 NH4H2PO4 + 3.0 NaNO3 + 0.5 Ta2O5 == 1.0 Na3TaP2O9 + 1.0 H2O + 3.333 NH3 + 3.083 O2 + 1.667 NO2\\
4&	NH4H2PO4, Na2CO3, Ta2O3	&2 NH4H2PO4 + 1.5 Na2CO3 + 0.5 O2 + 0.5 Ta2O3 == 1 Na3TaP2O9 + 1.5 CO2 + 3 H2O + 2 NH3 \\
5&	NaH2PO4, NaHO, Ta2O5	&2 NaH2PO4 + 1 NaHO + 0.5 Ta2O5 == 1 Na3TaP2O9 + 2.5 H2O \\
6&	(NH4)2HPO4, Na2CO3, Ta2O5	&2 (NH4)2HPO4 + 1.5 Na2CO3 + 0.5 Ta2O5 == 1 Na3TaP2O9 + 1.5 CO2 + 3 H2O + 4 NH3 \\
7&	Na2CO3, NaH2PO4, Ta2O5&	0.5 Na2CO3 + 2 NaH2PO4 + 0.5 Ta2O5 == 1 Na3TaP2O9 + 0.5 CO2 + 2 H2O \\
8&	(NH4)3PO4, Na2CO3, Ta2O5&	2 (NH4)3PO4 + 1.5 Na2CO3 + 0.5 Ta2O5 == 1 Na3TaP2O9 + 1.5 CO2 + 3 H2O + 6 NH3 \\
\hline \hline
\end{tabular}
\end{table}

\begin{table}[H]
\tiny
\caption{\label{tab:s4_Na4HfP2O9} Precursor recommendations for \ce{Na4HfP2O9} from the solid-state synthesis science analyzer (S4), as distributed by \href{https://matsynthesis.org}{matsynthesis.org} \cite{TextMinedSynthesisRecipes}.}
\begin{tabular}{lcc}
\hline \hline
\textbf{Ranking} & \textbf{Precursors} & \textbf{Reaction} \\
\hline
1&	HfO2, NH4H2PO41, Na2CO3&	1 HfO2 + 2 NH4H2PO4 + 2 Na2CO3 == 1 Na4HfP2O9 + 2 CO2 + 3 H2O + 2 NH3\\
2&	(NH4)2HPO41, HfO2, Na2CO3&	2 (NH4)2HPO4 + 1 HfO2 + 2 Na2CO3 == 1 Na4HfP2O9 + 2 CO2 + 3 H2O + 4 NH3\\
3&	HfO2, Na4P2O7	&1 HfO2 + 1 Na4P2O7 == 1 Na4HfP2O9\\
4&	HfO2, Na2CO3, NaH2PO4	&1 HfO2 + 1 Na2CO3 + 2 NaH2PO4 == 1 Na4HfP2O9 + 1 CO2 + 2 H2O\\
5&	(NH4)3PO4, HfO2, Na2CO3	&2 (NH4)3PO4 + 1 HfO2 + 2 Na2CO3 == 1 Na4HfP2O9 + 2 CO2 + 3 H2O + 6 NH3\\
6&	HfO2, Na2CO3, NaPO3&	1 HfO2 + 1 Na2CO3 + 2 NaPO3 == 1 Na4HfP2O9 + 1 CO2\\
7&	HfO2, NH4H2PO41, NaNO3&	1.0 HfO2 + 2.0 NH4H2PO4 + 4.0 NaNO3 == 1.0 Na4HfP2O9 + 1.0 H2O + 3.333 NH3 + 3.333 O2 + 2.667 NO2\\
8&	HfO2, NH4H2PO41, Na3PO4&	1 HfO2 + 0.667 NH4H2PO4 + 1.333 Na3PO4 == 1 Na4HfP2O9 + 1 H2O + 0.667 NH3\\
9&	HfO2, NH4H2PO41, NaHO	&1 HfO2 + 2 NH4H2PO4 + 4 NaHO == 1 Na4HfP2O9 + 5 H2O + 2 NH3\\
\hline \hline
\end{tabular}
\end{table}

\begin{table}[H]
\tiny
\caption{\label{tab:s4_Na4SnP2O9} Precursor recommendations for \ce{Na4SnP2O9} from the solid-state synthesis science analyzer (S4), as distributed by \href{https://matsynthesis.org}{matsynthesis.org} \cite{TextMinedSynthesisRecipes}.}
\begin{tabular}{lcc}
\hline \hline
\textbf{Ranking} & \textbf{Precursors} & \textbf{Reaction} \\
\hline
1&	NH4H2PO41, Na2CO3, SnO2&	2 NH4H2PO4 + 2 Na2CO3 + 1 SnO2 == 1 Na4SnP2O9 + 2 CO2 + 3 H2O + 2 NH3\\
2&	(NH4)2HPO41, Na2CO3, SnO2&	2 (NH4)2HPO4 + 2 Na2CO3 + 1 SnO2 == 1 Na4SnP2O9 + 2 CO2 + 3 H2O + 4 NH3\\
3&	Na4P2O7, SnO2	&1 Na4P2O7 + 1 SnO2 == 1 Na4SnP2O9\\
4&	Na2CO3, NaH2PO4, SnO	&1 Na2CO3 + 2 NaH2PO4 + 0.5 O2 + 1 SnO == 1 Na4SnP2O9 + 1 CO2 + 2 H2O\\
5&	Na2CO3, NaPO3, SnO	&1 Na2CO3 + 2 NaPO3 + 0.5 O2 + 1 SnO == 1 Na4SnP2O9 + 1 CO2\\
6&	Na2CO3, NaH2PO4, Sn(CO2)2&	1 Na2CO3 + 2 NaH2PO4 + 1 O2 + 1 Sn(CO2)2 == 1 Na4SnP2O9 + 3 CO2 + 2 H2O\\
7&	NH4H2PO41, NaNO3, SnO2&	2.0 NH4H2PO4 + 4.0 NaNO3 + 1.0 SnO2 == 1.0 Na4SnP2O9 + 1.0 H2O + 3.333 NH3 + 3.333 O2 + 2.667 NO2\\
8&	Na2CO3, NaH2PO4, SnO2	&1 Na2CO3 + 2 NaH2PO4 + 1 SnO2 == 1 Na4SnP2O9 + 1 CO2 + 2 H2O\\
9&	NH4H2PO41, NaHO, SnO	&2 NH4H2PO4 + 4 NaHO + 0.5 O2 + 1 SnO == 1 Na4SnP2O9 + 5 H2O + 2 NH3\\
10&	(NH4)2HPO41, Na2CO3, SnO	&2 (NH4)2HPO4 + 2 Na2CO3 + 0.5 O2 + 1 SnO == 1 Na4SnP2O9 + 2 CO2 + 3 H2O + 4 NH3\\
\hline \hline
\end{tabular}
\end{table}

\begin{table}[H]
\tiny
\caption{\label{tab:s4_Na3VP2O9} Precursor recommendations for \ce{Na3VP2O9} from the solid-state synthesis science analyzer (S4), as distributed by \href{https://matsynthesis.org}{matsynthesis.org} \cite{TextMinedSynthesisRecipes}.}
\begin{tabular}{lcc}
\hline \hline
\textbf{Ranking} & \textbf{Precursors} & \textbf{Reaction} \\
\hline
1&	NH4H2PO41, Na2CO3, V2O5	&2 NH4H2PO4 + 1.5 Na2CO3 + 0.5 V2O5 == 1 Na3VP2O9 + 1.5 CO2 + 3 H2O + 2 NH3\\
2&	NH4H2PO41, Na2CO3, V2O3	&2 NH4H2PO4 + 1.5 Na2CO3 + 0.5 O2 + 0.5 V2O3 == 1 Na3VP2O9 + 1.5 CO2 + 3 H2O + 2 NH3\\
3&	(NH4)2HPO41, Na2CO3, VH4NO3&	2 (NH4)2HPO4 + 1.5 Na2CO3 + 1 VH4NO3 == 1 Na3VP2O9 + 1.5 CO2 + 3.5 H2O + 5 NH3\\
4&	NH4H2PO41, NaHO, VH4NO3&	2 NH4H2PO4 + 3 NaHO + 1 VH4NO3 == 1 Na3VP2O9 + 5 H2O + 3 NH3\\
5&	NH4H2PO41, Na4P2O7, V2O3&	0.5 NH4H2PO4 + 0.75 Na4P2O7 + 0.5 O2 + 0.5 V2O3 == 1 Na3VP2O9 + 0.75 H2O + 0.5 NH3\\
6&	(NH4)2HPO41, Na2CO3, V2O5	&2 (NH4)2HPO4 + 1.5 Na2CO3 + 0.5 V2O5 == 1 Na3VP2O9 + 1.5 CO2 + 3 H2O + 4 NH3\\
7&	Na2CO3, NaH2PO4, V2O5&	0.5 Na2CO3 + 2 NaH2PO4 + 0.5 V2O5 == 1 Na3VP2O9 + 0.5 CO2 + 2 H2O\\
8&	Na2CO3, NaH2PO4, V2O3&	0.5 Na2CO3 + 2 NaH2PO4 + 0.5 O2 + 0.5 V2O3 == 1 Na3VP2O9 + 0.5 CO2 + 2 H2O\\
9&	Na2O2, NaH2PO4, V2O3&	0.5 Na2O2 + 2 NaH2PO4 + 0.25 O2 + 0.5 V2O3 == 1 Na3VP2O9 + 2 H2O\\
10&	NH4H2PO41, Na2CO3, VH4NO3	&2 NH4H2PO4 + 1.5 Na2CO3 + 1 VH4NO3 == 1 Na3VP2O9 + 1.5 CO2 + 3.5 H2O + 3 NH3\\
\hline \hline
\end{tabular}
\end{table}

\begin{table}[H]
\tiny
\caption{\label{tab:s4_Na5AlP2O9} Precursor recommendations for \ce{Na5AlP2O9} from the solid-state synthesis science analyzer (S4), as distributed by \href{https://matsynthesis.org}{matsynthesis.org} \cite{TextMinedSynthesisRecipes}.}
\begin{tabular}{lcc}
\hline \hline
\textbf{Ranking} & \textbf{Precursors} & \textbf{Reaction} \\
\hline
1&	Al2O3, NH4H2PO41, Na2CO3&	0.5 Al2O3 + 2 NH4H2PO4 + 2.5 Na2CO3 == 1 Na5AlP2O9 + 2.5 CO2 + 3 H2O + 2 NH3\\
2&	(NH4)2HPO41, Al2O3, Na2CO3&	2 (NH4)2HPO4 + 0.5 Al2O3 + 2.5 Na2CO3 == 1 Na5AlP2O9 + 2.5 CO2 + 3 H2O + 4 NH3\\
3&	Al2O3, Na2CO3, NaPO3	&0.5 Al2O3 + 1.5 Na2CO3 + 2 NaPO3 == 1 Na5AlP2O9 + 1.5 CO2\\
4&	Al2O3, NH4H2PO41, NaNO3	&0.5 Al2O3 + 2.0 NH4H2PO4 + 5.0 NaNO3 == 1.0 Na5AlP2O9 + 1.0 H2O + 3.333 NH3 + 3.583 O2 + 3.667 NO2\\
5&	(NH4)2HPO41, Al(HO)3, Na2CO3&	2 (NH4)2HPO4 + 1 Al(HO)3 + 2.5 Na2CO3 == 1 Na5AlP2O9 + 2.5 CO2 + 4.5 H2O + 4 NH3\\
6&	Al2O3, Na2CO3, NaH2PO4	&0.5 Al2O3 + 1.5 Na2CO3 + 2 NaH2PO4 == 1 Na5AlP2O9 + 1.5 CO2 + 2 H2O\\
7&	Al2O3, Na2CO3, Na4P2O7&	0.5 Al2O3 + 0.5 Na2CO3 + 1 Na4P2O7 == 1 Na5AlP2O9 + 0.5 CO2\\
8&	Al2O3, NH4H2PO41, Na3PO4&	0.5 Al2O3 + 0.333 NH4H2PO4 + 1.667 Na3PO4 == 1 Na5AlP2O9 + 0.5 H2O + 0.333 NH3\\
9&	Al2O3, NaH2PO4, NaHO&	0.5 Al2O3 + 2 NaH2PO4 + 3 NaHO == 1 Na5AlP2O9 + 3.5 H2O\\
10&	Al(HO)3, NaHO, NaPO3	&1 Al(HO)3 + 3 NaHO + 2 NaPO3 == 1 Na5AlP2O9 + 3 H2O\\
\hline \hline
\end{tabular}
\end{table}

\begin{table}[H]
\tiny
\caption{\label{tab:s4_Na4FeP2O9} Precursor recommendations for \ce{Na4FeP2O9} from the solid-state synthesis science analyzer (S4), as distributed by \href{https://matsynthesis.org}{matsynthesis.org} \cite{TextMinedSynthesisRecipes}.}
\begin{tabular}{lcc}
\hline \hline
\textbf{Ranking} & \textbf{Precursors} & \textbf{Reaction} \\
\hline
1&	Fe2O3, NH4H2PO41, Na2CO3&	0.5 Fe2O3 + 2 NH4H2PO4 + 2 Na2CO3 + 0.25 O2 == 1 Na4FeP2O9 + 2 CO2 + 3 H2O + 2 NH3\\
2&	(NH4)2HPO41, Fe2O3, Na2CO3	&2 (NH4)2HPO4 + 0.5 Fe2O3 + 2 Na2CO3 + 0.25 O2 == 1 Na4FeP2O9 + 2 CO2 + 3 H2O + 4 NH3\\
3&	FeC2O4, Na2CO3, NaH2PO4&	1 FeC2O4 + 1 Na2CO3 + 2 NaH2PO4 + 1 O2 == 1 Na4FeP2O9 + 3 CO2 + 2 H2O\\
4&	Fe(NO3)3, Na4P2O7	&1 Fe(NO3)3 + 1 Na4P2O7 + 0.25 O2 + 3 [OH-] == 1 Na4FeP2O9 + 1.5 H2O + 3 [NO3-]\\
5&	FeC2O4, NH4H2PO41, Na2CO3&	1 FeC2O4 + 2 NH4H2PO4 + 2 Na2CO3 + 1 O2 == 1 Na4FeP2O9 + 4 CO2 + 3 H2O + 2 NH3\\
6&	FePO4, NH4H2PO41, Na2CO3&	1 FePO4 + 1 NH4H2PO4 + 2 Na2CO3 + 0.25 O2 == 1 Na4FeP2O9 + 2 CO2 + 1.5 H2O + 1 NH3\\
7&	FePO4, Na2CO3, Na4P2O7&	1 FePO4 + 1 Na2CO3 + 0.5 Na4P2O7 + 0.25 O2 == 1 Na4FeP2O9 + 1 CO2\\
8&	(NH4)2HPO41, FePO4, Na2CO3&	1 (NH4)2HPO4 + 1 FePO4 + 2 Na2CO3 + 0.25 O2 == 1 Na4FeP2O9 + 2 CO2 + 1.5 H2O + 2 NH3\\
9&	Fe(NO3)3, Na2CO3, NaH2PO4	&1 Fe(NO3)3 + 1 Na2CO3 + 2 NaH2PO4 + 0.25 O2 + 3 [OH-] == 1 Na4FeP2O9 + 1 CO2 + 3.5 H2O + 3 [NO3-]\\
10&	CH3COONa, FeC2O4, NH4H2PO4 &4 CH3COONa + 1 FeC2O4 + 2 NH4H2PO4 + 5 [OH-] == 1 Na4FeP2O9 + 4 H2O + 2 NH3 + 1 O2 + 5 [CH3COO-]\\
\hline \hline
\end{tabular}
\end{table}

\begin{table}[H]
\tiny
\caption{\label{tab:s4_Na4GaP2O9} Precursor recommendations for \ce{Na4GaP2O9} from the solid-state synthesis science analyzer (S4), as distributed by \href{https://matsynthesis.org}{matsynthesis.org} \cite{TextMinedSynthesisRecipes}.}
\begin{tabular}{lcc}
\hline \hline
\textbf{Ranking} & \textbf{Precursors} & \textbf{Reaction} \\
\hline
1&	Ga2O3, NH4H2PO41, Na2CO3&	0.5 Ga2O3 + 2 NH4H2PO4 + 2 Na2CO3 + 0.25 O2 == 1 Na4GaP2O9 + 2 CO2 + 3 H2O + 2 NH3\\
2&	(NH4)2HPO41, Ga2O3, Na2CO3	&2 (NH4)2HPO4 + 0.5 Ga2O3 + 2 Na2CO3 + 0.25 O2 == 1 Na4GaP2O9 + 2 CO2 + 3 H2O + 4 NH3\\
3&	Ga2O3, Na2CO3, NaH2PO4	&0.5 Ga2O3 + 1 Na2CO3 + 2 NaH2PO4 + 0.25 O2 == 1 Na4GaP2O9 + 1 CO2 + 2 H2O\\
4&	Ga(NO3)3, Na4P2O7	&1 Ga(NO3)3 + 1 Na4P2O7 + 0.25 O2 + 3 [OH-] == 1 Na4GaP2O9 + 1.5 H2O + 3 [NO3-]\\
5&	Ga2O3, NH4H2PO41, Na3PO4	&0.5 Ga2O3 + 0.667 NH4H2PO4 + 1.333 Na3PO4 + 0.25 O2 == 1 Na4GaP2O9 + 1 H2O + 0.667 NH3\\
6&	Ga(NO3)3, NH4H2PO41, NaNO3	&1.0 Ga(NO3)3 + 2.0 NH4H2PO4 + 4.0 NaNO3 == 1.0 Na4GaP2O9 + 1.0 H2O + 3.333 NH3 + 3.833 O2 + 5.667 NO2\\
7&	Ga(NO3)3, Na2CO3, NaPO3	&1 Ga(NO3)3 + 1 Na2CO3 + 2 NaPO3 + 0.25 O2 + 3 [OH-] == 1 Na4GaP2O9 + 1 CO2 + 1.5 H2O + 3 [NO3-]\\
8&	(NH4)3PO4, Ga(NO3)3, Na2CO3	&2.0 (NH4)3PO4 + 1.0 Ga(NO3)3 + 2.0 Na2CO3 == 1.0 Na4GaP2O9 + 1.0 H2O + 2.0 CO2 + 7.333 NH3 + 2.833 O2 + 1.667 NO2\\
\hline \hline
\end{tabular}
\end{table}

\begin{table}[H]
\tiny
\caption{\label{tab:s4_Na5InP2O9} Precursor recommendations for \ce{Na5InP2O9} from the solid-state synthesis science analyzer (S4), as distributed by \href{https://matsynthesis.org}{matsynthesis.org} \cite{TextMinedSynthesisRecipes}.}
\begin{tabular}{lcc}
\hline \hline
\textbf{Ranking} & \textbf{Precursors} & \textbf{Reaction} \\
\hline
1&	In2O3, NH4H2PO41, Na2CO3&	0.5 In2O3 + 2 NH4H2PO4 + 2.5 Na2CO3 == 1 Na5InP2O9 + 2.5 CO2 + 3 H2O + 2 NH3\\
2&	In2O3, Na2CO3, Na4P2O7	&0.5 In2O3 + 0.5 Na2CO3 + 1 Na4P2O7 == 1 Na5InP2O9 + 0.5 CO2\\
3&	(NH4)2HPO41, In2O3, Na2CO3&	2 (NH4)2HPO4 + 0.5 In2O3 + 2.5 Na2CO3 == 1 Na5InP2O9 + 2.5 CO2 + 3 H2O + 4 NH3\\
4&	In2O3, Na2CO3, NaH2PO4&	0.5 In2O3 + 1.5 Na2CO3 + 2 NaH2PO4 == 1 Na5InP2O9 + 1.5 CO2 + 2 H2O\\
5&	In2O3, NH4H2PO41, NaHO&	0.5 In2O3 + 2 NH4H2PO4 + 5 NaHO == 1 Na5InP2O9 + 5.5 H2O + 2 NH3\\
6&	In2O3, Na2CO3, NaPO3	&0.5 In2O3 + 1.5 Na2CO3 + 2 NaPO3 == 1 Na5InP2O9 + 1.5 CO2\\
7&	In2O3, NH4H2PO41, NaNO3	&0.5 In2O3 + 2.0 NH4H2PO4 + 5.0 NaNO3 == 1.0 Na5InP2O9 + 1.0 H2O + 3.333 NH3 + 3.583 O2 + 3.667 NO2\\
8&	(NH4)3PO4, In2O3, Na2CO3	&2 (NH4)3PO4 + 0.5 In2O3 + 2.5 Na2CO3 == 1 Na5InP2O9 + 2.5 CO2 + 3 H2O + 6 NH3\\
\hline \hline
\end{tabular}
\end{table}

\begin{table}[H]
\tiny
\caption{\label{tab:s4_Na4CrP2O9} Precursor recommendations for \ce{Na4CrP2O9} from the solid-state synthesis science analyzer (S4), as distributed by \href{https://matsynthesis.org}{matsynthesis.org} \cite{TextMinedSynthesisRecipes}.}
\begin{tabular}{lcc}
\hline \hline
\textbf{Ranking} & \textbf{Precursors} & \textbf{Reaction} \\
\hline
1&	Cr2O3, NH4H2PO41, Na2CO3	&0.5 Cr2O3 + 2 NH4H2PO4 + 2 Na2CO3 + 0.25 O2 == 1 Na4CrP2O9 + 2 CO2 + 3 H2O + 2 NH3\\
2&	CrO3, Na2CO3, NaH2PO4	&1 CrO3 + 1 Na2CO3 + 2 NaH2PO4 == 1 Na4CrP2O9 + 1 CO2 + 2 H2O + 0.5 O2\\
3&	Cr(NO3)3, Na4P2O7	&1 Cr(NO3)3 + 1 Na4P2O7 + 0.25 O2 + 3 [OH-] == 1 Na4CrP2O9 + 1.5 H2O + 3 [NO3-]\\
4&	(NH4)2HPO41, Cr(NO3)3, Na2CO3	&2.0 (NH4)2HPO4 + 1.0 Cr(NO3)3 + 2.0 Na2CO3 == 1.0 Na4CrP2O9 + 1.0 H2O + 2.0 CO2 + 5.333 NH3 + 2.833 O2 + 1.667 NO2\\
5&	CrO2, Na2CO3, NaH2PO4&	1 CrO2 + 1 Na2CO3 + 2 NaH2PO4 == 1 Na4CrP2O9 + 1 CO2 + 2 H2O\\
6&	CrO2, Na2O, NaPO3	&1 CrO2 + 1 Na2O + 2 NaPO3 == 1 Na4CrP2O9\\
7&	Cr(NO3)3, Na2CO3, NaPO3	&1 Cr(NO3)3 + 1 Na2CO3 + 2 NaPO3 + 0.25 O2 + 3 [OH-] == 1 Na4CrP2O9 + 1 CO2 + 1.5 H2O + 3 [NO3-]\\
8&	CrO2, Na2O2, NaH2PO4&	1 CrO2 + 1 Na2O2 + 2 NaH2PO4 == 1 Na4CrP2O9 + 2 H2O + 0.5 O2\\
9&	Cr(NO3)3, NaHO, NaPO3&	1 Cr(NO3)3 + 2 NaHO + 2 NaPO3 + 0.25 O2 + 3 [OH-] == 1 Na4CrP2O9 + 2.5 H2O + 3 [NO3-]\\
10&	CrO2, NH4H2PO41, NaNO3&	1.0 CrO2 + 2.0 NH4H2PO4 + 4.0 NaNO3 == 1.0 Na4CrP2O9 + 1.0 H2O + 3.333 NH3 + 3.333 O2 + 2.667 NO2\\
\hline \hline
\end{tabular}
\end{table}

\begin{table}[H]
\tiny
\caption{\label{tab:s4_Na5ScP2O9} Precursor recommendations for \ce{Na5ScP2O9} from the solid-state synthesis science analyzer (S4), as distributed by \href{https://matsynthesis.org}{matsynthesis.org} \cite{TextMinedSynthesisRecipes}.}
\begin{tabular}{lcc}
\hline \hline
\textbf{Ranking} & \textbf{Precursors} & \textbf{Reaction} \\
\hline
1&	NH4H2PO41, Na2CO3, Sc2O3&	2 NH4H2PO4 + 2.5 Na2CO3 + 0.5 Sc2O3 == 1 Na5ScP2O9 + 2.5 CO2 + 3 H2O + 2 NH3\\
2&	NH4H2PO41, Na3PO4, Sc2O3&	0.333 NH4H2PO4 + 1.667 Na3PO4 + 0.5 Sc2O3 == 1 Na5ScP2O9 + 0.5 H2O + 0.333 NH3\\
3&	(NH4)2HPO41, Na2CO3, Sc2O3	&2 (NH4)2HPO4 + 2.5 Na2CO3 + 0.5 Sc2O3 == 1 Na5ScP2O9 + 2.5 CO2 + 3 H2O + 4 NH3\\
4&	Na2CO3, Na4P2O7, Sc2O3&	0.5 Na2CO3 + 1 Na4P2O7 + 0.5 Sc2O3 == 1 Na5ScP2O9 + 0.5 CO2\\
5&	Na2CO3, NaH2PO4, Sc2O3&	1.5 Na2CO3 + 2 NaH2PO4 + 0.5 Sc2O3 == 1 Na5ScP2O9 + 1.5 CO2 + 2 H2O\\
6&	Na2CO3, NaPO3, Sc2O3&	1.5 Na2CO3 + 2 NaPO3 + 0.5 Sc2O3 == 1 Na5ScP2O9 + 1.5 CO2\\
7&	NH4H2PO41, NaNO3, Sc2O3&	2.0 NH4H2PO4 + 5.0 NaNO3 + 0.5 Sc2O3 == 1.0 Na5ScP2O9 + 1.0 H2O + 3.333 NH3 + 3.583 O2 + 3.667 NO2\\
8&	NH4H2PO41, NaHO, Sc2O3	&2 NH4H2PO4 + 5 NaHO + 0.5 Sc2O3 == 1 Na5ScP2O9 + 5.5 H2O + 2 NH3\\
\hline \hline
\end{tabular}
\end{table}

\begin{table}[H]
\tiny
\caption{\label{tab:s4_Na4MoP2O9} Precursor recommendations for \ce{Na4MoP2O9} from the solid-state synthesis science analyzer (S4), as distributed by \href{https://matsynthesis.org}{matsynthesis.org} \cite{TextMinedSynthesisRecipes}.}
\begin{tabular}{lcc}
\hline \hline
\textbf{Ranking} & \textbf{Precursors} & \textbf{Reaction} \\
\hline
1&	MoO3, NH4H2PO41, Na2CO3&	1 MoO3 + 2 NH4H2PO4 + 2 Na2CO3 == 1 Na4MoP2O9 + 2 CO2 + 3 H2O + 2 NH3 + 0.5 O2\\
2&	(NH4)2HPO41, MoO3, Na2CO3	&2 (NH4)2HPO4 + 1 MoO3 + 2 Na2CO3 == 1 Na4MoP2O9 + 2 CO2 + 3 H2O + 4 NH3 + 0.5 O2\\
3&	MoO3, Na4P2O7	&1 MoO3 + 1 Na4P2O7 == 1 Na4MoP2O9 + 0.5 O2\\
4&	MoO3, Na2CO3, NaH2PO4	&1 MoO3 + 1 Na2CO3 + 2 NaH2PO4 == 1 Na4MoP2O9 + 1 CO2 + 2 H2O + 0.5 O2\\
5&	(NH4)2HPO41, Mo7H24(NO4)6, Na2CO3	&2 (NH4)2HPO4 + 0.143 Mo7H24(NO4)6 + 2 Na2CO3 == 1 Na4MoP2O9 + 2 CO2 + 3.429 H2O + 4.857 NH3 + 0.5 O2\\
6&	MoO3, NH4H2PO41, NaHO	&1 MoO3 + 2 NH4H2PO4 + 4 NaHO == 1 Na4MoP2O9 + 5 H2O + 2 NH3 + 0.5 O2\\
7&	Mo7H24(NO4)6, Na2CO3, NaH2PO4&	0.143 Mo7H24(NO4)6 + 1.0 Na2CO3 + 2.0 NaH2PO4 == 1.0 Na4MoP2O9 + 2.701 H2O + 1.0 CO2 + 0.675 NH3 + 0.182 O2 + 0.182 NO2\\
8&	MoO3, Na2CO3, NaPO3&	1 MoO3 + 1 Na2CO3 + 2 NaPO3 == 1 Na4MoP2O9 + 1 CO2 + 0.5 O2\\
\hline \hline
\end{tabular}
\end{table}

\begin{table}[H]
\tiny
\caption{\label{tab:s4_Na4PbP2O9} Precursor recommendations for \ce{Na4PbP2O9} from the solid-state synthesis science analyzer (S4), as distributed by \href{https://matsynthesis.org}{matsynthesis.org} \cite{TextMinedSynthesisRecipes}.}
\begin{tabular}{lcc}
\hline \hline
\textbf{Ranking} & \textbf{Precursors} & \textbf{Reaction} \\
\hline
1&	NH4H2PO41, Na2CO3, PbO&	2 NH4H2PO4 + 2 Na2CO3 + 0.5 O2 + 1 PbO == 1 Na4PbP2O9 + 2 CO2 + 3 H2O + 2 NH3\\
2&	(NH4)2HPO41, Na2CO3, PbO&	2 (NH4)2HPO4 + 2 Na2CO3 + 0.5 O2 + 1 PbO == 1 Na4PbP2O9 + 2 CO2 + 3 H2O + 4 NH3\\
3&	Na4P2O7, PbO	&1 Na4P2O7 + 0.5 O2 + 1 PbO == 1 Na4PbP2O9\\
4&	Na2CO3, NaH2PO4, PbO	&1 Na2CO3 + 2 NaH2PO4 + 0.5 O2 + 1 PbO == 1 Na4PbP2O9 + 1 CO2 + 2 H2O\\
5&	NH4H2PO41, NaHO, PbO	&2 NH4H2PO4 + 4 NaHO + 0.5 O2 + 1 PbO == 1 Na4PbP2O9 + 5 H2O + 2 NH3\\
6&	Na2CO3, NaPO3, PbO	&1 Na2CO3 + 2 NaPO3 + 0.5 O2 + 1 PbO == 1 Na4PbP2O9 + 1 CO2\\
7&	NH4H2PO41, NaNO3, PbO	&2.0 NH4H2PO4 + 4.0 NaNO3 + 1.0 PbO == 1.0 Na4PbP2O9 + 1.0 H2O + 3.333 NH3 + 2.833 O2 + 2.667 NO2\\
8&	NH4H2PO41, Na3PO4, PbO	&0.667 NH4H2PO4 + 1.333 Na3PO4 + 0.5 O2 + 1 PbO == 1 Na4PbP2O9 + 1 H2O + 0.667 NH3\\
9&	CH3COONa, NH4H2PO41, Pb(NO3)2&	4.0 CH3COONa + 2.0 NH4H2PO4 + 1.0 Pb(NO3)2 + 3.333 NO2 == 1.0 Na4PbP2O9 + 1.0 H2O + 8.0 CO2 + 7.333 NH3 + 1.333 O2\\
\hline \hline
\end{tabular}
\end{table}

\begin{table}[H]
\tiny
\caption{\label{tab:s4_Na4MnP2O9} Precursor recommendations for \ce{Na4MnP2O9} from the solid-state synthesis science analyzer (S4), as distributed by \href{https://matsynthesis.org}{matsynthesis.org} \cite{TextMinedSynthesisRecipes}.}
\begin{tabular}{lcc}
\hline \hline
\textbf{Ranking} & \textbf{Precursors} & \textbf{Reaction} \\
\hline
1&	MnO2, NH4H2PO41, Na2CO3	&1 MnO2 + 2 NH4H2PO4 + 2 Na2CO3 == 1 Na4MnP2O9 + 2 CO2 + 3 H2O + 2 NH3\\
2&	MnCO3, Na4P2O7	&1 MnCO3 + 1 Na4P2O7 + 0.5 O2 == 1 Na4MnP2O9 + 1 CO2\\
3&	MnO, Na2CO3, NaH2PO4&	1 MnO + 1 Na2CO3 + 2 NaH2PO4 + 0.5 O2 == 1 Na4MnP2O9 + 1 CO2 + 2 H2O\\
4&	(NH4)2HPO41, MnO2, Na2CO3&	2 (NH4)2HPO4 + 1 MnO2 + 2 Na2CO3 == 1 Na4MnP2O9 + 2 CO2 + 3 H2O + 4 NH3\\
5&	(NH4)2HPO41, Mn2O3, Na2CO3	&2 (NH4)2HPO4 + 0.5 Mn2O3 + 2 Na2CO3 + 0.25 O2 == 1 Na4MnP2O9 + 2 CO2 + 3 H2O + 4 NH3\\
6&	Mn3O4, Na4P2O7	&0.333 Mn3O4 + 1 Na4P2O7 + 0.333 O2 == 1 Na4MnP2O9\\
7&	Mn2O3, Na2CO3, NaH2PO4&	0.5 Mn2O3 + 1 Na2CO3 + 2 NaH2PO4 + 0.25 O2 == 1 Na4MnP2O9 + 1 CO2 + 2 H2O\\
8&	MnCO3, Na2CO3, NaH2PO4	&1 MnCO3 + 1 Na2CO3 + 2 NaH2PO4 + 0.5 O2 == 1 Na4MnP2O9 + 2 CO2 + 2 H2O\\
9&	MnO2, Na4P2O7	1& MnO2 + 1 Na4P2O7 == 1 Na4MnP2O9\\
10&	Mn2O3, Na2O2, NaH2PO4	&0.5 Mn2O3 + 1 Na2O2 + 2 NaH2PO4 == 1 Na4MnP2O9 + 2 H2O + 0.25 O2\\
\hline \hline
\end{tabular}
\end{table}

\begin{table}[H]
\tiny
\caption{\label{tab:s4_Na5GeP2O9} Precursor recommendations for \ce{Na5GeP2O9} from the solid-state synthesis science analyzer (S4), as distributed by \href{https://matsynthesis.org}{matsynthesis.org} \cite{TextMinedSynthesisRecipes}.}
\begin{tabular}{lcc}
\hline \hline
\textbf{Ranking} & \textbf{Precursors} & \textbf{Reaction} \\
\hline
1&	GeO2, NH4H2PO41, Na2CO3&	1 GeO2 + 2 NH4H2PO4 + 2.5 Na2CO3 == 1 Na5GeP2O9 + 2.5 CO2 + 3 H2O + 2 NH3 + 0.25 O2\\
2&	GeO2, Na2CO3, Na4P2O7	&1 GeO2 + 0.5 Na2CO3 + 1 Na4P2O7 == 1 Na5GeP2O9 + 0.5 CO2 + 0.25 O2\\
3&	(NH4)2HPO41, GeO2, Na2CO3&	2 (NH4)2HPO4 + 1 GeO2 + 2.5 Na2CO3 == 1 Na5GeP2O9 + 2.5 CO2 + 3 H2O + 4 NH3 + 0.25 O2\\
4&	GeO2, Na2CO3, NaH2PO4&	1 GeO2 + 1.5 Na2CO3 + 2 NaH2PO4 == 1 Na5GeP2O9 + 1.5 CO2 + 2 H2O + 0.25 O2\\
5&	(NH4)3PO4, GeO2, Na2CO3	&2 (NH4)3PO4 + 1 GeO2 + 2.5 Na2CO3 == 1 Na5GeP2O9 + 2.5 CO2 + 3 H2O + 6 NH3 + 0.25 O2\\
6&	GeO2, Na2CO3, NaPO3&	1 GeO2 + 1.5 Na2CO3 + 2 NaPO3 == 1 Na5GeP2O9 + 1.5 CO2 + 0.25 O2\\
7&	GeO2, NH4H2PO41, NaHO	&1 GeO2 + 2 NH4H2PO4 + 5 NaHO == 1 Na5GeP2O9 + 5.5 H2O + 2 NH3 + 0.25 O2\\
\hline \hline
\end{tabular}
\end{table}

\begin{table}[H]
\tiny
\caption{\label{tab:s4_Na4ZrP2O9} Precursor recommendations for \ce{Na4ZrP2O9} from the solid-state synthesis science analyzer (S4), as distributed by \href{https://matsynthesis.org}{matsynthesis.org} \cite{TextMinedSynthesisRecipes}.}
\begin{tabular}{lcc}
\hline \hline
\textbf{Ranking} & \textbf{Precursors} & \textbf{Reaction} \\
\hline
1&	NH4H2PO41, Na2CO3, ZrO2	&2 NH4H2PO4 + 2 Na2CO3 + 1 ZrO2 == 1 Na4ZrP2O9 + 2 CO2 + 3 H2O + 2 NH3\\
2&	(NH4)2HPO41, Na2CO3, ZrO2&	2 (NH4)2HPO4 + 2 Na2CO3 + 1 ZrO2 == 1 Na4ZrP2O9 + 2 CO2 + 3 H2O + 4 NH3\\
3&	(NH4)3PO4, Na2CO3, ZrO2	&2 (NH4)3PO4 + 2 Na2CO3 + 1 ZrO2 == 1 Na4ZrP2O9 + 2 CO2 + 3 H2O + 6 NH3\\
4&	Na4P2O7, ZrO2	&1 Na4P2O7 + 1 ZrO2 == 1 Na4ZrP2O9\\
5&	Na2CO3, NaH2PO4, ZrO2	&1 Na2CO3 + 2 NaH2PO4 + 1 ZrO2 == 1 Na4ZrP2O9 + 1 CO2 + 2 H2O\\
6&	Na2CO3, NaPO3, ZrO2	&1 Na2CO3 + 2 NaPO3 + 1 ZrO2 == 1 Na4ZrP2O9 + 1 CO2\\
7&	NH4H2PO41, NaNO3, ZrO2&	2.0 NH4H2PO4 + 4.0 NaNO3 + 1.0 ZrO2 == 1.0 Na4ZrP2O9 + 1.0 H2O + 3.333 NH3 + 3.333 O2 + 2.667 NO2\\
8&	NH4H2PO41, NaHO, ZrO2	&2 NH4H2PO4 + 4 NaHO + 1 ZrO2 == 1 Na4ZrP2O9 + 5 H2O + 2 NH3\\
\hline \hline
\end{tabular}
\end{table}

\begin{table}[H]
\tiny
\caption{\label{tab:s4_Na4TiP2O9} Precursor recommendations for \ce{Na4TiP2O9} from the solid-state synthesis science analyzer (S4), as distributed by \href{https://matsynthesis.org}{matsynthesis.org} \cite{TextMinedSynthesisRecipes}.}
\begin{tabular}{lcc}
\hline \hline
\textbf{Ranking} & \textbf{Precursors} & \textbf{Reaction} \\
\hline
1&	NH4H2PO41, Na2CO3, TiO2	&2 NH4H2PO4 + 2 Na2CO3 + 1 TiO2 == 1 Na4TiP2O9 + 2 CO2 + 3 H2O + 2 NH3\\
2&	(NH4)2HPO41, Na2CO3, TiO2	&2 (NH4)2HPO4 + 2 Na2CO3 + 1 TiO2 == 1 Na4TiP2O9 + 2 CO2 + 3 H2O + 4 NH3\\
3&	Na2CO3, NaH2PO4, TiO2&	1 Na2CO3 + 2 NaH2PO4 + 1 TiO2 == 1 Na4TiP2O9 + 1 CO2 + 2 H2O\\
4&	Na4P2O7, TiO2	&1 Na4P2O7 + 1 TiO2 == 1 Na4TiP2O9\\
5&	NH4H2PO41, NaHO, TiO2	&2 NH4H2PO4 + 4 NaHO + 1 TiO2 == 1 Na4TiP2O9 + 5 H2O + 2 NH3\\
6&	Na2CO3, NaPO3, TiO2	&1 Na2CO3 + 2 NaPO3 + 1 TiO2 == 1 Na4TiP2O9 + 1 CO2\\
7&	NH4H2PO41, NaNO3, TiO2	&2.0 NH4H2PO4 + 4.0 NaNO3 + 1.0 TiO2 == 1.0 Na4TiP2O9 + 1.0 H2O + 3.333 NH3 + 3.333 O2 + 2.667 NO2\\
\hline \hline
\end{tabular}
\end{table}

\subsection{Reaction Network}

Reaction network output for the \ce{Na4SnP2O9} target

\begin{table}[H]
\tiny
\caption{\label{tab:supp_rxn_network} Precursor sets recommended for \ce{Na4SnP2O9} by reaction network \cite{McDermott_reactionNetwork_model, Matt_Selectivity_Model}.}
\begin{tabular}{cccccccc}
\hline \hline
\textbf{Ranking} &	\textbf{Reaction} &	\textbf{Energy}	& \textbf{Added Elements}	& \textbf{Separable} &	\textbf{Primary Competition}	& \textbf{Secondary Competition}	& \textbf{Cost}  \\
\hline
51	& Na2SnO3 + 2 NaPO3 $\rightarrow$ Na4SnP2O9 &	-0.10123	&	&TRUE 	&-0.01905 &	0.01604	&-0.01148 \\
62&	Na4P2O7 + SnO2 $\rightarrow$ Na4SnP2O9	&-0.01734	&	&TRUE	&-0.017386	&0&	-0.00956\\
120&	Na2PHO4 + 0.5 SnO $\rightarrow$ 0.5 Na4SnP2O9 + 0.5 H2	&-0.05420	&H	&TRUE	&-0.02700 &	0.03218 &	-0.00309\\
733	&0.3333 Na2Sn3O7 + 0.6667 Na5P3O10 $\rightarrow$ Na4SnP2O9	&-0.048830 &		&TRUE	&-0.00686	&0.02979	&0.00544\\
2840&	P2H8O9 + Na4SnS4 $\rightarrow$ Na4SnP2O9 + 4 H2S&	-0.09984	&H-S	&TRUE	&0.00159 &	0.06242	&0.01882\\
4700&	Na4P2O7 + SnH4(Cl2O)2 $\rightarrow$ Na4SnP2O9 + 4 HCl	&-0.02153	&Cl-H	&TRUE&	0.02818&	0.03585 &	0.026659\\
5378&	2 NaPH3NO3 + Na2SnO3 $\rightarrow$ Na4SnP2O9 + 3 H2 + N2	& -0.12238	&H-N	&TRUE	&-0.01820 &	0.11024	&0.029177\\
7620&	Na2P2H2O7 + Na2SnO2 $\rightarrow$ Na4SnP2O9 + H2	&-0.15264	&H &	TRUE	&0.00269	&0.11294	&0.036772\\
8983&	2 PH4NO3 + Na4SnO3 $\rightarrow$ Na4SnP2O9 + 4 H2 + N2&	-0.23024 &	H-N	&TRUE	&0.00880	&0.13330 &	0.04092\\
9922&	Na2PO3F + 0.5 Li2SnO3 $\rightarrow$ 0.5 Na4SnP2O9 + LiF&	-0.11954	&F-Li	&TRUE	&0.00375 &	0.12007	&0.04377\\
14111&	Na2SnO3 + 2 NaPH3NO3 $\rightarrow$ Na4SnP2O9 + 2 H3N	&-0.07163 &	H-N	&TRUE&	0.03255 &	0.11024 &	0.05709\\
14606&	Na4SnO4 + K2P2O5F2 $\rightarrow$ Na4SnP2O9 + 2 KF	&-0.22289	&F-K	&TRUE	&0.00584	&0.17443 &	0.05883\\
17061&	Na2PO3F + 0.5 K2SnO3 $\rightarrow$ KF + 0.5 Na4SnP2O9&	-0.13778 &	F-K	&TRUE&	0.00533	&0.17593	&0.06779\\
19421&	KSnO2 + Na4P2O7 $\rightarrow$ Na4SnP2O9 + K	&0.04561	&K	&TRUE&	0.09258	&0.07144 &	0.07837\\
19882&	0.5 Na8SnO6 + 0.5 Sn(PO3)4 $\rightarrow$ Na4SnP2O9	&-0.38202	& &	TRUE	&-0.00164	&0.26620	&0.08085\\
20740&	SnP2O7 + 2 Na2O $\rightarrow$ Na4SnP2O9	&-0.37956	&	&TRUE	&0.05431	&0.22056	&0.08574\\
21126&	2 LiPH4NO3F + Na4SnO3 $\rightarrow$ Na4SnP2O9 + 2 LiF + N2 + 4 H2	&-0.23516	&F-H-Li-N&	TRUE	&0.01416 &	0.23382 & 0.08807\\
22448&	2 PH6NO3 + Na4SnO3 $\rightarrow$ Na4SnP2O9 + 6 H2 + N2	&-0.27945	&H-N&	TRUE	&0.00146	&0.27711	&0.09741\\
22815&	KPO3 + 0.5 Na4SnO3 $\rightarrow$ 0.5 Na4SnP2O9 + K	&-0.02608 &	K	&TRUE	&0.15454 &	0.07438	&0.10041\\
23396	& K2SnO2 + Na4P2O7 $\rightarrow$ Na4SnP2O9 + 2 K&	0.07852 &	K	&TRUE	&0.16460	&0.05207	&0.10535\\
23428	&K2SnP2O7 + 2 Na2O $\rightarrow$ Na4SnP2O9 + 2 K	&-0.07141	&K&	TRUE	&0.17181&	0.07859 &	0.10554\\
23598&	Na2PHO3 + 0.5 K2SnO3 $\rightarrow$ KH + 0.5 Na4SnP2O9	&-0.07983 &	H-K	&TRUE	&0.07467	&0.18061 &	0.10689 \\
23668&	Na4SnO3 + 2 P(HO)3 $\rightarrow$ Na4SnP2O9 + 3 H2	&-0.35459	&H&	TRUE	&-0.00966	&0.32742&	0.10753\\
23706&	PO3 + 0.5 Na4SnO3 $\rightarrow$ 0.5 Na4SnP2O9&	-0.65107	& &	TRUE& -0.04599 &	0.43016 &0.10777\\
26854&	Na8SnO6 + LiSn(PO3)4 $\rightarrow$ Li + 2 Na4SnP2O9&	-0.29865&	Li	&TRUE	&0.06154	&0.31466	&0.13942\\
27384&	Li2SnO3 + 2 Na2PHO3 $\rightarrow$ 2 LiH + Na4SnP2O9&	0.03878&	H-Li	&TRUE	&0.18233 &	0.13246	&0.14553\\
27903&	LiPO3 + 0.5 Na4SnO3 $\rightarrow$ Li + 0.5 Na4SnP2O9&	-0.024226&	Li&	TRUE	&0.21281 &	0.12985	&0.15177\\
28371&	Na8SnO6 + Li2Sn(PO3)4 $\rightarrow$ 2 Li + 2 Na4SnP2O9	&-0.20352&	Li	&TRUE	&0.12053	&0.27603	&0.15810\\
29850&	LiSnP2O7 + 2 Na2O $\rightarrow$ Li + Na4SnP2O9	&-0.24540&	Li	&TRUE	&0.15153	&0.30555&	0.18115\\
29983&	Li2SnP2O7 + 2 Na2O $\rightarrow$ 2 Li + Na4SnP2O9	&-0.07628 &	Li&	TRUE	&0.23294	&0.19185	&0.18353\\
30232&	Na2SO3 + 0.5 SnP2Cl8O3 $\rightarrow$ 0.5 Na4SnP2O9 + SCl4	&0.05270	&Cl-S	&TRUE	&0.21466	&0.19137	&0.18798\\
30516&	P2O5 + Na4SnO4 $\rightarrow$ Na4SnP2O9 &	-0.51695	&	&TRUE&	0.02213	&0.52151	&0.19294\\
34880&	Na2CO3 + 0.5 SnP2Cl8O3 $\rightarrow$ CCl4 + 0.5 Na4SnP2O9&	0.11201 &	C-Cl	&TRUE&	0.44134	&0.38161	&0.38153\\

\hline \hline
\end{tabular}
\end{table}

\section{Precursor Sourcing and Preparation Information}

Precursors utilized within synthesis experiments in this work. All precursors are bought as powders if possible, and small chunks if not (\textit{i.e.} \ce{Na4P2O9}).

\subsection{Purchased Precursor Information}
\begin{table}[H]
\tiny
\caption{\label{tab:precursor-info}Precursor information, including the CAS identification number, vendor, and reported purity. Any notes, such as whether a material was sold as a nanoparticle of a maximum size, is reported in the notes section.}
\begin{tabular}{lccccc}
\hline \hline
\textbf{Precursor Name} & \textbf{Formula} & \textbf{CAS Number} & \textbf{Vendor} & \textbf{Purity} & \textbf{Notes} \\
\hline
Sodium carbonate & \ce{Na2CO3}	& 
497-19-8  &   Sigma-Aldrich & $\geq$99.5\% &  \\
Sodium hexametaphosphate & \ce{(NaPO3)6}	& 68915-31-1  &  Sigma-Aldrich & 96\%&crystalline, +200 mesh\\
Pentasodium tripolyphosphate & \ce{Na5P3O10}	&  7758-29-4 &   Sigma-Aldrich  & 85\%  & Technical Grade  \\
Disodium hydrogen phosphate & \ce{Na2HPO4}	& 7558-79-4  & Sigma-Aldrich  & $\geq$99.0\% & anhydrous\\
Tetrasodium pyrophosphate decahydrate & \ce{Na4P2O7}$\cdot$\ce{H2O} & 13472-36-1  &   Sigma-Aldrich&  $>$99\%&  vacuum dried to become \ce{Na4P2O7} \\
\hline
Ammonium phosphate monobasic & \ce{NH4H2PO4}   & 7722-76-1   &   Sigma-Aldrich&   $>$99.99\%&  \\
Ammonium phosphate dibasic & \ce{(NH4)2HPO4}   &  7783-28-0  &   Sigma-Aldrich&   $>$99.99\%&  \\
\hline
Aluminum (III) oxide & \ce{Al2O3}   &  1344-28-1  & Sigma-Aldrich  & 99.5\%  & \\
Chromium (III) oxide & \ce{Cr2O3}   &  1308-38-9  & Sigma-Aldrich  & 99.9\%  & \\
Gallium (III) oxide & \ce{Ga2O3}   &  12024-21-4  &  Sigma-Aldrich &  $>$99.99\% & \\
Germanium(IV) oxide & \ce{GeO2}   &  1310-53-8  &  Sigma-Aldrich & 99.998\%  & \\
Hafnium (IV) oxide & \ce{HfO2}   & 12055-23-1   &  Sigma-Aldrich & 98\%  & \\
Iron (III) oxide & \ce{Fe2O3}   &  1309-37-1  & Sigma-Aldrich  &  $\geq$96\% & $<$5\,$\mu$m particle size\\
Indium (III) oxide & \ce{In2O3}   &  1312-43-2  &  Sigma-Aldrich & 99.99\% & $<$0.01\,mmHg (25$\circ$C) \\
Lead (IV) oxide & \ce{PbO2}   &  1309-60-0  &  BeanTown Chemical & $\geq$97.0\%  & \\
Manganese (II, III) oxide & \ce{Mn3O4}   &  
1317-35-7  & Sigma-Aldrich  &  97\% & \\
Molybdenum (IV) oxide & \ce{MoO2}   &  18868-43-4  &  Sigma-Aldrich &  99\% & \\
Tantalum (V) oxide & \ce{Ta2O5}   & 1314-61-0   &   Sigma-Aldrich &   99.99\%& $<$20um particle size\\
Tin (IV) oxide & \ce{SnO2}   &  18282-10-5  &   Sigma-Aldrich &   99.99\%&  \\
Tin (IV) oxide & \ce{SnO2}   &  18282-10-5  &   Sigma-Aldrich &   99.99\%& nanoparticle (300 mesh) \\
Titanium (IV) oxide & \ce{TiO2}   &  1317-70-0  & Sigma-Aldrich  &  99.\% & anatase\\
Ruthenium (IV) oxide & \ce{RuO2}   &  12036-10-1  & Sigma-Aldrich  &  99.9\% & \\
Vanadium (V) oxide & \ce{V2O5}   & 1314-62-1 & Sigma-Aldrich  &  $\geq$98\% &  \\
Zirconium (IV) oxide & \ce{ZrO2}   &   1314-23-4 & Sigma-Aldrich  &  99\% & 5\,$\mu$m particle size\\
\hline \hline
\end{tabular}
\end{table}

\subsection{\ce{Na2SnO3} preparation and characterization}
\ce{Na2SnO3} is prepared by mixing stoichiometric amounts of \ce{Na2CO3} and \ce{SnO2} with method I. Samples are heated in air at 5\,$^\circ$C/min up to 750\,$^\circ$C, and held at their dwell temperature for 12 hours. The samples are then allowed to cool naturally in the furnace. All \ce{Na2SnO3} is confirmed with powdered XRD prior to use in further experimentation.

\begin{figure}[H]
    \centering
    \includegraphics[width=1.0\linewidth]{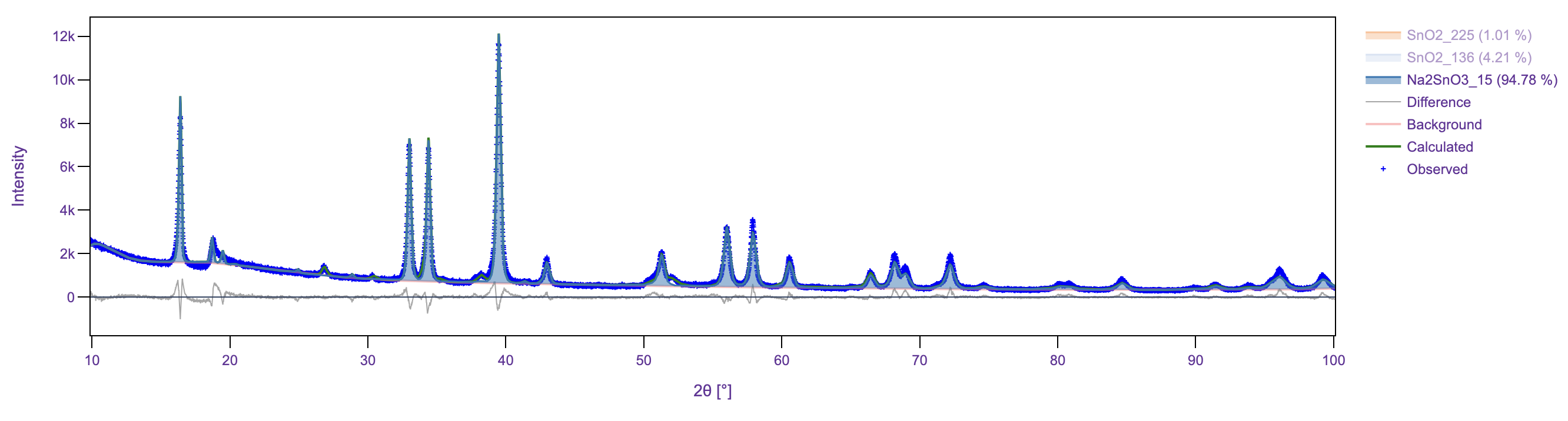}
    \caption{Refinement of a powdered XRD pattern for the prepared precursor of \ce{Na2SnO3}.}
    \label{fig:na2sno3_xrd_refinement}
\end{figure}

The powder XRD pattern and fitting is shown below (\autoref{fig:na2sno3_xrd_refinement}). When fitting the pattern with the \ce{Na2SnO3} and \ce{SnO2} phases, there are no unattributed peaks, no additional peaks from the calculated pattern, and there is an RWP of 8.51\% with refinement.

\subsection{\ce{Na4P2O7}$\cdot$10\ce{H2O} Drying and Characterization}
Na4P2O7 put into an uncapped plastic vial and allowed to dry at 70\,$^\circ$C under vacuum for at least 24 hours. The vial is taken out and capped immediately before weighing. After weighing, the \ce{Na4P2O7} precursor is put back into the vacuum oven.

To confirm any additional water content which could contribute to inaccurate weighing, thermogravimetric analysis was performed on the precursor after drying. No weight loss was observed up to 300\,$^\circ$C, confirming that no \ce{H2O} units remain after drying.

To confirm that significant drying does not break down the polyphosphate, we performed XRD on the dried precursor. No additional peaks are found beyond those fit to a calculated curve refined for orthorhombic \ce{Na4P2O7}. The RWP is 9.05\%.

\begin{figure}[H]
    \centering
    \includegraphics[width=0.75\linewidth]{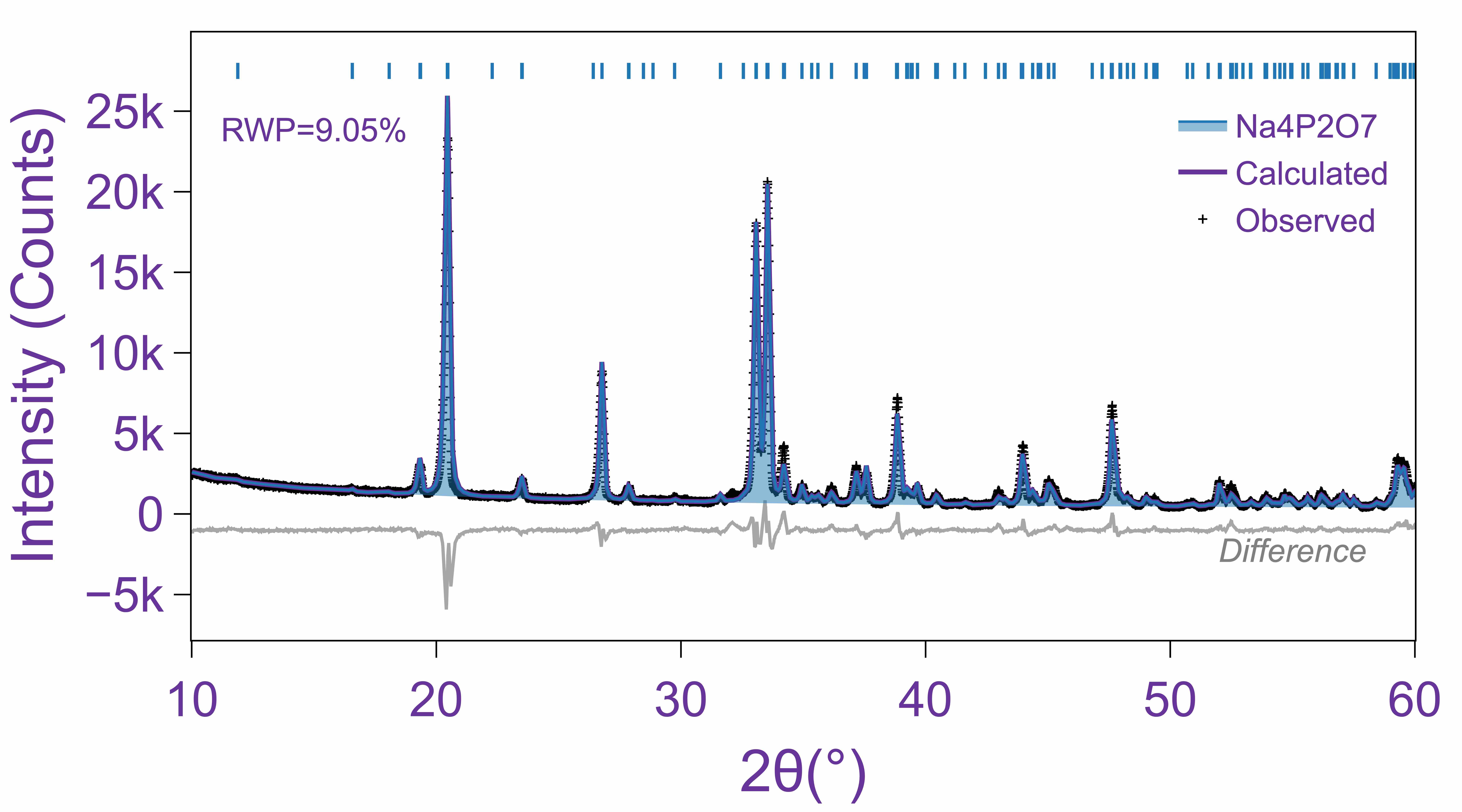}
    \caption{Refinement of XRD of \ce{Na4P2O7} precursor after vacuum drying.}
    \label{fig:xrd_na4p2o7}
\end{figure}

\section{Precursor Reactions}

\subsection{Precursor Reaction Interfaces}

\subsection{Sodium Phosphate Precursor Reactions}

\begin{table}[H]
\caption{\label{tab:sodium_phosphate_pairwise_PM} Table on the reactions of sodium phosphate reactions at specific temperatures. All precursors are ground for 10 minutes with a pestle and mortar first. The powder is then transferred to a crucible and then heated at a rate of 5$^\circ$C per minute. Samples are held at the maximum dwell temperature for 12 hours. Recovered powder is characterized by XRD.}
\begin{tabular}{lcccc}
\hline \hline
{} & \multicolumn{4}{c}{\textbf{Product found after dwell}} \\
\textbf{Precursor(s)} & \textbf{80\,$^\circ$C}  & \textbf{150\,$^\circ$C} & \textbf{300\,$^\circ$C} & \textbf{500\,$^\circ$C} \\
\hline
\ce{Na2CO3} +  &  \ce{Na2CO3} +  &  \ce{Na2CO3} + &  \ce{Na5P3O10} +  & \ce{Na5P3O10} + \\
 \ce{NH4H2PO4} & \ce{NH4H2PO4} &  \ce{Na4P2O7} &  \ce{Na3PO4} & \ce{Na3PO4}\\
 \hline
\ce{Na2HPO4} & \ce{Na2HPO4}&  \ce{Na2HPO4}&  \ce{Na4P2O7}& \ce{Na4P2O7}\\
\hline
\ce{Na4P2O7} & \ce{Na4P2O7}&  \ce{Na4P2O7}&  \ce{Na4P2O7}& \ce{Na4P2O7} + \\
{} & {} &  {} &  {} & \ce{Na3PO4}\\
\hline \hline
\end{tabular}
\end{table}

\begin{table}[H]
\caption{\label{tab:sodium_phosphate_pairwise_wetspinmix} Table on the reactions of sodium phosphate reactions at specific temperatures. All precursors are weighed and transfered to a plastic vial with 10 zirconnia balls and wet spin mixed for 10 minutes. The slurry is then transferred to a crucible and heated to evaporate ethanol at 80\,$^\circ$C. The sample is then heated at a rate of 5$^\circ$C per minute. Samples are held at the maximum dwell temperature for 12 hours. Recovered powder is characterized by XRD.}
\begin{tabular}{lcccc}
\hline \hline
{} & \multicolumn{4}{c}{\textbf{Product found after dwell}} \\
\textbf{Precursor(s)} & \textbf{80\,$^\circ$C}  & \textbf{150\,$^\circ$C} & \textbf{300\,$^\circ$C} & \textbf{500\,$^\circ$C} \\
\hline
\ce{Na2CO3} +  &\ce{Na4P2O7} &  \ce{Na4P2O7} &  \ce{Na4P2O7} & \ce{Na4P2O7} +\\
 \ce{NH4H2PO4} & {} &  {} &  {} & \ce{Na3PO4}\\
 \hline
\ce{Na2HPO4} & \ce{Na2HPO4}&  \ce{Na2HPO4}&  \ce{Na4P2O7}& \ce{Na4P2O7}\\
\hline
\ce{Na4P2O7} & \ce{Na4P2O7}&  \ce{Na4P2O7}&  \ce{Na4P2O7}& \ce{Na4P2O7} + \\
{} & {} &  {} &  {} & \ce{Na3PO4}\\
\hline \hline
\end{tabular}
\end{table}

\section{Synthesis Results of High Throughput A Site Substitution}

\begin{table}[H]
\caption{\label{tab:Acation_synthesis_results} Synthesis results for A site candidate in the NAP prototype with the \ce{NH4H2PO4}, \ce{Na2CO3}, and \ce{A_yO_z} precursor set. Reactions are completed in the self-driving A-Lab. Sample results are characterized by Rietveld refinement of powder XRD patterns. Systems with hard bulk solids found will be labeled HBS.}
\begin{tabular}{lccc}
\hline \hline
\textbf{Target} & \textbf{E$_{\mathrm{hull}}$} (meV/atom)  & \textbf{Dwell Temperatures ($^\circ$C} & \textbf{Synthesized} \\
\hline
\ce{Na5AlP2O9} & {25.9} &  {750-1100} &  \ce{AlPO4}, \ce{Na2O2}, HBS \\ 
\ce{Na4TiP2O9} & {2.2} &  {650-950} &  \ce{Na4TiP2O7}, \ce{TiO2}, \ce{Na4P2O7} \\ %
\ce{Na3VP2O9} & {26.3} &  {750-950} &  \ce{NaVO3}, \ce{NaV2O4}, \ce{VP2O7}, HBS \\ %
\ce{Na4CrP2O9} & {19.0} &  {750-1000} &  \ce{Na4P2O7}, \ce{Cr2O3} \\ 
\ce{Na4MnP2O9} & {8.3} &  {750-950} &  \ce{MnO2}, HBS \\ 
\ce{Na4FeP2O9} & {23.4} &  {650-1000} &  \ce{NaFeO3}\ce{Fe4O5}, \ce{Fe2O3}, \ce{P2O5}, \ce{Na3PO4}, HBS \\ 
\ce{Na4GaP2O9} & {23.2} &  {750-1000} &  \ce{NaGaO2}, \ce{Ga2O3}\\ 
\ce{Na4GeP2O9} & {6.7} &  {650-1000} &  \ce{GeO2}, \ce{Na4P2O7}, HBS \\ 
\ce{Na5YP2O9} & {37.9} &  {750-1000} &  \ce{Y2O3}, \ce{Na4P2O7} \\ 
\ce{Na4ZrP2O9} & {3.0} &  {750-1100} &  \ce{ZrO2}, \ce{Na4P2O7} \\ 
\ce{Na4MoP2O9} & {17.0} &  {750-950} &  {HBS} \\ 
\ce{Na4SnP2O9} & {0.4} &  {600-1100} &  \ce{Na4SnP2O7}, \ce{TiO2}, \ce{Na4P2O7} \\ %
\ce{Na4HfP2O9} & {0} &  {750-1100} &  \ce{HfO2}, \ce{Na4P2O7} \\ 
\ce{Na3TaP2O9} & {19.8} &  {750-1000} &  \ce{Na3TaP2O9}, \ce{Ta2O5}, \ce{Na4P2O7} \\ 
\hline \hline 
\end{tabular}
\end{table}

\section{Chosen Precursor Sets for \ce{Na4SnP2O9}}

\begin{table}[H]
\tiny
\caption{\label{tab:nsp_precursor_full_list} Attempted precursor sets for \ce{Na4SnP2O9}. Successful means that any amount of NSP was ever found from the reaction. All reactions are completed with solid state synthesis method I (either ethanol mixing by hand or A-Lab preparation). Reaction energies are calculated at 900\,$^\circ$C utilizing DFT total electronic energies to approximate 0\,K enthalpy, with temperature effects included with methods by C. Bartel, et al. \cite{Gibbs_Bartel}. Gases were fit to experimental data from NIST Janaf Tables \cite{NIST_Janaf}.}
\begin{tabular}{cccccc}
\hline \hline
Precursor Set &\textbf{Precursor 1} & \textbf{Precursor 2} & \textbf{Precursor 3} &\textbf{Reaction Energy (meV/atom)} & \textbf{Target \%} \\
\hline
R1&\ce{Na2CO3}&   \ce{NH4H2PO4}  &  \ce{SnO2}  & -109.9 &  88.65\\      
R2&\ce{Na2CO3}&   \ce{NH4H2PO4}  &  \ce{Sn}  & -213.2 &  67.71\\        
R3&\ce{Na2CO3}&   \ce{(NH4)2HPO4}  &  \ce{SnO2}  & -95.3 &  84.35\\       
R4&\ce{Na2CO3}&   \ce{(NH4)2HPO4}  &  \ce{Sn}  & -181.0 &  90.24\\      
R5&\ce{Sn3(PO4)2} & \ce{Na2CO3}&   \ce{NH4H2PO4} & -174.6 &  5.68\\    
R6&\ce{Na5P3O10} &  \ce{Na2CO3} & \ce{SnO2}   & -60.6 &  0.03\\      
R7&\ce{Na2HPO4}&    \ce{SnO2}  & - & -46.6 &  1.76\\                  
R8&\ce{Na2HPO4}&    \ce{Sn}  & -& -259.6 &  91.49\\                    
R9&\ce{Na4P2O7}&    \ce{SnO2}  & - & -18.7 &  62.94\\                 
R10&\ce{Na4P2O7}&    \ce{Sn}  & - & -242.5 &  82.12\\                  
R11 &\ce{NaPO3} &  \ce{Na2SnO3} &  &  -104.1 & 36.74\\      
\hline \hline
\end{tabular}
\end{table}

\subsection{Regrinding Test}

Two sample series was subjected to six re-grind and re-heat steps. Precursors were stoichiometric amounts of \ce{Na2CO3}, \ce{NH4H2PO4}, and \ce{SnO2}. Samples were ground in a pestle and mortar for at least 10 minutes before each heating step.  The heating profile is defined as a 5\,$^\circ$C/min ramp up to 950\,$^\circ$C, and held at their dwell temperature for 12 hours. Additionally, the first sample undergoes a 150\,$^\circ$C low temperature anneal. At the end of the dwell, the sample is allowed to cool naturally. XRD is performed prior to a re-grind step.

\begin{figure}[H]
    \centering
    \includegraphics[width=0.75\linewidth]{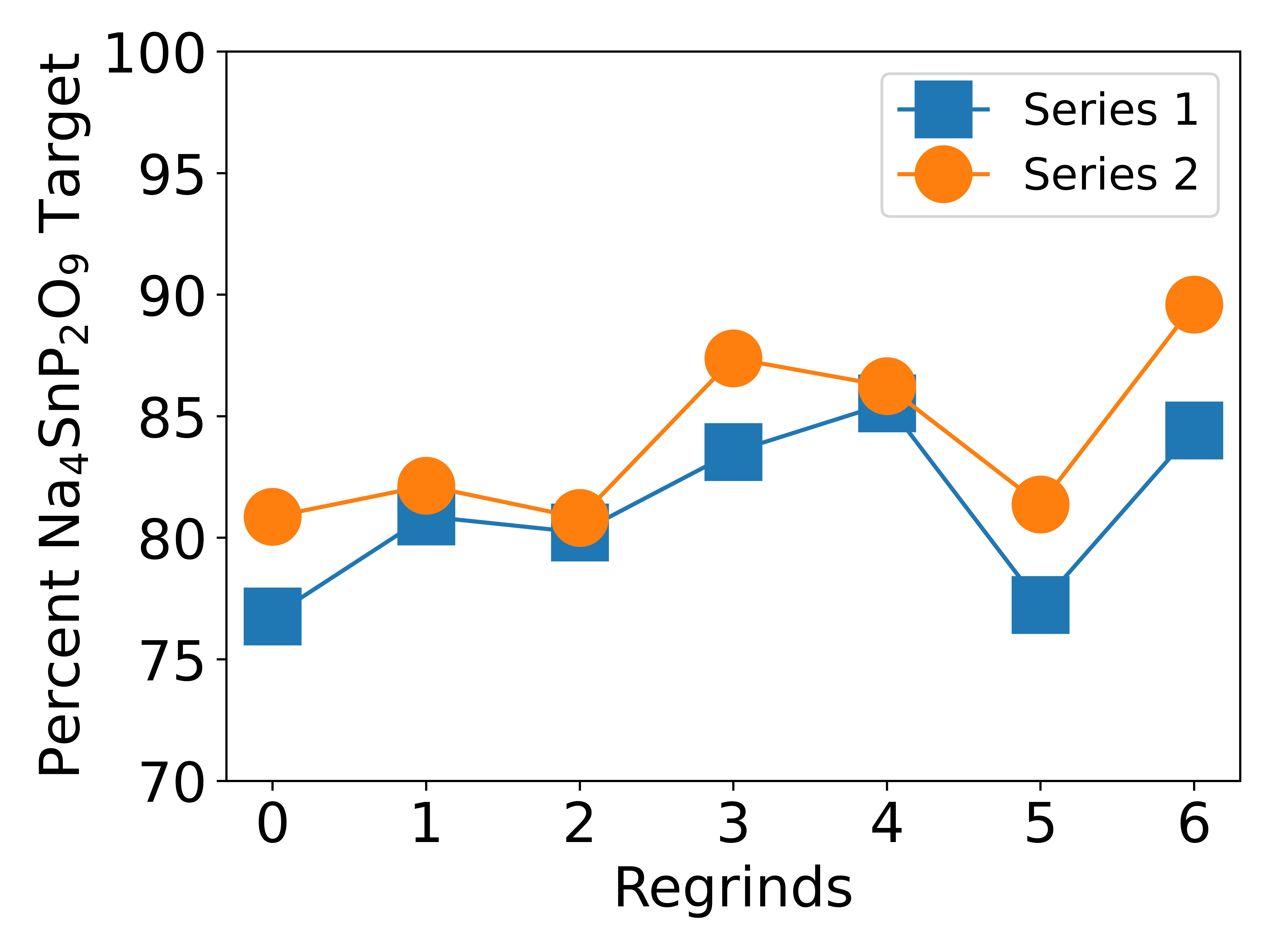}
    \caption{Two sample series' featuring six regrinds. Regrind 0 is the initial sample made from precursors.}
    \label{fig:regrind}
\end{figure}

\autoref{fig:regrind} shows the percent target recovered during a series of regrinding and reheating steps for \ce{Na4SnP2O9}. Percent target is found from Reitveld refinement with Dara. \ce{Na4P2O7} and \ce{SnO2} impurities were identified for both sample series over the course of all regrind and reheat steps. Though over time. There wasn't a distinct increase in target weight or decrease in relative impurity peak intensity with each regrind step.

\subsection{Washing of Target for Increased Purity}
\begin{figure}[H]
    \centering
    \includegraphics[width=0.75\linewidth]{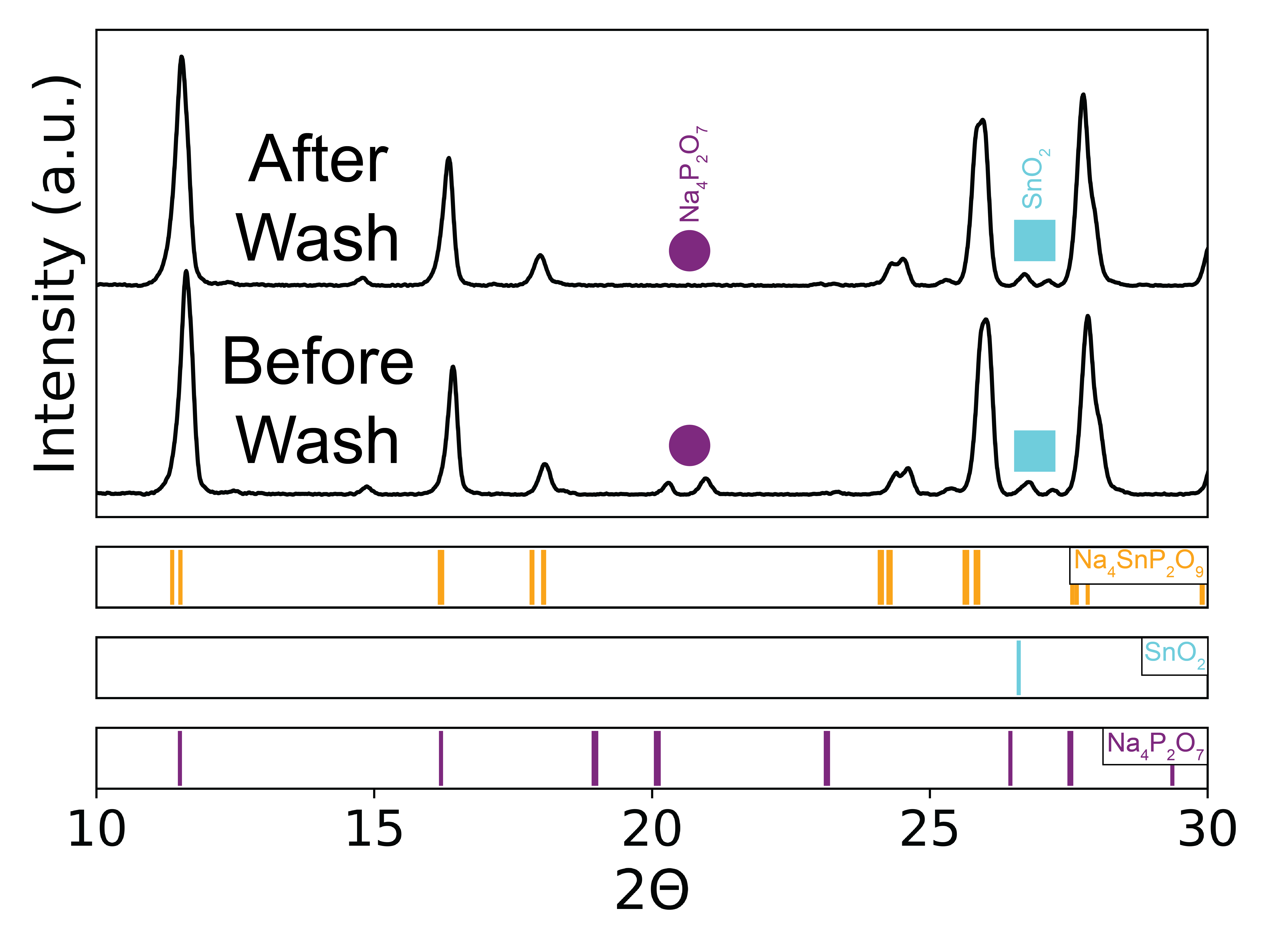}
    \caption{Sample washed with water and then ethanol, and then dried. Only sodium pyrophosphate associated powder XRD peaks are shown to decrease.}
    \label{fig:wash}
\end{figure}

We find that washing is not effective to remove all relevant impurities for \ce{Na4SnP2O9}. An approximately 76.0\% pure sample by Reitveld refinement was synthesis via solid state synthesis with mixing method I utilizing \ce{Na2CO3}, \ce{SnO2}, and \ce{NH4H2PO4} precursors. Impurities in \autoref{fig:wash} were identified by powder XRD as \ce{SnO2} (approximately 2.88\%) and \ce{Na4P2O7} (approximately 21.2\%). Approximately 1.5\,g samples were washed with 45\,mL of water twice, followed by one wash with 40\,mL. Samples were then dried in a vacuum oven at 70\,$^\circ$C for at least 24 hours. Approximately 500\,mg of powder remained after washing. Characterization via XRD found remaining \ce{SnO2} in all samples. For that shown in \autoref{fig:wash}, 3.4\% \ce{SnO2} was found to remain. The target appeared to be insoluble and unaffected by washing. Final sample purity was recorded as 96.6\%.

\section{Additional Microscopy Images}

\subsection{TEM-EDS}

\begin{figure}[H]
    \centering
    \includegraphics[width=0.75\linewidth]{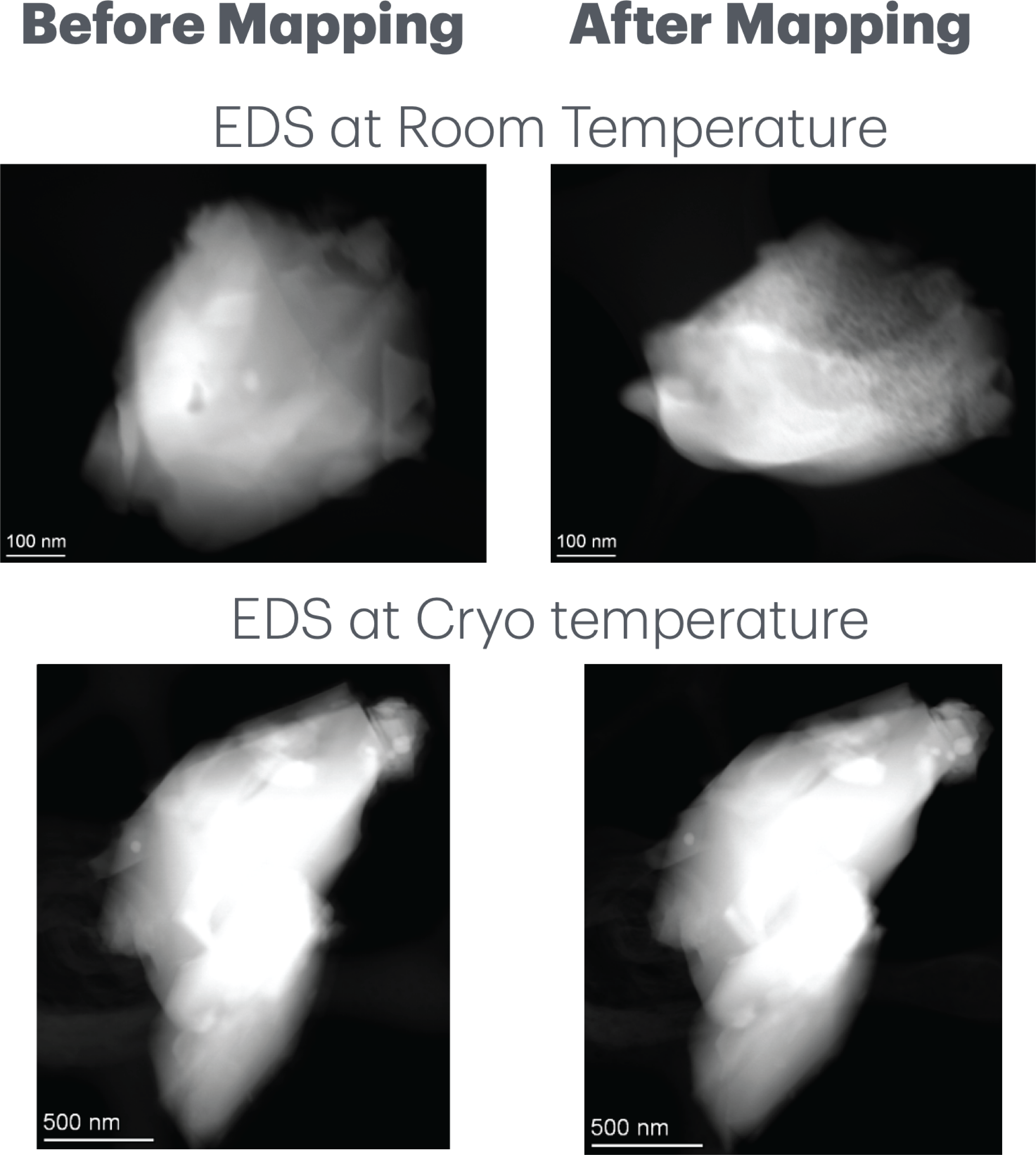}
    \caption{Demonstration of the beam damage of the particle at the room temperature. A significant difference in the STEM-HAADF micrograph is seen before and after EDS mapping at the room temperature. However, at cryogenic temperatures ($\approx$87\,K) no significant changes to the particle morphology is observed after and before EDS mapping.}
    \label{fig:eds-beam-damage}
\end{figure}

\begin{figure}[H]
    \centering
    \includegraphics[width=1.0\linewidth]{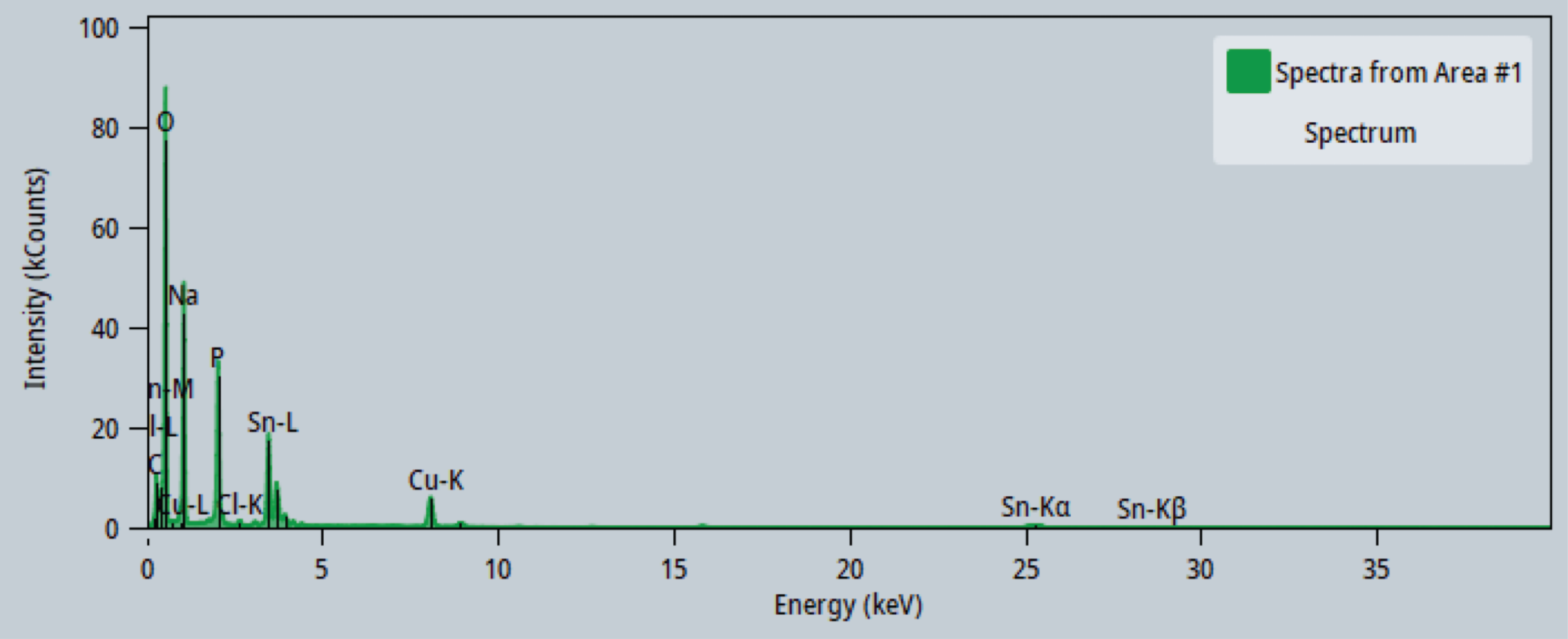}
    \caption{Summed EDS spectrum of the particle with the different elemental edges marked.}
    \label{fig:EDS_spectra_summation}
\end{figure}

\begin{figure}[H]
    \centering
    \includegraphics[width=0.75\linewidth]{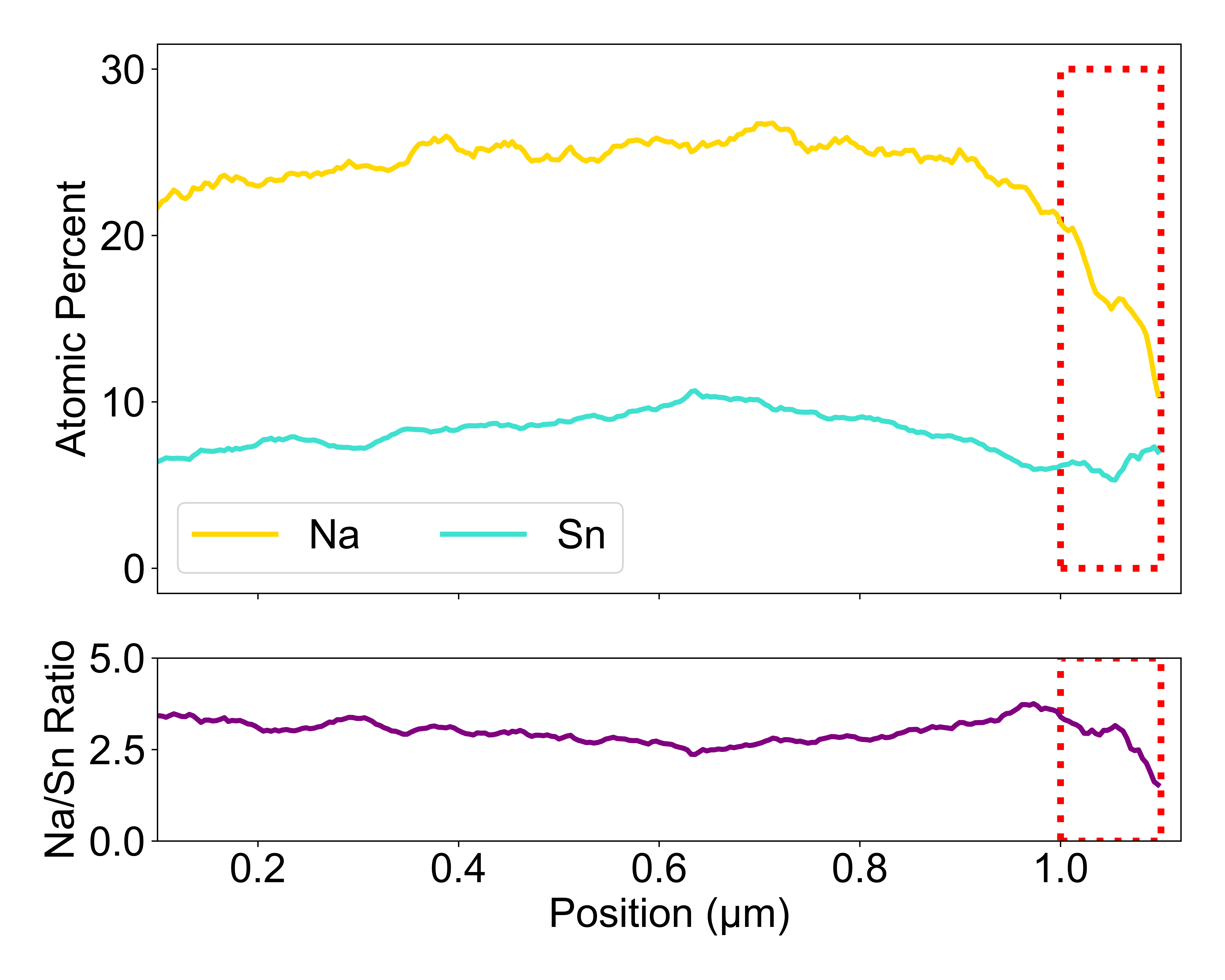}
    \caption{Full composition line profile of Na and Sn.}
    \label{fig:full_line_profile_nasn}
\end{figure}

\begin{figure}[H]
    \centering
    \includegraphics[width=0.75\linewidth]{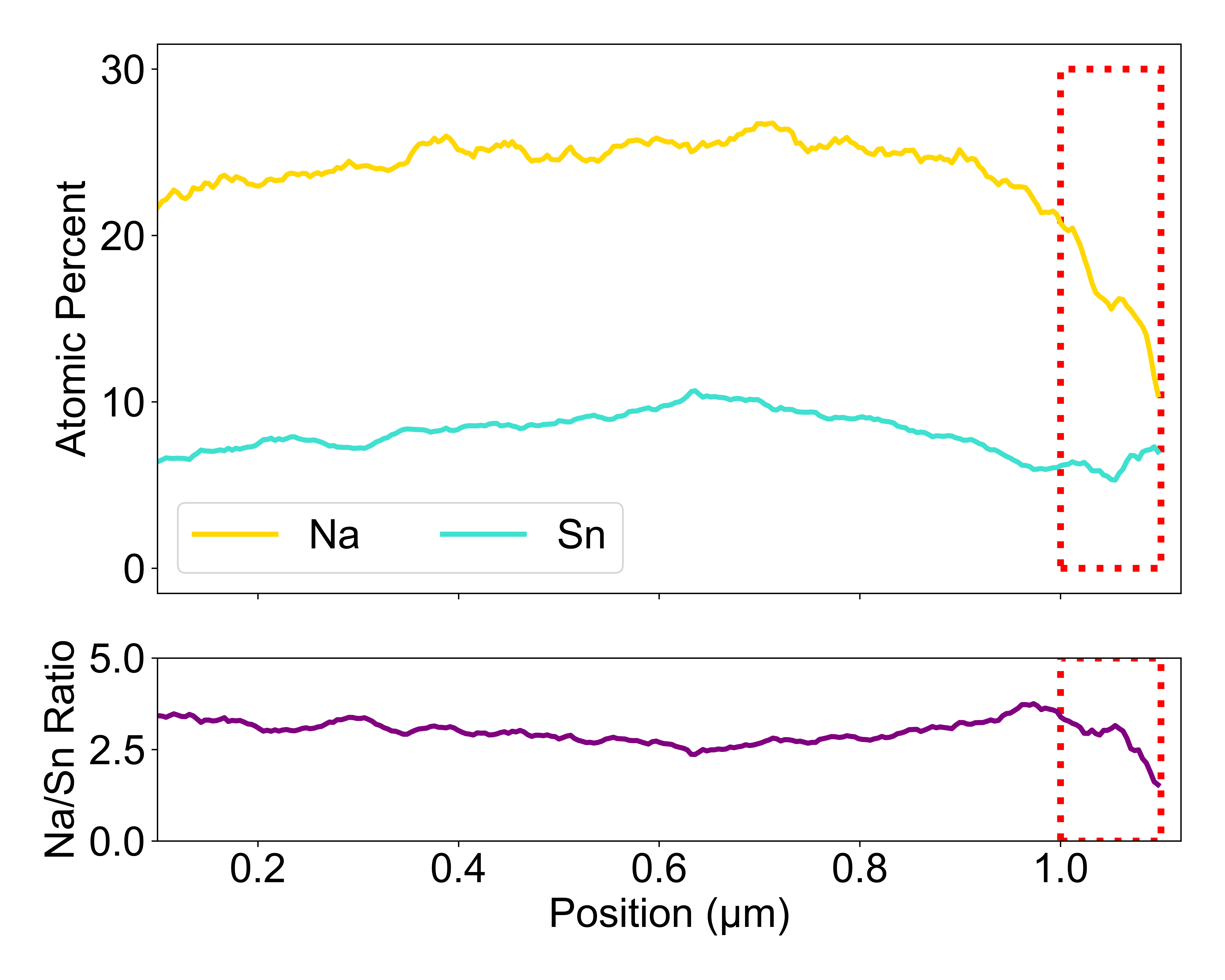}
    \caption{Composition line profile of Na and Sn with a rolling average of 10 periods ($\approx$3.8\,nm).}
    \label{fig:additional_line_profile}
\end{figure}

\subsection{SEM-EDS}

\begin{figure}[H]
    \centering
    \includegraphics[width=0.75\linewidth]{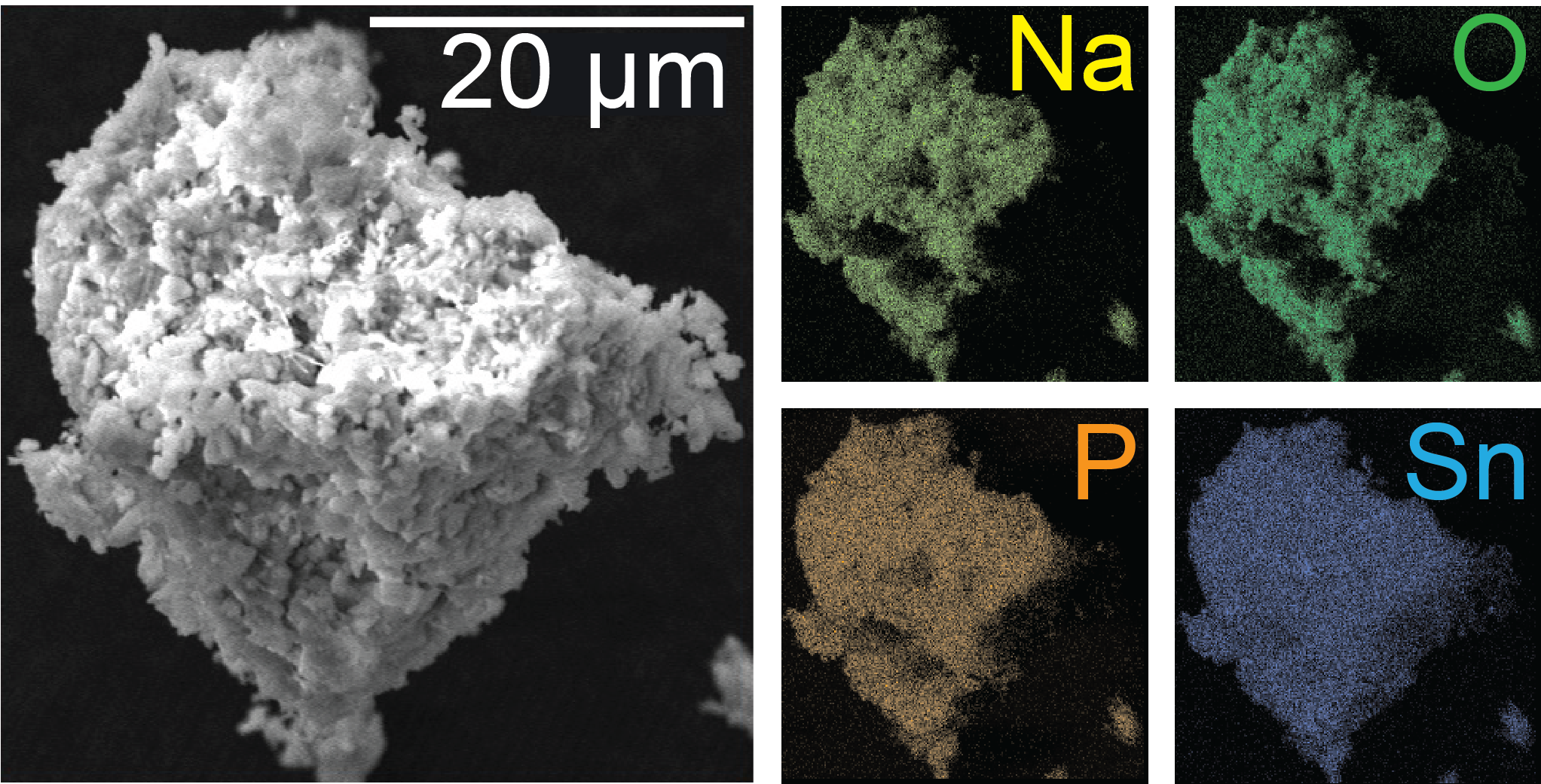}
    \caption{SEM image and EDS mapping of phase pure \ce{Na4SnP2O9}.}
    \label{fig:sem-eds}
\end{figure}

\begin{figure}[H]
    \centering
    \includegraphics[width=0.75\linewidth]{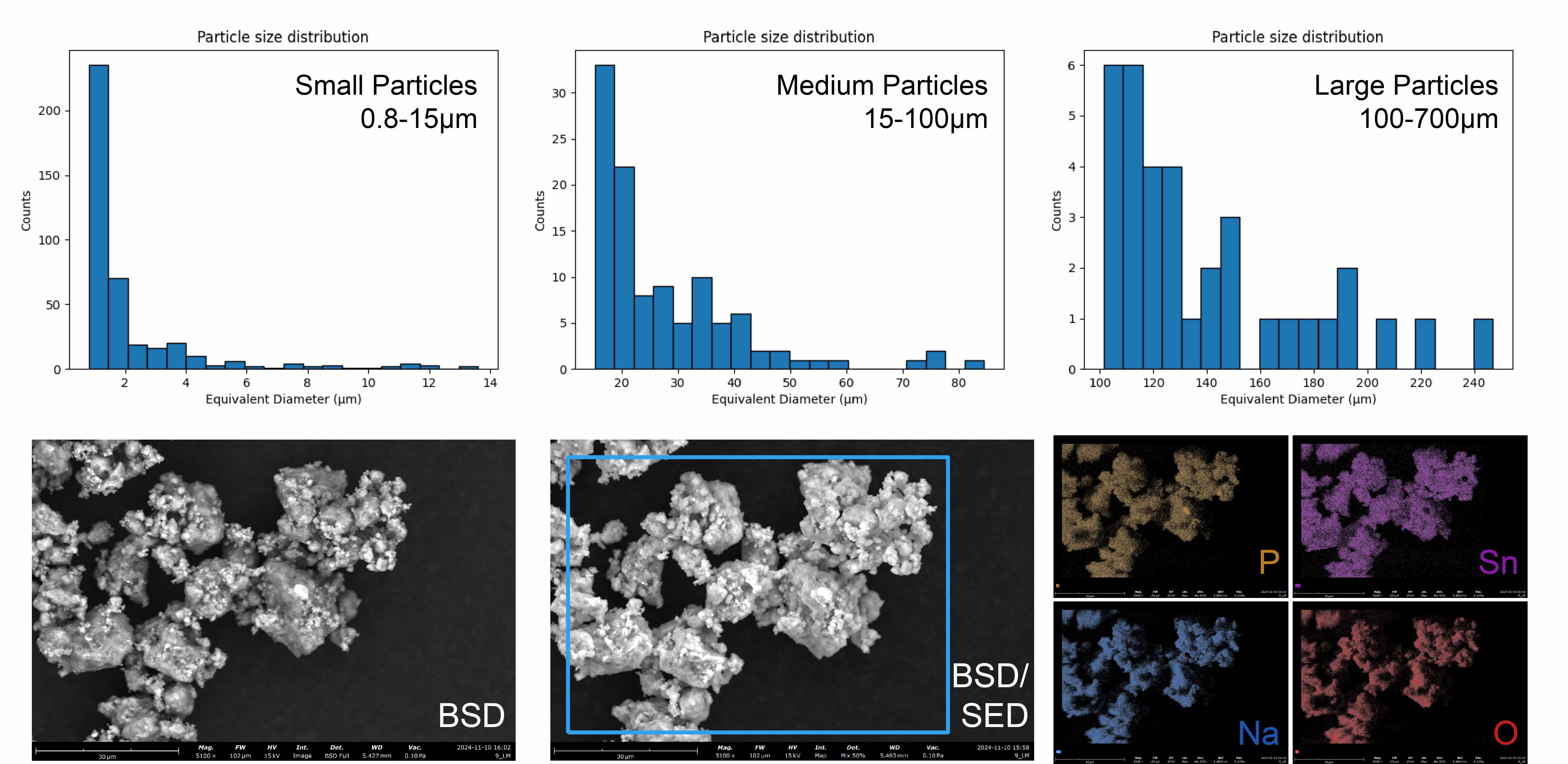}
    \caption{SEM imaging, EDS mapping, and particle size distributions for ball milled nanoscale \ce{SnO2} and \ce{Na4P2O7}.}
    \label{fig:eds-particlesize-327}
\end{figure}

\begin{figure}[H]
    \centering
    \includegraphics[width=0.75\linewidth]{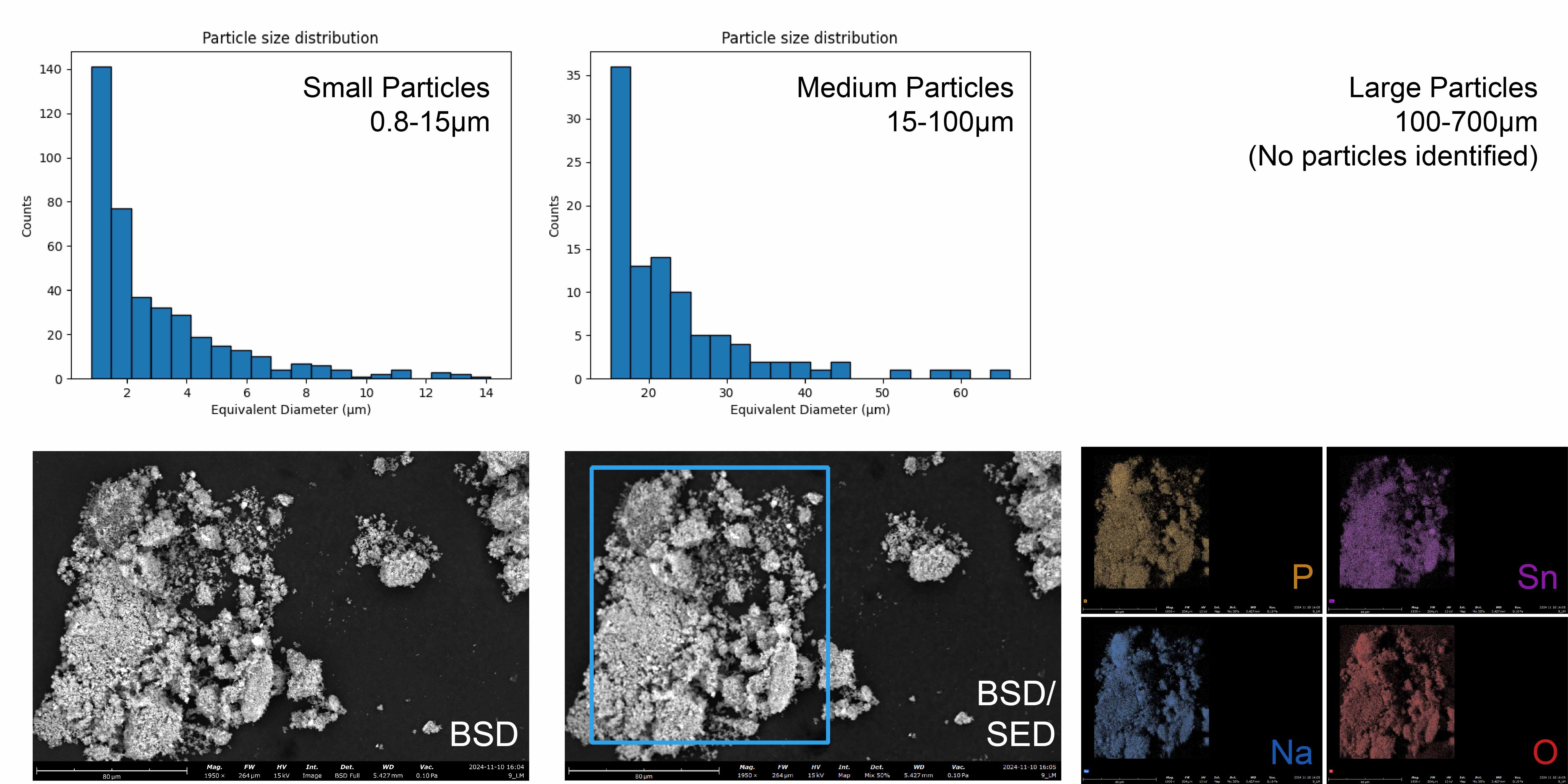}
    \caption{SEM imaging, EDS mapping, and particle size distributions for pestled and mortared nanoscale \ce{SnO2} and \ce{Na4P2O7}.}
    \label{fig:eds-particlesize-328}
\end{figure}

\begin{figure}[H]
    \centering
    \includegraphics[width=0.75\linewidth]{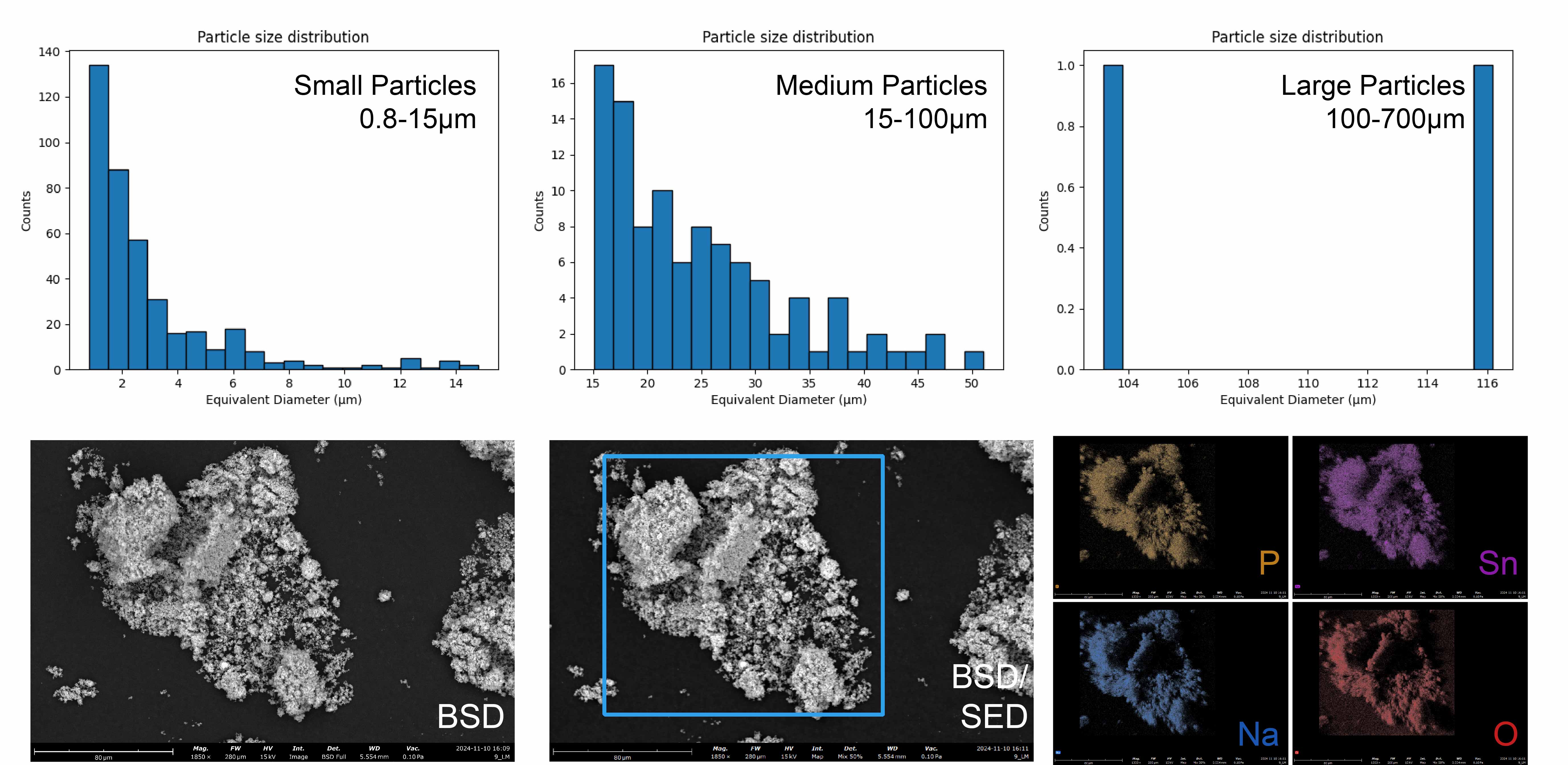}
    \caption{SEM imaging, EDS mapping, and particle size distributions for wet ball milled nanoscale \ce{SnO2} and \ce{Na4P2O7}.}
    \label{fig:eds-particlesize-329}
\end{figure}

\begin{figure}[H]
    \centering
    \includegraphics[width=0.75\linewidth]{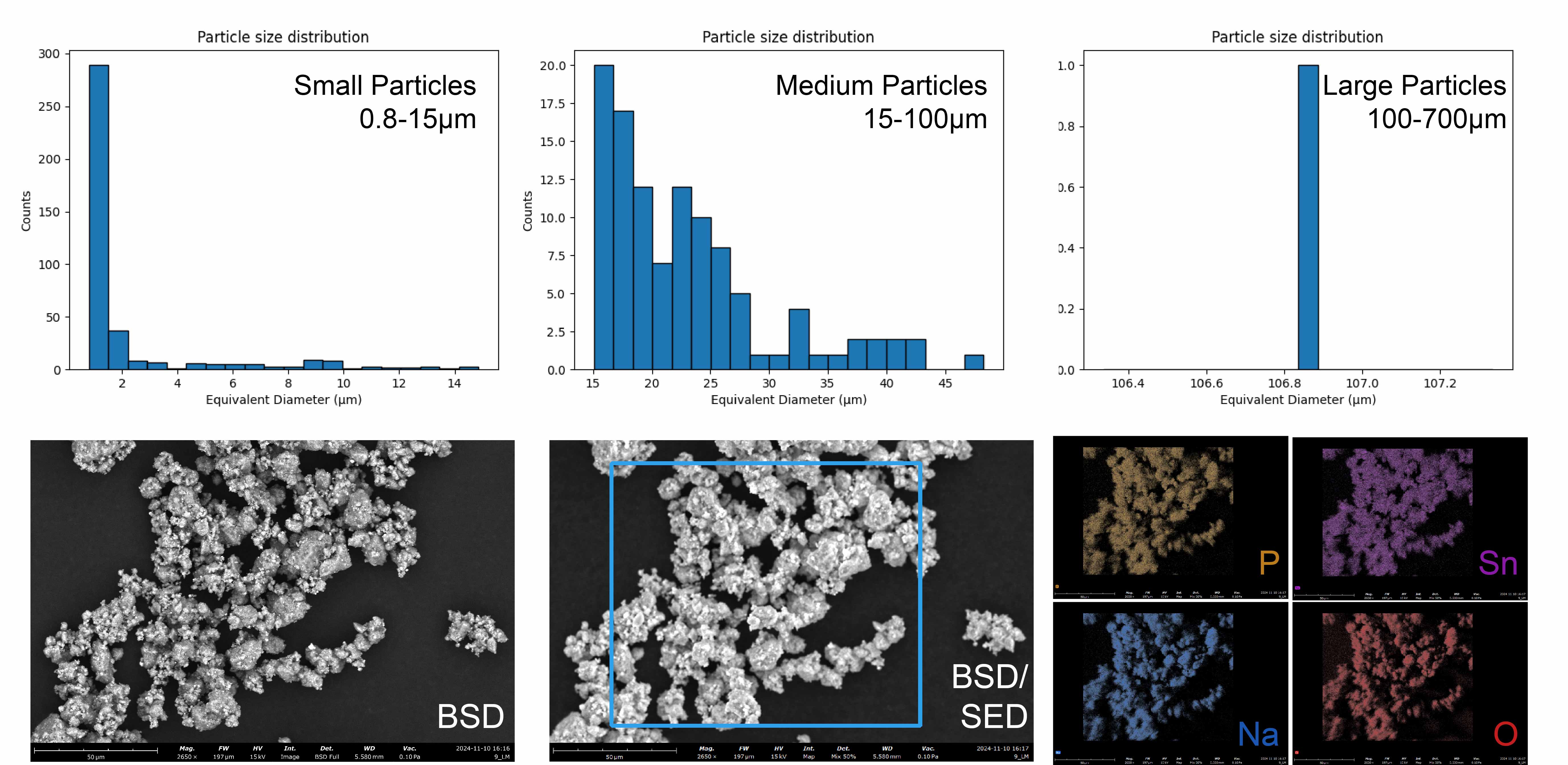}
    \caption{SEM imaging, EDS mapping, and particle size distributions for ball milled micron \ce{SnO2} and \ce{Na4P2O7}.}
    \label{fig:eds-particlesize-337}
\end{figure}

\begin{figure}[H]
    \centering
    \includegraphics[width=0.75\linewidth]{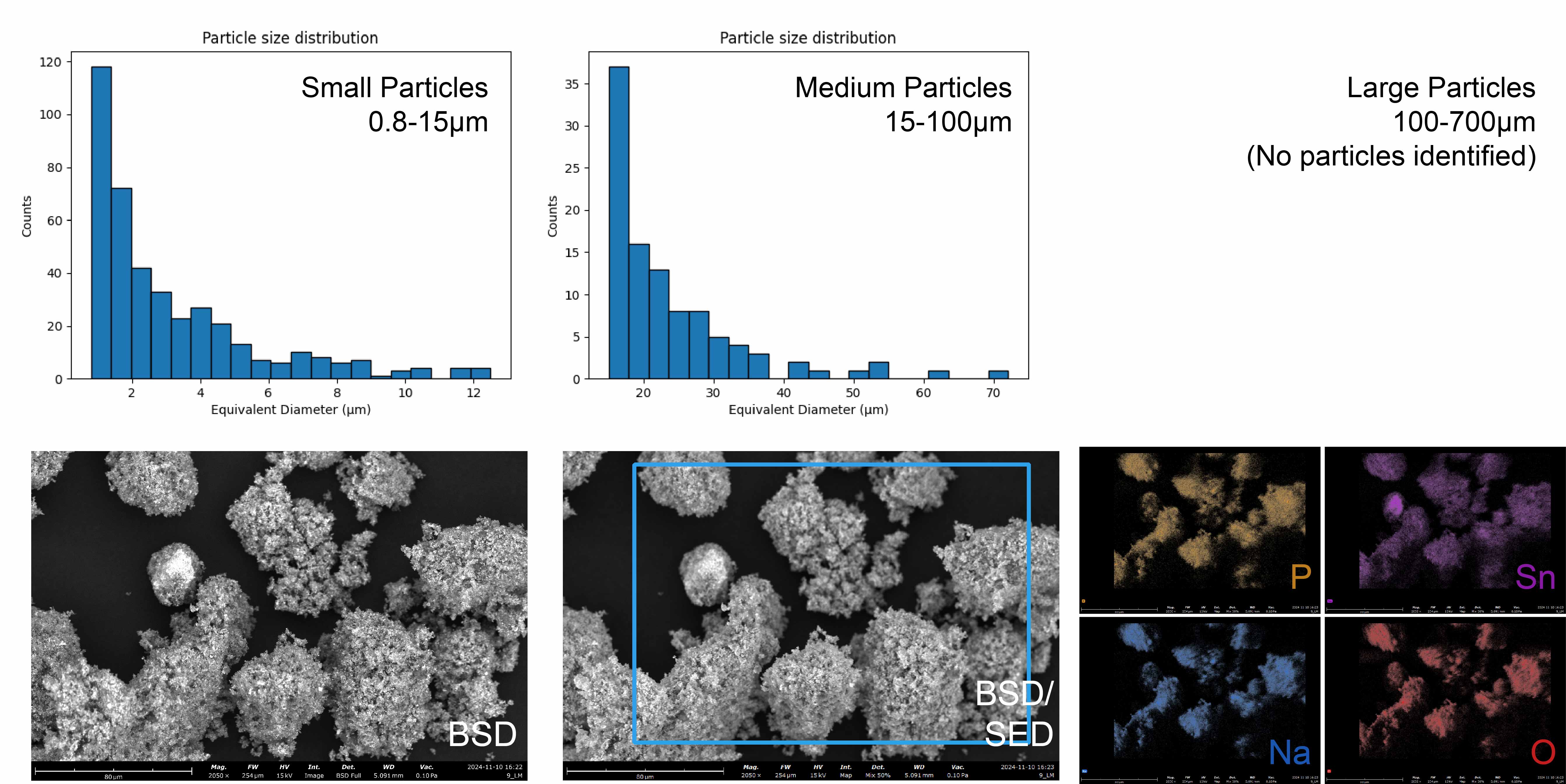}
    \caption{SEM imaging, EDS mapping, and particle size distributions for pestled and mortared micron \ce{SnO2} and \ce{Na4P2O7}.}
    \label{fig:eds-particlesize-338}
\end{figure}

\begin{figure}[H]
    \centering
    \includegraphics[width=0.75\linewidth]{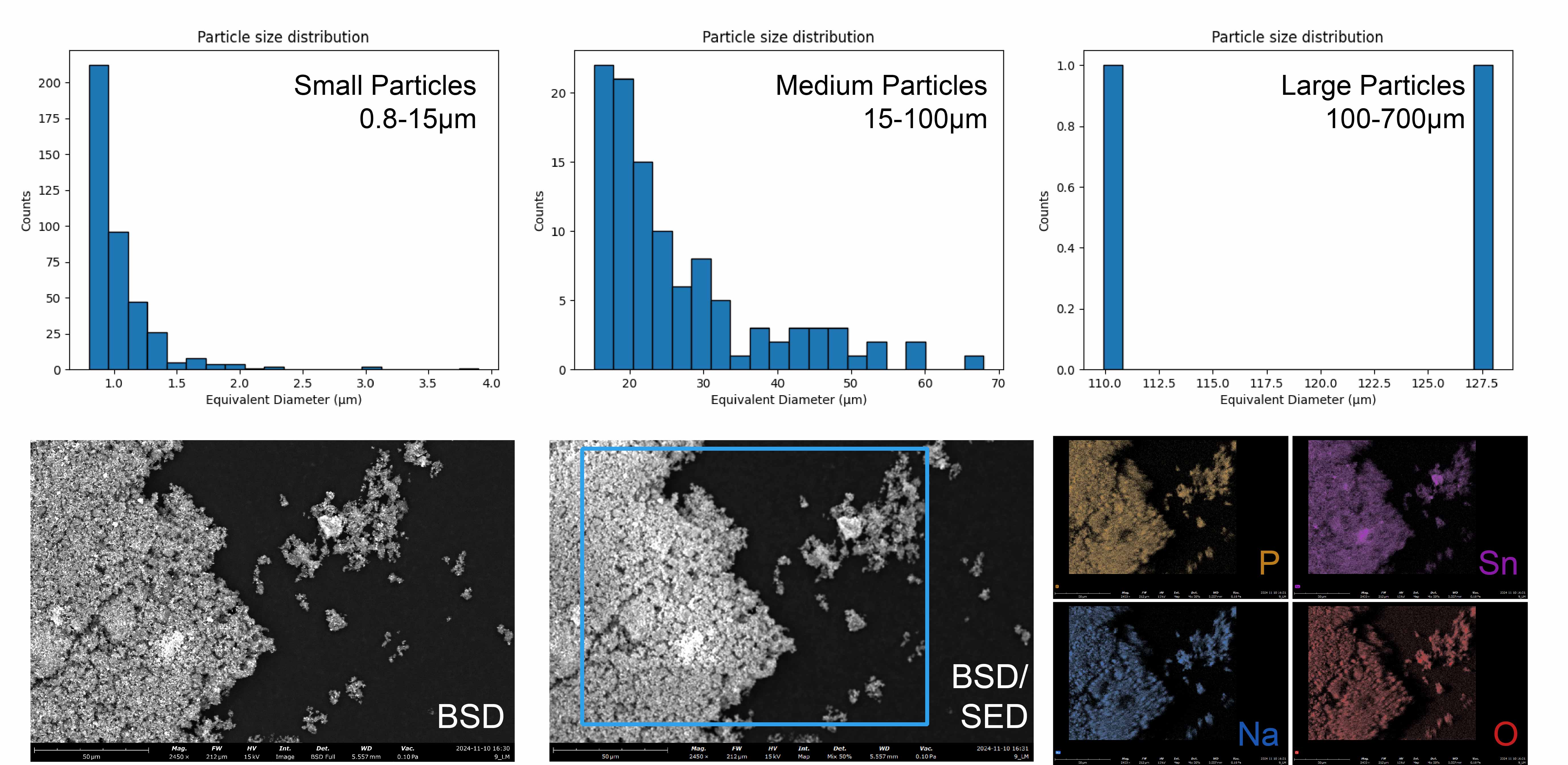}
    \caption{SEM imaging, EDS mapping, and particle size distributions for wet ball milled micron \ce{SnO2} and \ce{Na4P2O7}.}
    \label{fig:eds-particlesize-339}
\end{figure}

\begin{figure}[H]
    \centering
    \includegraphics[width=0.75\linewidth]{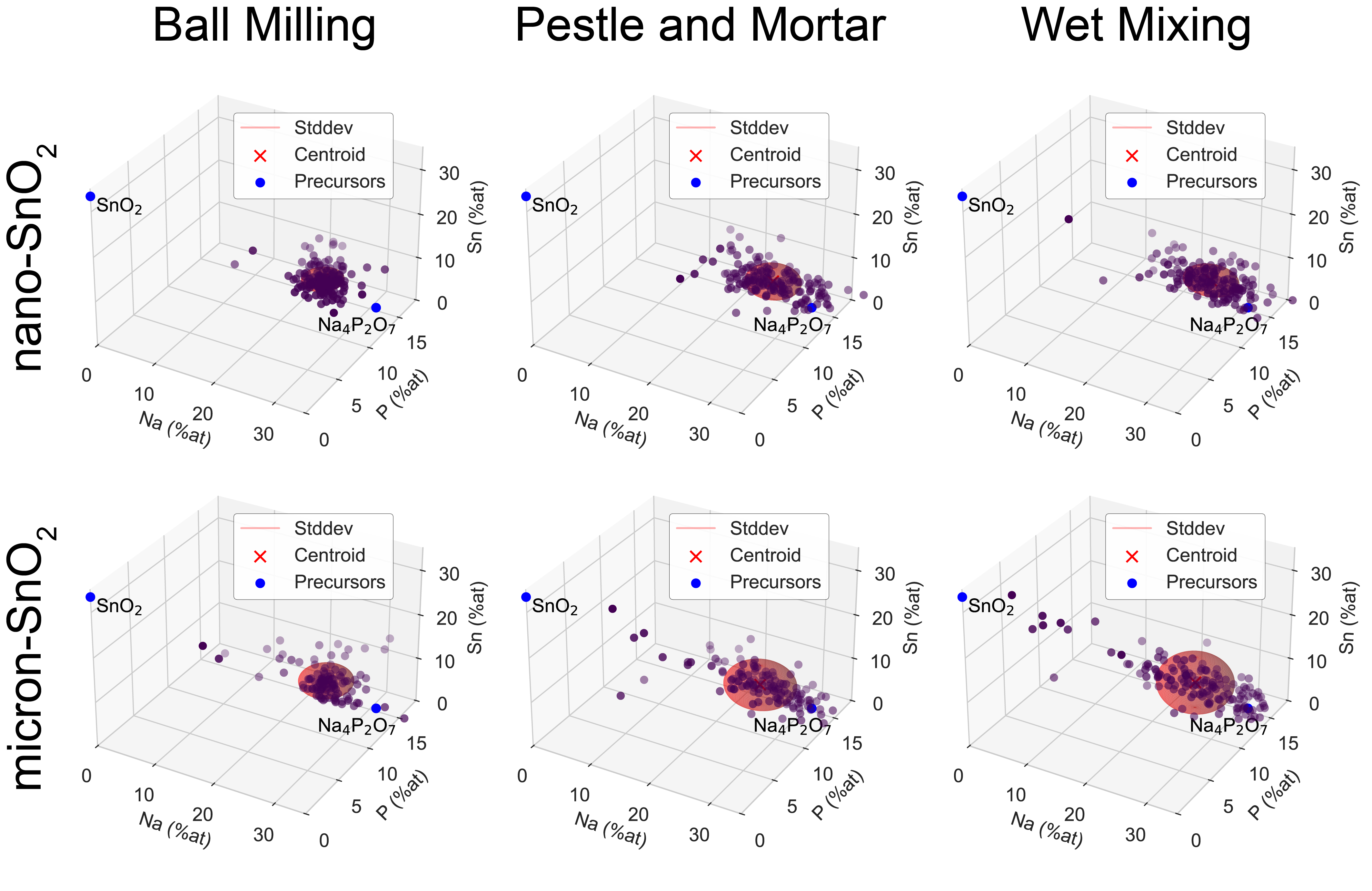}
    \caption{3D particle distributions used for particle composition analysis.}
    \label{fig:nsp-3d-distributions}
\end{figure}

\begin{table}[H]
\caption{\label{tab:sem_eds_elemental_composition} Elemental composition of as-prepared samples made with different mixing methods and starting nanometer \ce{SnO2}. Compositions are taken with SEM-EDS and fit with AutoSEMEDS \cite{EDS_Fitting_Andrea_Method}. Compositions of roughly \ce{Na4SnP2O9} are confirmed.}
\begin{tabular}{lccccc}
\hline \hline
\textbf{Sample Number} & \textbf{Mixing Method}  & \textbf{Na Content} & \textbf{Sn Content} & \textbf{P Content} & \textbf{O Content}\\
\hline
327 & Ball Mill& 22.29 & 6.26 & 15.33& 56.11\\
328 & Pestle and Mortar& 22.88 & 7.00 & 14.06 & 56.06\\
329 & Wet Spin Mix& 23.68 & 7.12 & 13.94 & 55.25 \\
\hline \hline
\end{tabular}
\end{table}

\section{Additional XRD Refinements}

\subsection{Refinement of Ball Milled Precursors}

XRD refinement of ball milled \ce{Na4P2O7} and \ce{SnO2}. RWP is found to be 11.52\%.

\begin{figure}[H]
    \centering
    \includegraphics[width=0.75\linewidth]{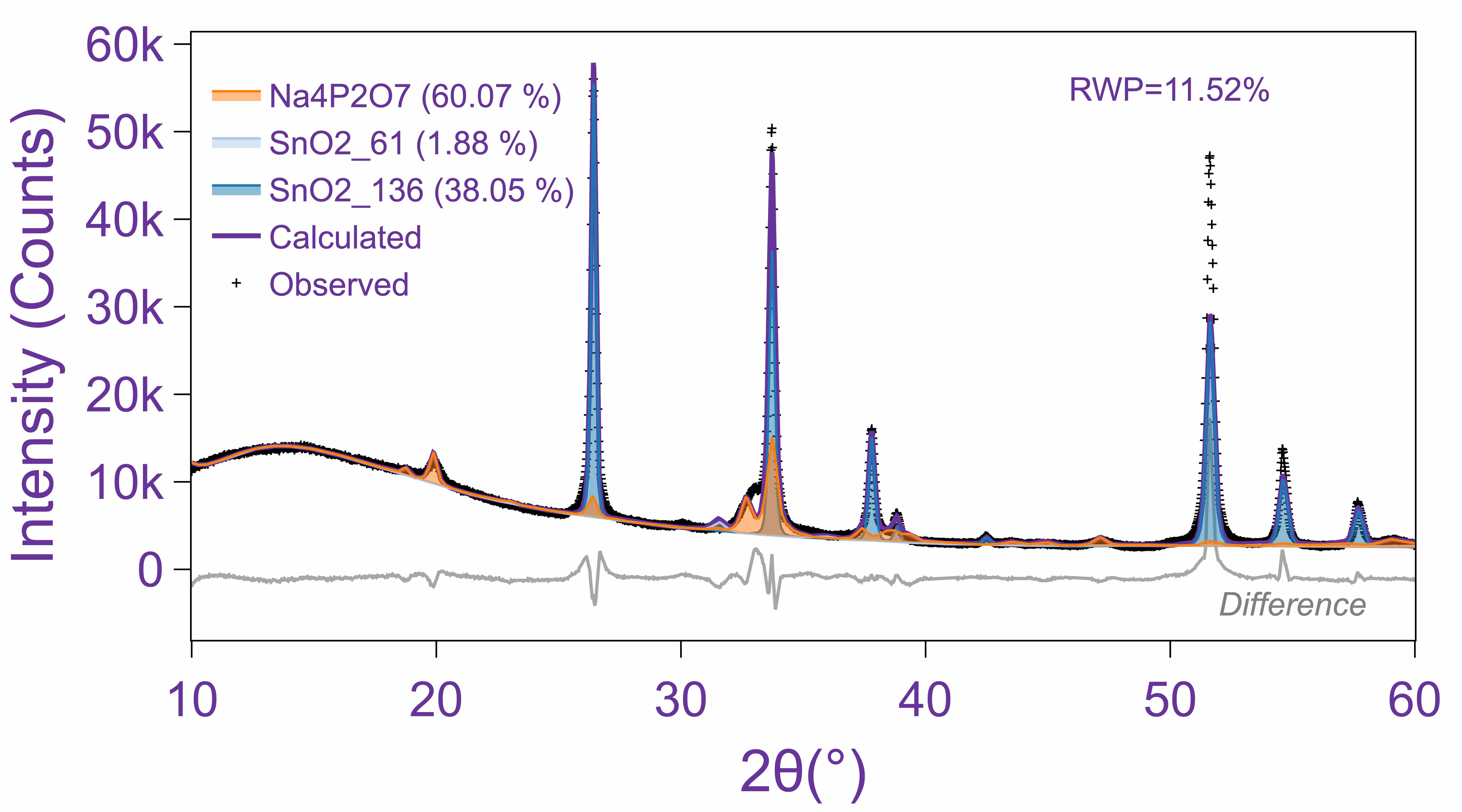}
    \caption{XRD Refinement of ball milled \ce{Na4P2O7} and \ce{SnO2}. Pattern is taken on a Rigaku Miniflex.}
    \label{fig:xrd_refinement_sno2_na4p2o7_ballmilled}
\end{figure}

\subsection{XRD Refinement of Phase Pure Target by Dara without Target Structure Included}

\begin{figure}[H]
    \centering
    \includegraphics[width=0.75\linewidth]{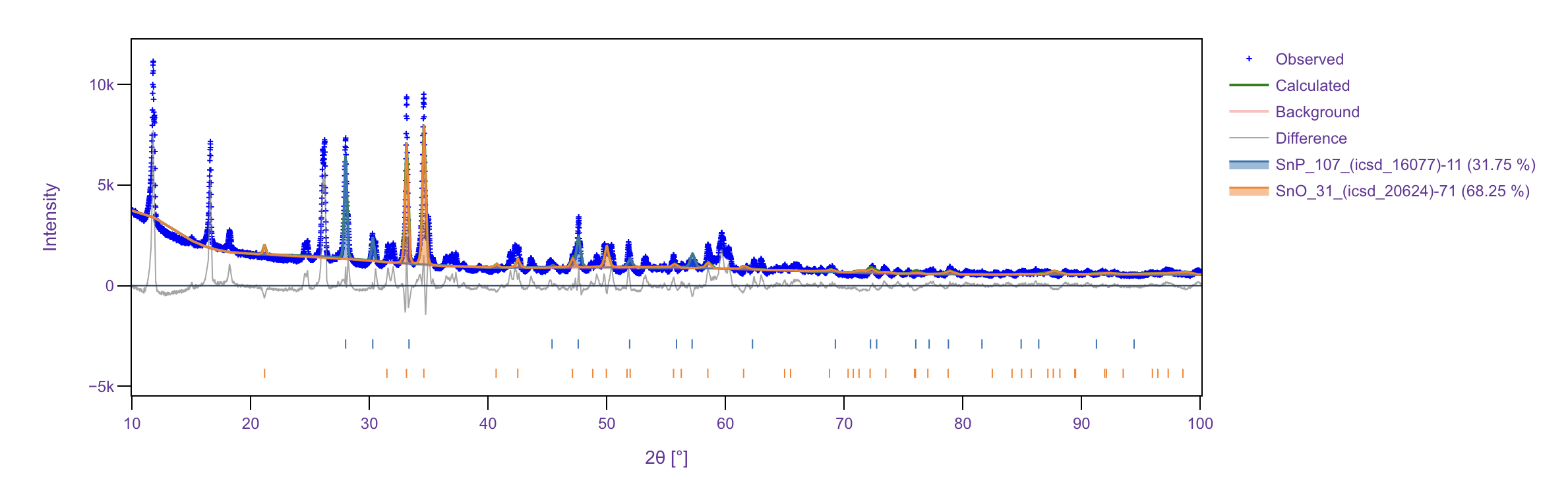}
    \caption{Dara executed on hypothesized phase pure \ce{Na4SnP2O9}, however, no structure with the target composition is included in the search-match process. $R_{wp}$\,=\,24.82.}
    \label{fig:dara_no_nsp}
\end{figure}

To understand the contribution of \ce{Na4SnP2O9} to the XRD structure solution, refinement was attempted to the phase pure XRD pattern without the inclusion of the target, but with all other compounds in ICSD that contain sodium, tin, phosphorous, and/or oxygen. The best refinement is shown in \autoref{fig:dara_no_nsp}. It was found that a majority of the peaks are unmatched, including the first three high intensity peaks at low $2\theta$. Furthermore, the best refinement has an $R_{wp}$ of 24.82 \%. Finally, the phases identified are SnO and SnP, both of which are likely not synthesizable in air. SnO is Sn(II), which is unlikely to be synthesized in higher oxygen chemical potentials, as the more stable Sn(IV). Furthermore, SnP is reportedly synthesized at high pressure \cite{SnP_synthesis}. This supports the fact that the best known structure solution is the NSP phase.

\subsection{Refinement with Different Reference NSP Structure Models}

\subsection{\textit{in-situ} XRD During NSP Synthesis}

\begin{figure}[H]
    \centering
    \includegraphics[width=1.00\linewidth]{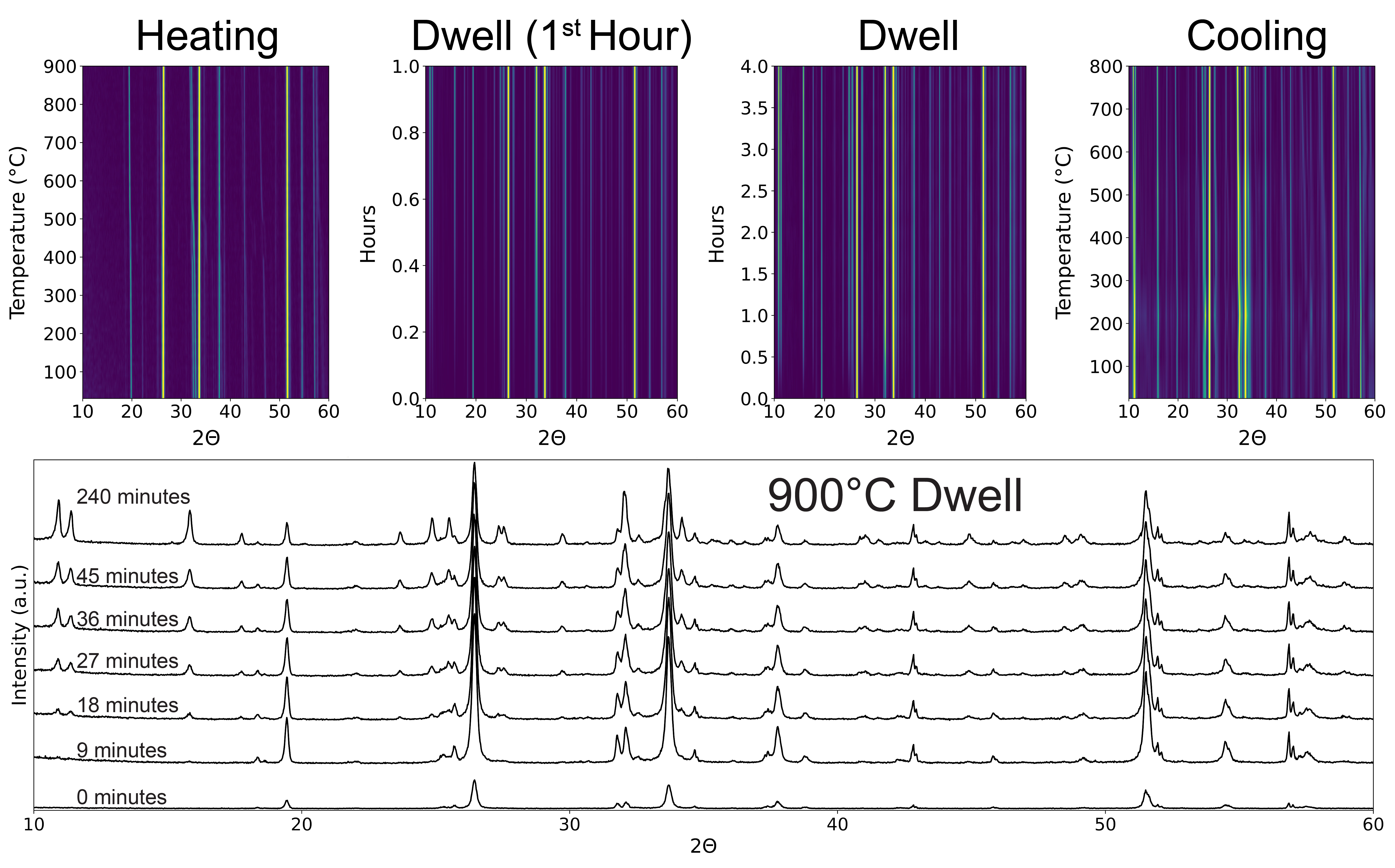}
    \caption{Heat maps showing the phase progression of the synthesis as a function of temperature. Samples are heated at 10$^\circ$ per minute between 2 minute XRD 2 minute. Heat maps created from XRD patterns are shown for the heating stages, throughout the four hour dwell at 900$^\circ$, and in the natural cooling regime.}
    \label{fig:heat-map-insitu}
\end{figure}

\subsection{Elevated Temperature NSP XRD}

An in-situ XRD was used to measure the XRD spectra of as-synthesized NSP at high temperatures. The room temperature XRD on the sample was recorded before and after heating, included as benchmark.

\begin{figure}[H]
    \centering
    \includegraphics[width=1.0\linewidth]{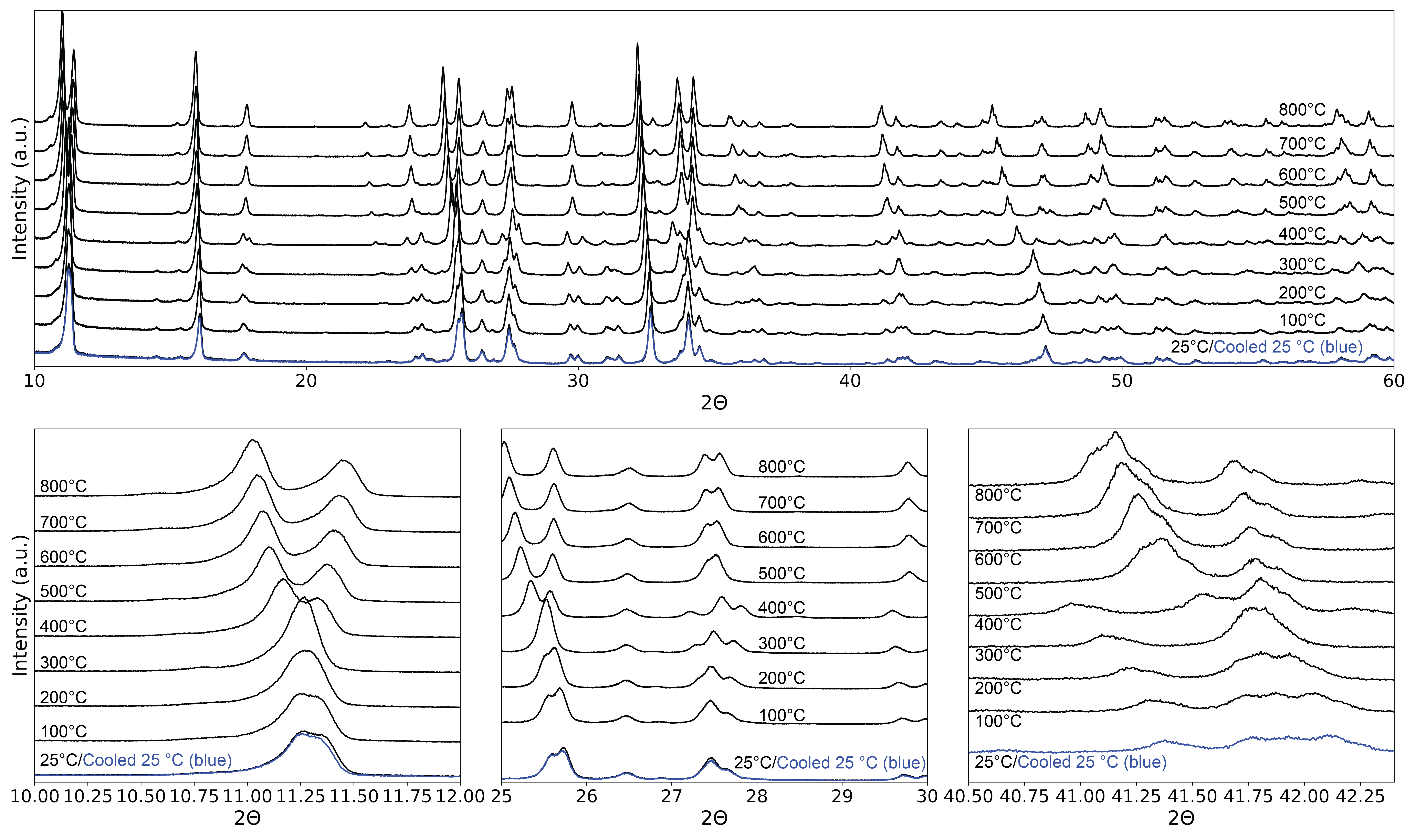}
    \caption{XRD of \ce{Na4SnP2O9} at non-ambient conditions. Significant peak changes are observed at a number of different high temperature points, including near 11\,$2\theta$, around 25\,$2\theta$, and between 30 and 35\,$2\theta$, and between 40 and 50\,$2\theta$. An overlaid dashed line in blue depicts the sample after heating when cooled to room temperature for benchmarking. Good agreement is found, which supports the hypothesis that observed diffraction changes are due to reversible structural changes rather than chemical reactions, such as degradation of NSP.}
    \label{fig:temperature_dep_xrd}
\end{figure}

\begin{figure}[H]
    \centering
    \includegraphics[width=0.75\linewidth]{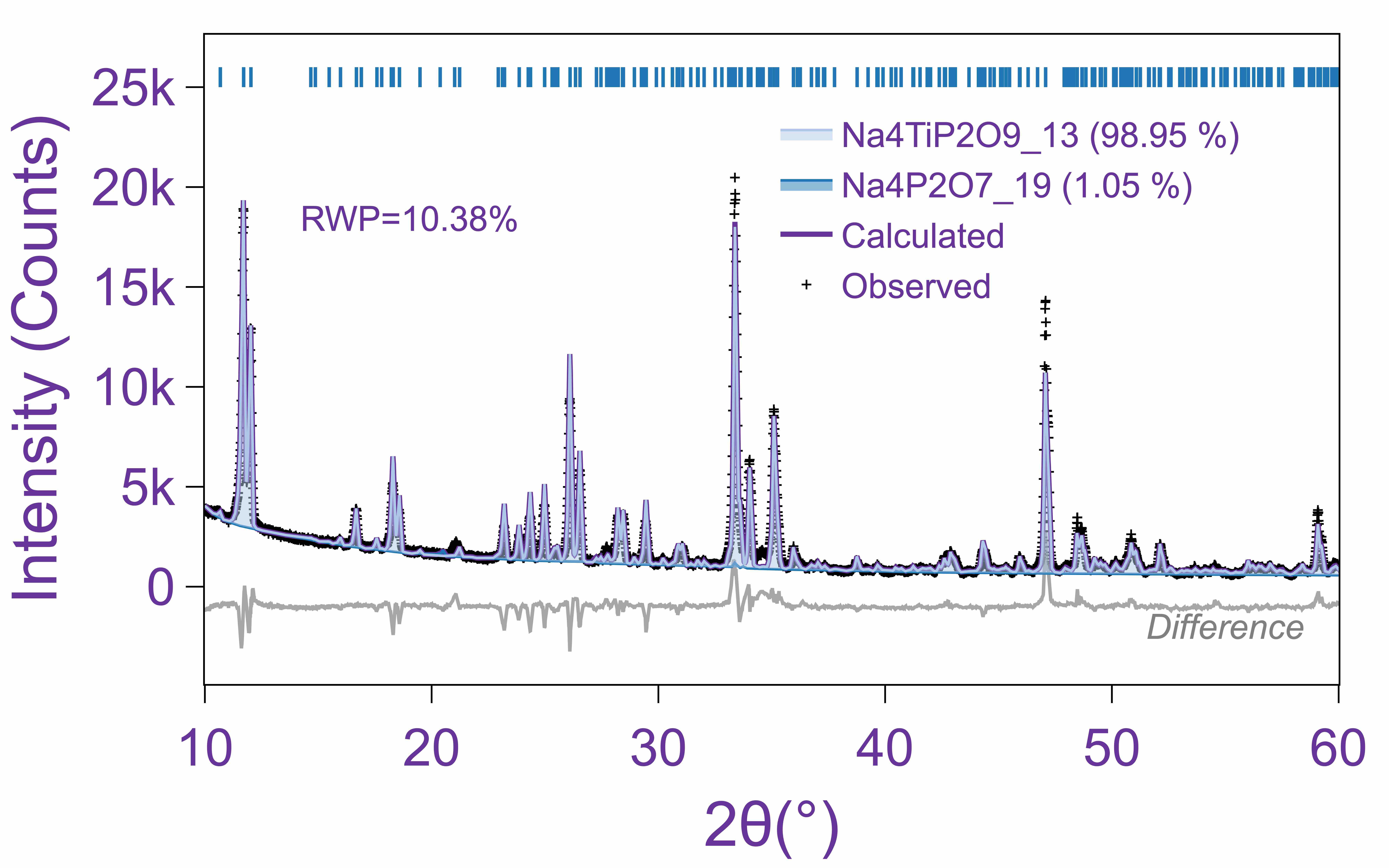}
    \caption{NTP solid state synthesis with A-Lab. The refinement of powder XRD indicates primarily monoclinic LT-NTP, with 1.05\% \ce{Na4P2O7} impurity.}
    \label{fig:ntp_xrd_refinement}
\end{figure}

\begin{figure}[H]
    \centering
    \includegraphics[width=0.75\linewidth]{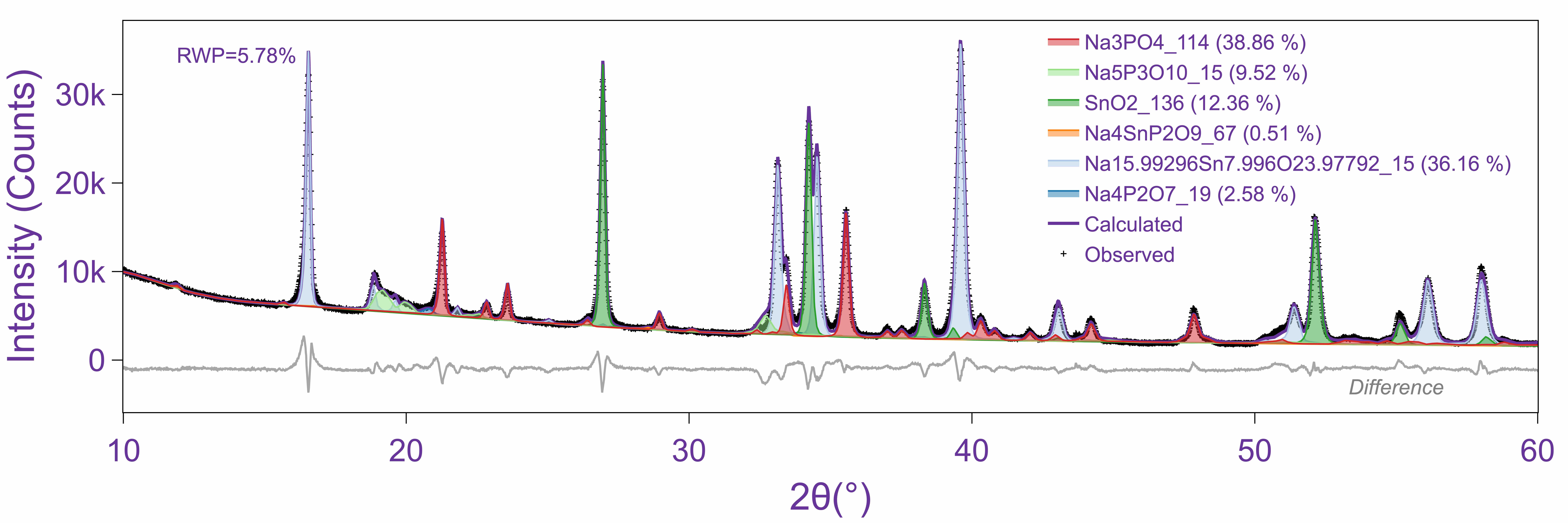}
    \caption{Attempted synthesis of NSP with \ce{Na5P3O10} as a precursor.}
    \label{fig:nsp_xrd_refinement_Na5P3O10}
\end{figure}

\begin{figure}[H]
    \centering
    \includegraphics[width=0.75\linewidth]{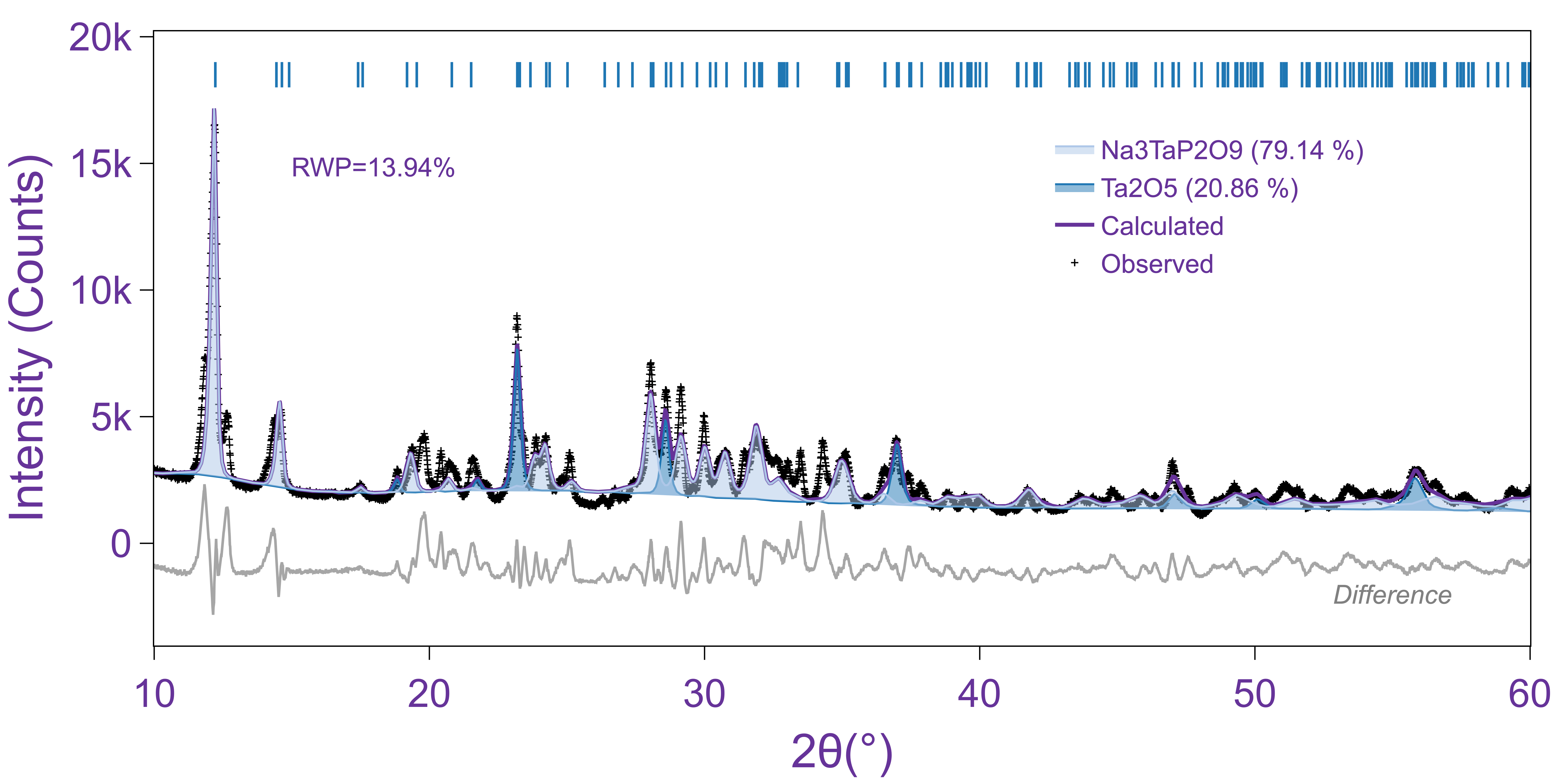}
    \caption{Attempted synthesis of \ce{Na3TaP2O9} with \ce{Na2CO3}, \ce{Ta2O5}, and \ce{NH4H2PO4} precursors. Sample was held at 750\,$^\circ$C for 12 hours. A phase with the target composition (ICSD 48829 \cite{Na3TaP2O9}, \textit{P2$_1$2$_1$2$_1$}) was found by Dara to be the best fit majority phase. This phase is symmetrically and topologically inequivalent to the NAP parent phase.}
    \label{fig:na3tap2o9_xrd}
\end{figure}

\section{Additional Electrochemical Impedance Spectroscopy Results}

\subsection{NSP Low temperature EIS Measurements}

\begin{figure}[H]
    \centering
    \includegraphics[width=0.75\linewidth]{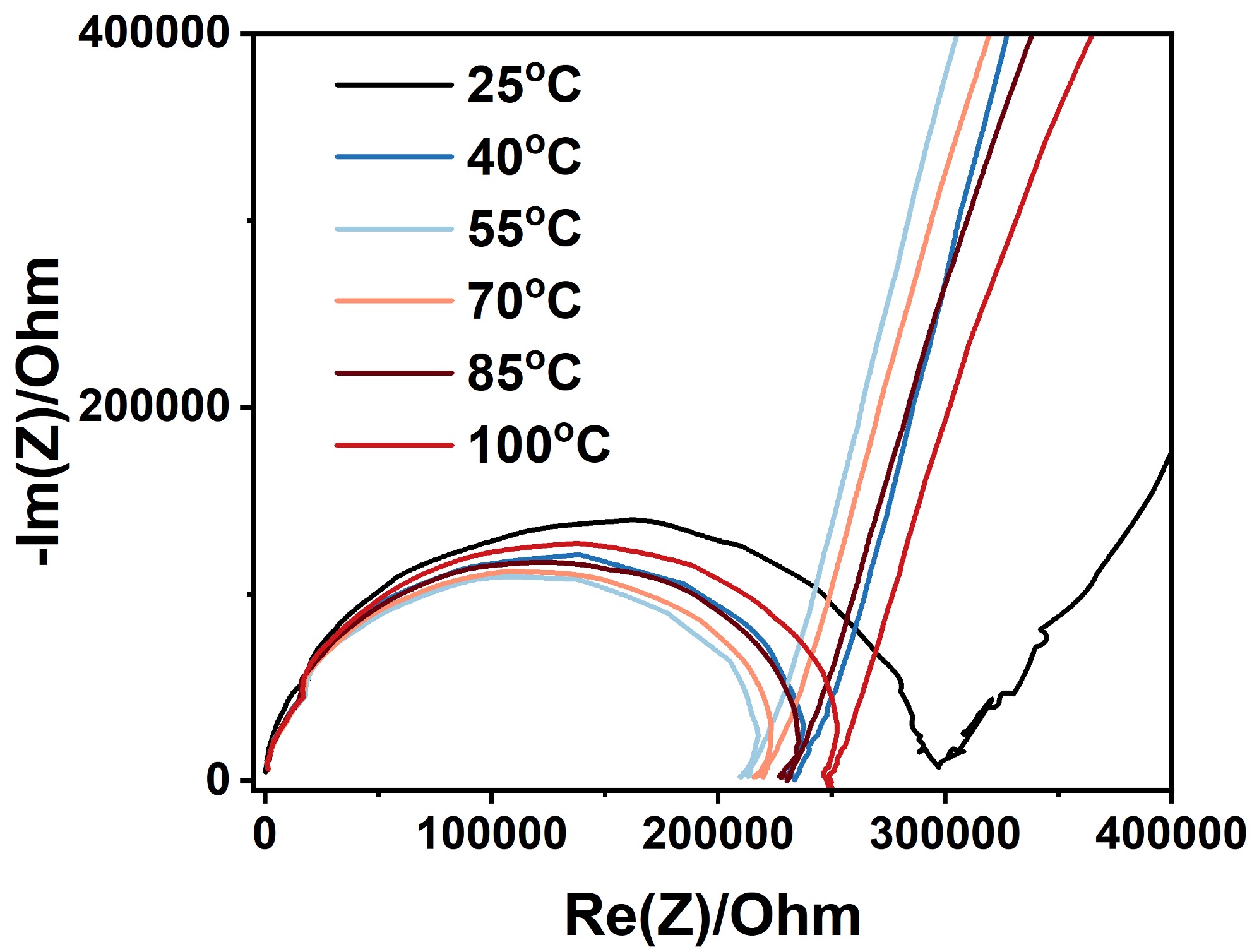}
    \caption{Temperature-dependent EIS measurements for \ce{Na4SnP2O9} the range of 25-100\,$^\circ$C. NSP is found to have quite a small conductivity. At 100\,$^\circ$C, the measured resistance becomes less precise. This is possibly due to some surface water boiling off.}
    \label{fig:nsp_exp_eis_lowT}
\end{figure}

\begin{figure}[H]
    \centering
    \includegraphics[width=0.75\linewidth]{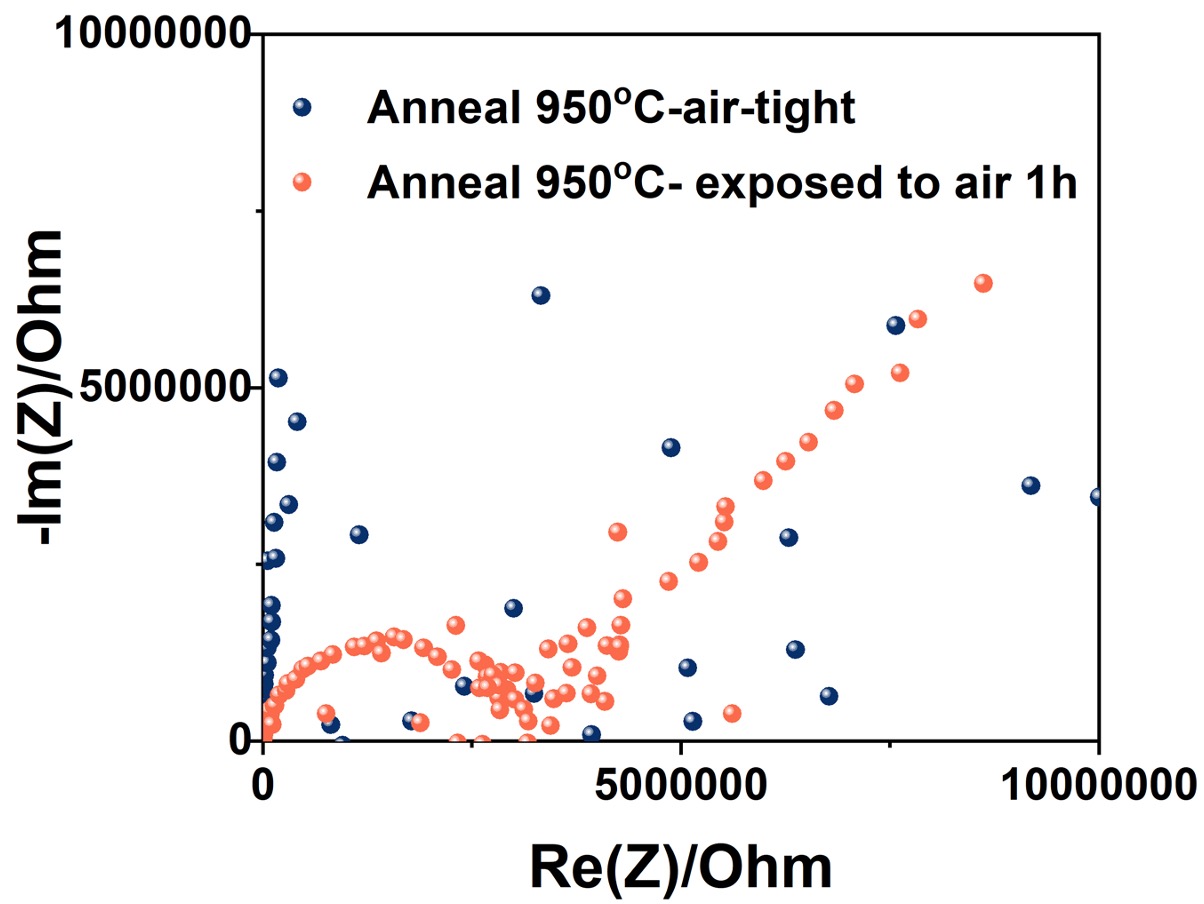}
    \caption{Room temperature EIS measurements for \ce{Na4SnP2O9}. Any measurable conductivity is only possible when the pellet is exposed to air. If the pellet is moved directly to an airtight chamber after heating, it is found to be ionically nonconductive. This supports the idea that any measured conductivity at low temperatures occurs from surface absorbed water.}
    \label{fig:nsp_exp_eis_airtight}
\end{figure}

\subsection{NTP High temperature EIS Measurements}

\begin{figure}[H]
    \centering
    \includegraphics[width=0.75\linewidth]{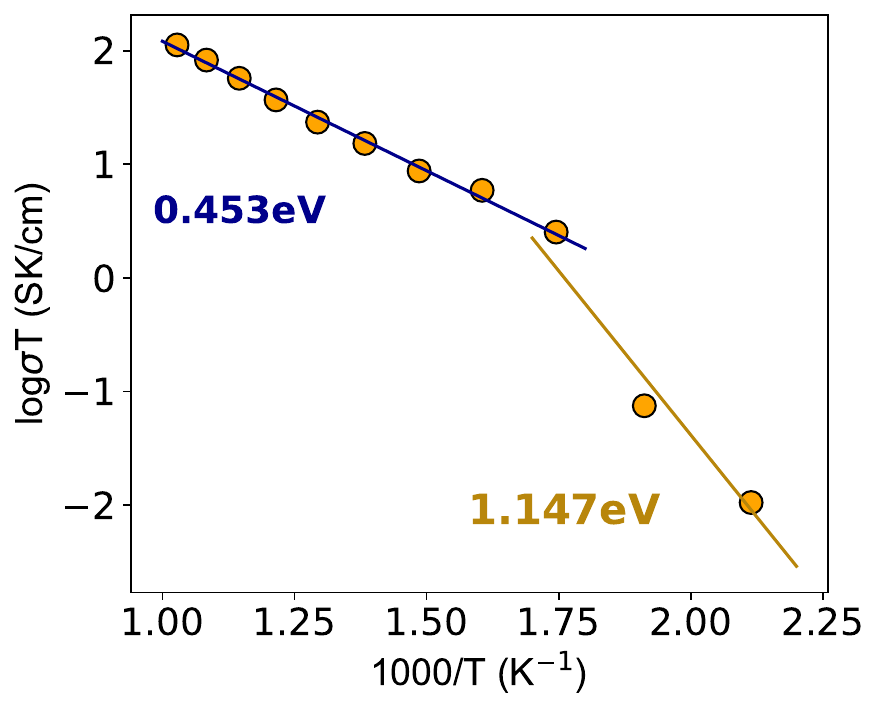}
    \caption{High temperature EIS measurements for \ce{Na4TiP2O9}. A conductivity transition is found near 300\,$^\circ$C (570\,K), in agreement with previous work \cite{NTP_1, NTP_original}.}
    \label{fig:ntp_exp_eis}
\end{figure}

\section{CHGNet Fine Tuning}

\begin{table}[H]
\caption{\label{tab:chgnet-finetuning-results} Fine tuning mean absolute errors for CHGNet \cite{CHGNet}.}
\begin{tabular}{lccccc}
\hline \hline
\textbf{A Site Cation} & \textbf{MAE Energy (meV/atom)}  & \textbf{MAE Force (meV/\AA)} & \textbf{MAE Stress (GPa)}\\
\hline
Al & 1 & 46 & 0.085 \\
Ga & 2 & 46 & 0.073 \\
In & 1 & 43 & 0.076 \\
Sc & 1 & 42 & 0.069 \\
\hline
Hf & 2 & 58 & 0.090\\
Sn & 2 & 48 & 0.074\\
Zr & 2 & 51 & 0.077\\
Cr & 2 & 66 & 0.107\\
Fe & 2 & 52 & 0.092\\
Ge & 2 & 49 & 0.090\\
Mn & 3 & 70 & 0.127\\
Pb & 2 & 61 & 0.097\\
Ti & 2 & 50 & 0.053\\
\hline
Ta & 2 & 62 & 0.095\\
V  & 3 & 75 & 0.116 \\
\hline \hline
\end{tabular}
\end{table}

\section{Direction-Dependent Sodium Diffusion}

Sodium displacement within CHGNet MD calculations \cite{CHGNet} for different NAP candidates. Sodium displacement is shown, broken down by direction, over a 2\,ns. All calculations were performed at 600\,$^\circ$C except for \ce{Na4ZnP2O9}, which was performed at 700\,$^\circ$C due to limited sodium diffusion at lower temperatures. The c-direction, which is perpendicular to the layers in the NAP structure, has the lowest sodium displacement for all compositions.

\begin{figure}[H]
    \centering
    \includegraphics[width=0.75\linewidth]{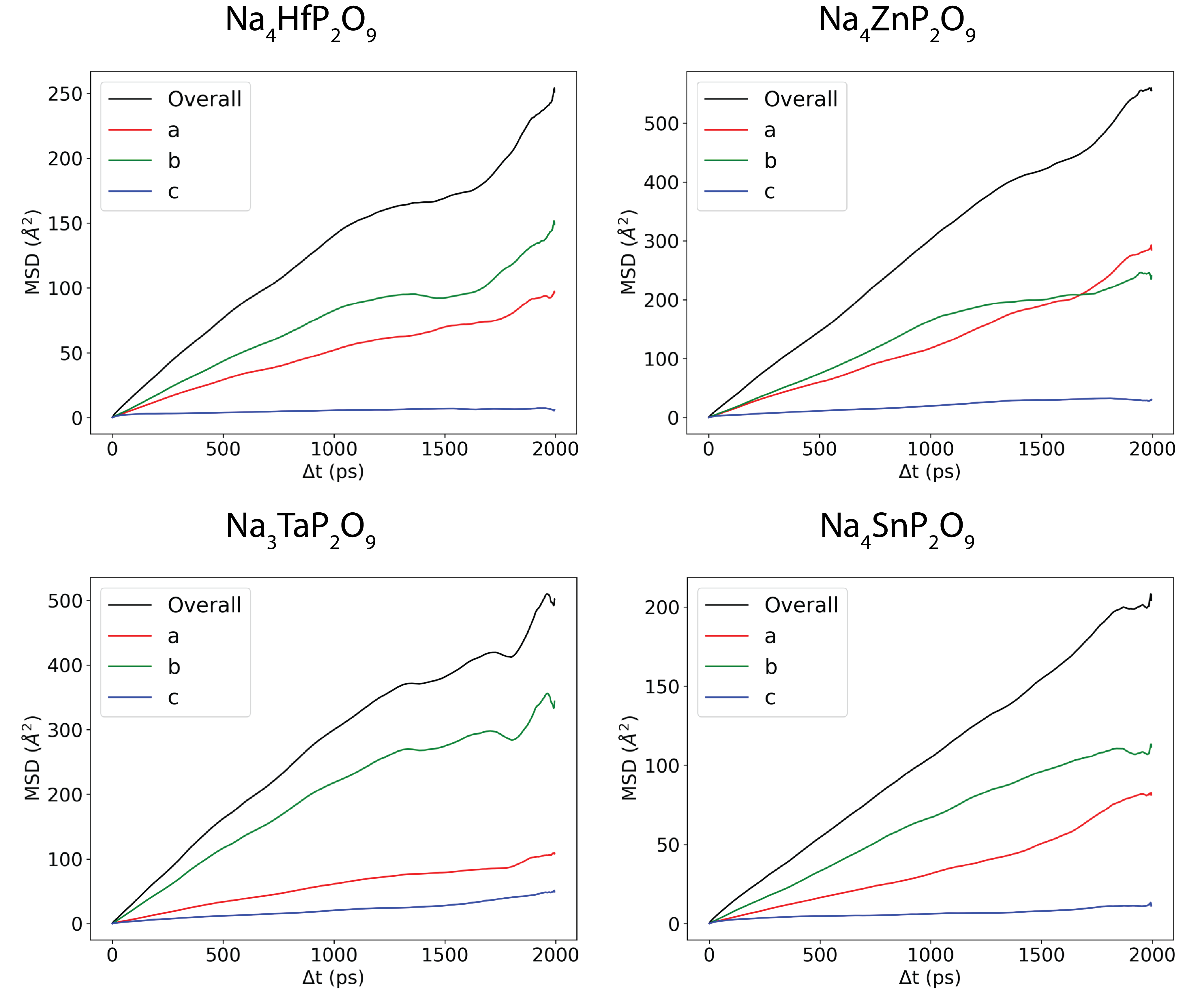}
    \caption{MSD analysis of Na for MD calculations. Calculations are broken down by direction. All calculations were performed at 600\,$^\circ$C except for \ce{Na4ZnP2O9}, which was performed at 700\,$^\circ$C.}
    \label{fig:c-direction-conductivity}
\end{figure}

\begin{figure}[H]
    \centering
    \includegraphics[width=0.75\linewidth]{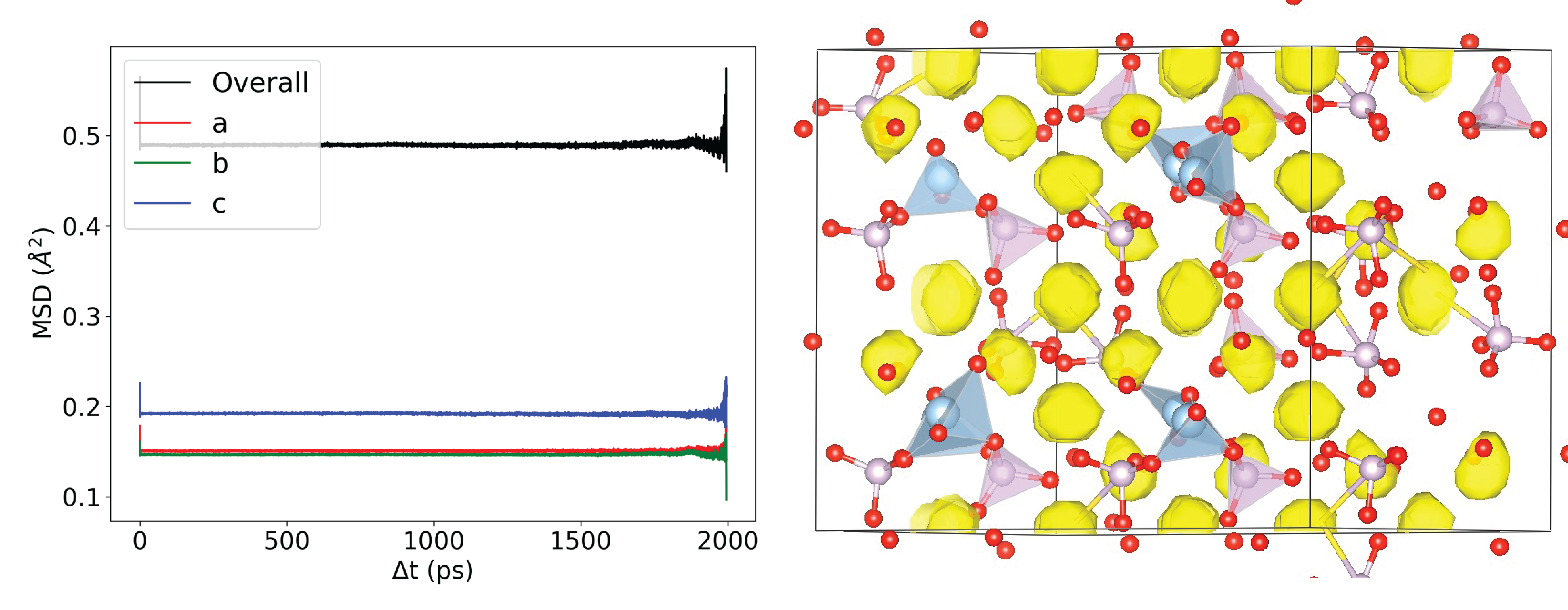}
    \caption{Output from an MD diffusivity calculation for \ce{Na5AlPO9} at 1000\,K. (left) Probability density (right) hop number. No hopping is seen over the 2\,ns timeframe.}
    \label{fig:Na5AlP2O9-diffusivity-calc}
\end{figure}

\begin{figure}[H]
    \centering
    \includegraphics[width=0.75\linewidth]{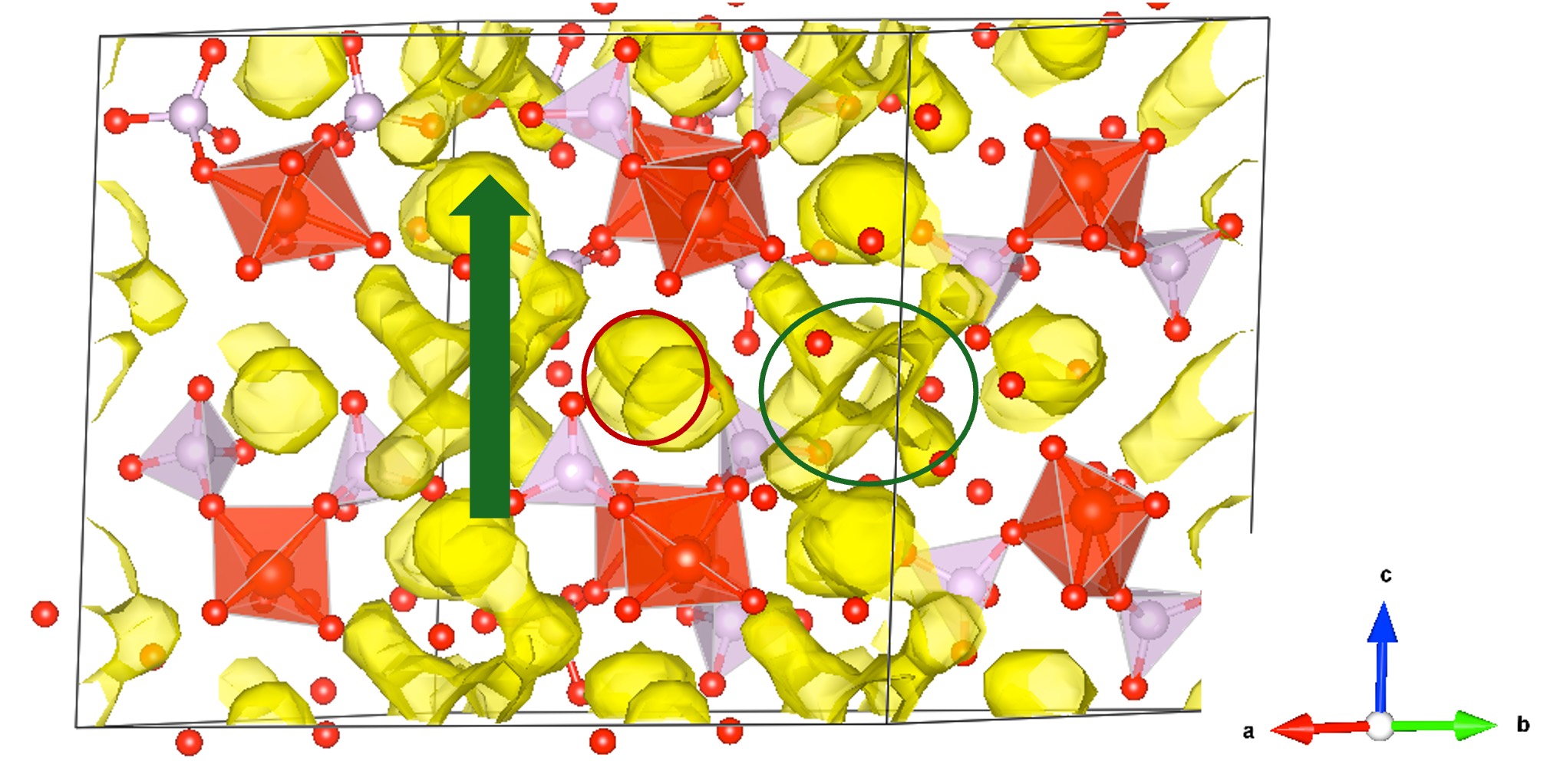}
    \caption{Probability density \ce{Na3VP2O9} from a 2\,ns diffusivity calculation at 700\,K.}
    \label{fig:na3vp2o9_probilitydensity}
\end{figure}

\begin{figure}[H]
    \centering
    \includegraphics[width=0.75\linewidth]{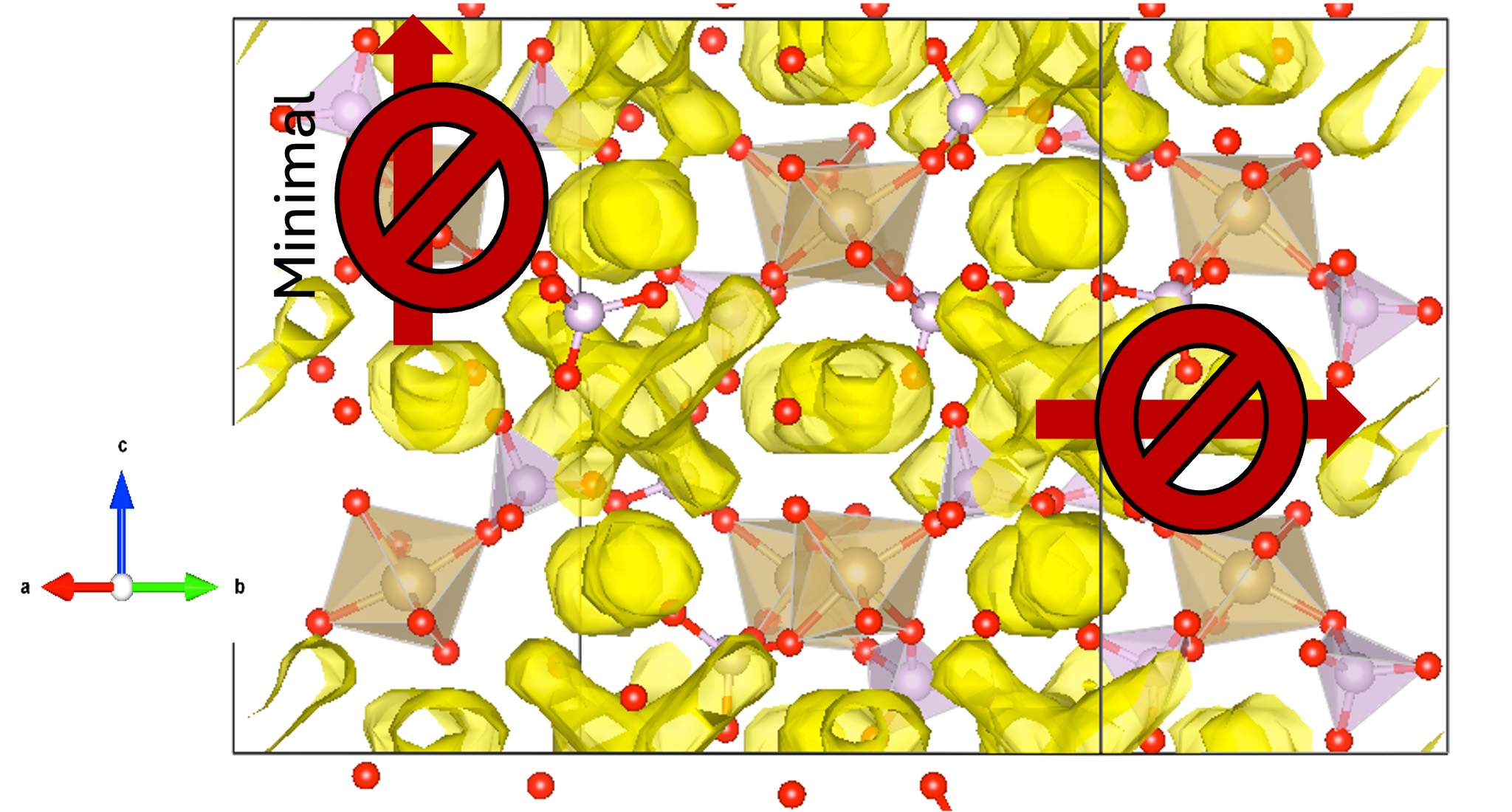}
    \caption{Probability density \ce{Na3TaP2O9} from a 2\,ns diffusivity calculation at 700\,K.}
    \label{fig:na3tap2o9_probilitydensity}
\end{figure}

\section{Na Site Occupation in NTP}

\begin{figure}[H]
    \centering
    \includegraphics[width=0.75\linewidth]{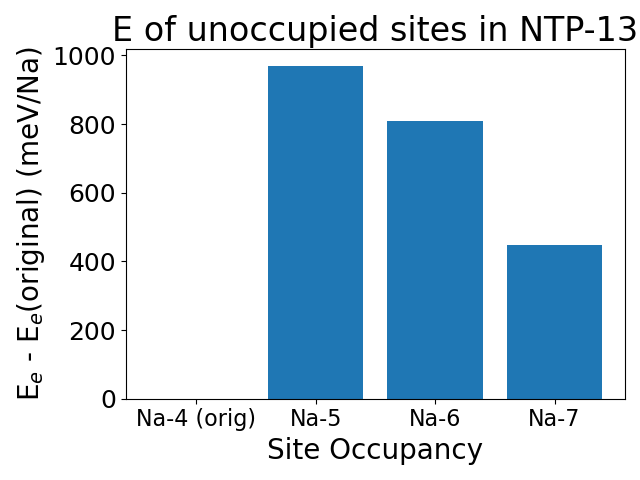}
    \caption{Occupied site energies for all inequivalent sodium sites within a 1D chain in monoclinic NTP.}
    \label{fig:ntp-sg13-site-occupation}
\end{figure}

\section{Phonon Calculations}

\begin{figure}[H]
    \centering
    \includegraphics[width=0.75\linewidth]{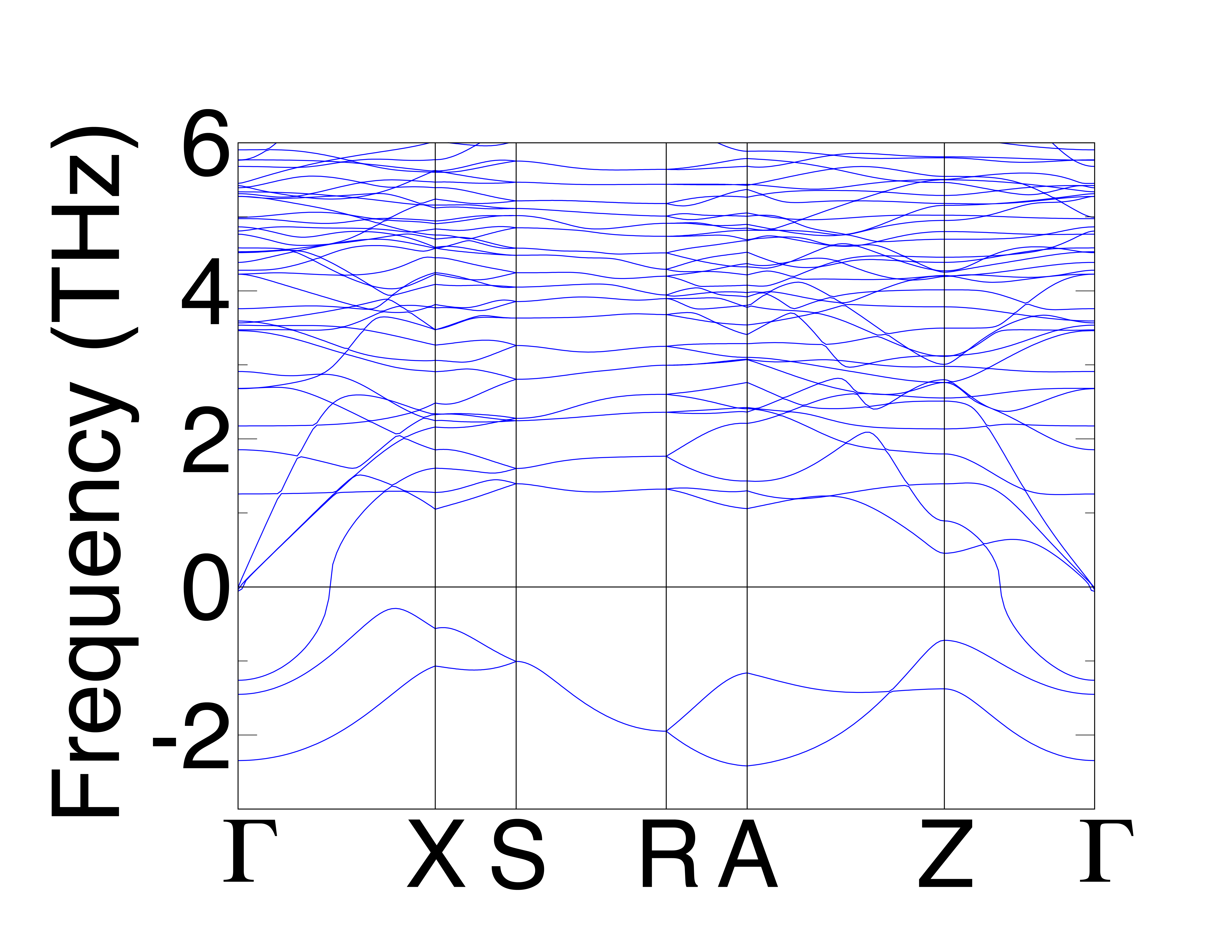}
    \caption{Phonon modulation of NSP in the orthorhombic \textit{Bmem} prototype.}
    \label{fig:phonon_nsp}
\end{figure}

\begin{figure}[H]
    \centering
    \includegraphics[width=0.75\linewidth]{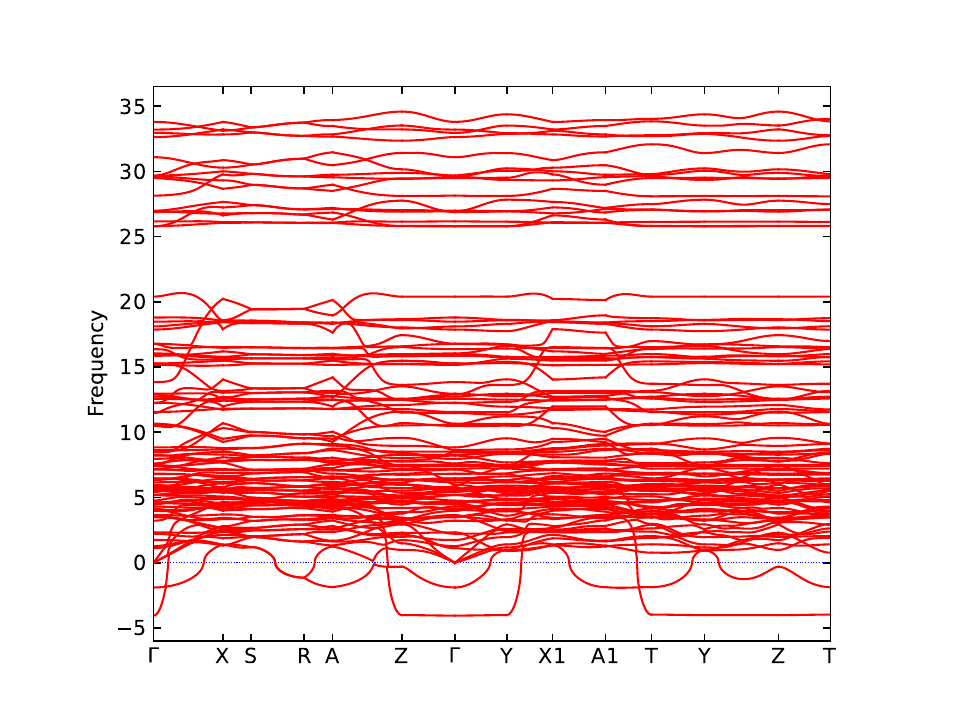}
    \caption{Phonon band structure through a complete Brillion zone path.}
    \label{fig:phonon_ntp_full}
\end{figure}

\begin{figure}[H]
    \centering
    \includegraphics[width=0.75\linewidth]{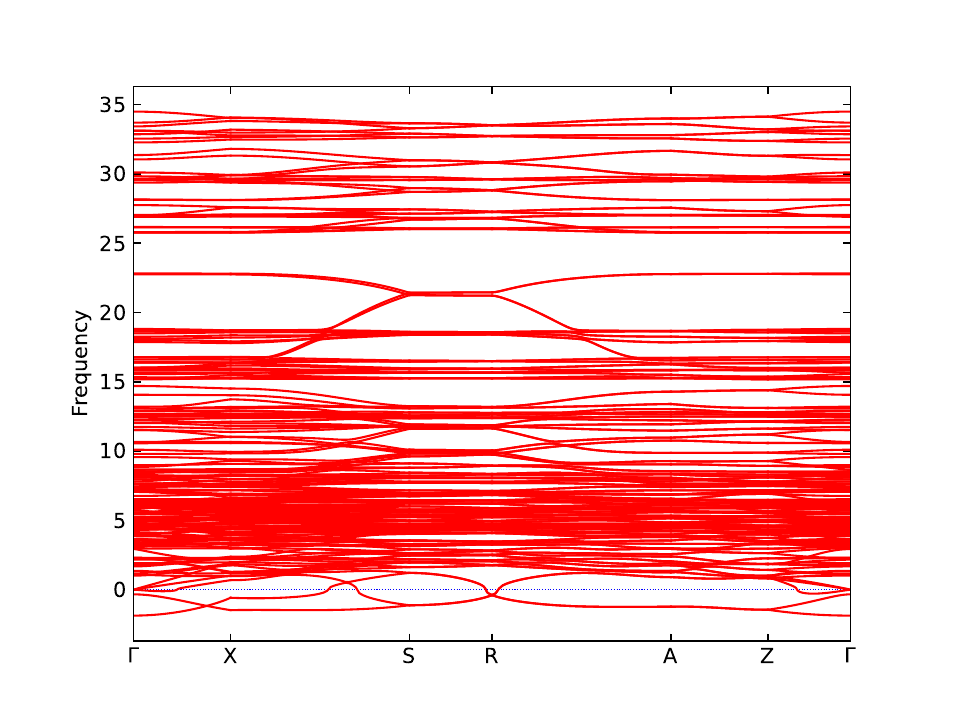}
    \caption{Phonon modulation of NTP by G the mode.}
    \label{fig:phonon_ntp_gmod}
\end{figure}

\begin{figure}[H]
    \centering
    \includegraphics[width=0.75\linewidth]{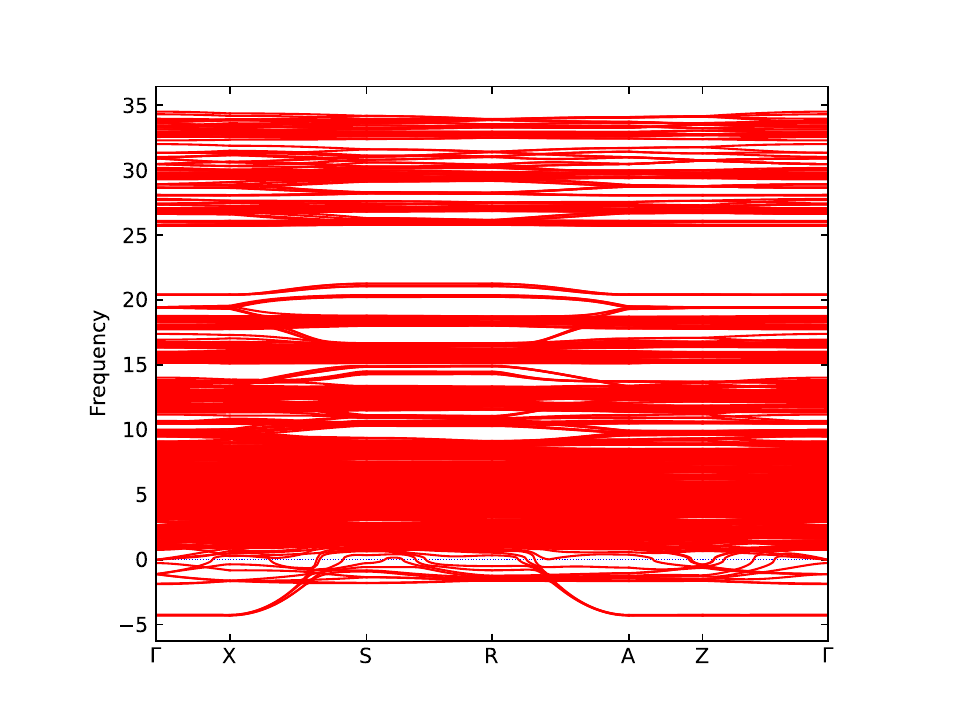}
    \caption{Phonon modulation of NTP by A the mode.}
    \label{fig:phonon_ntp_amod}
\end{figure}

\begin{figure}[H]
    \centering
    \includegraphics[width=0.75\linewidth]{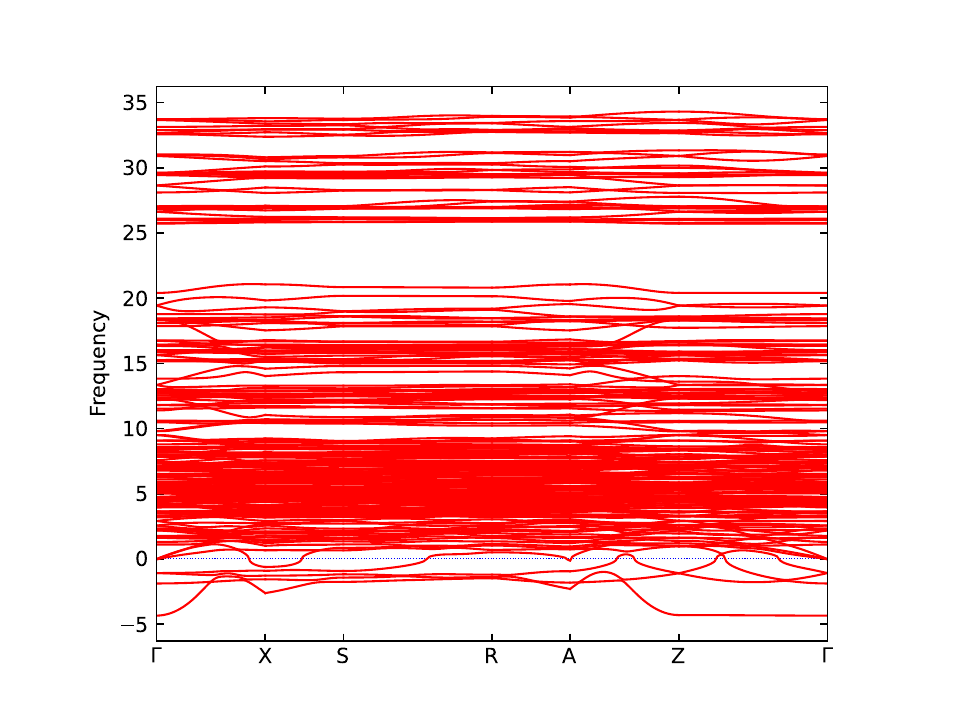}
    \caption{Phonon modulation of NTP by R the mode.}
    \label{fig:phonon_ntp_rmod}
\end{figure}

\begin{figure}[H]
    \centering
    \includegraphics[width=0.75\linewidth]{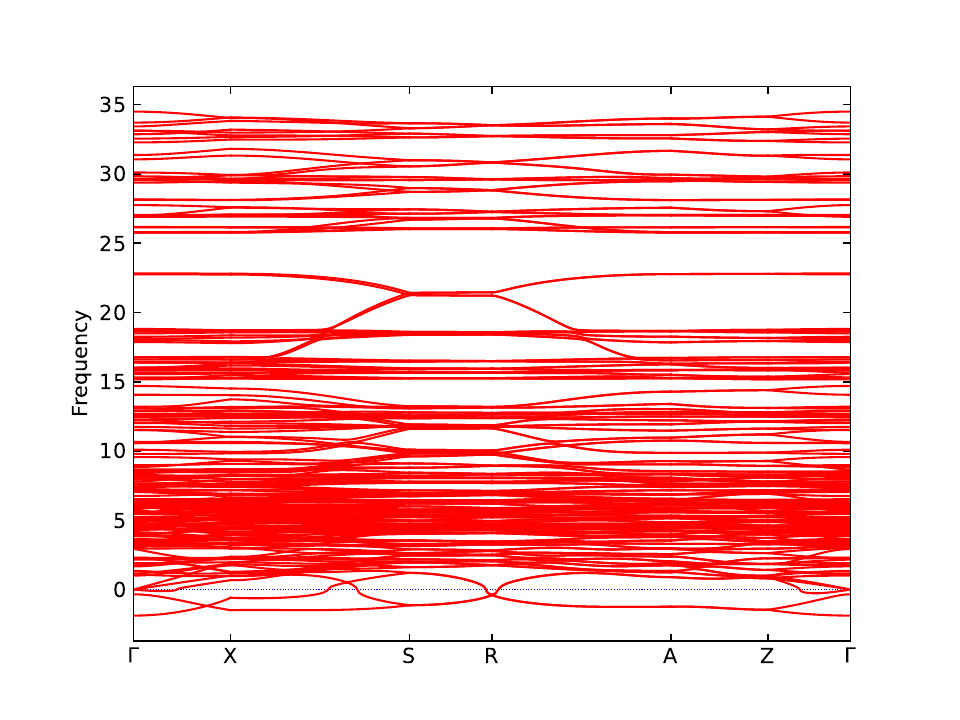}
    \caption{Phonon modulation of NTP by Z the mode.}
    \label{fig:phonon_ntp_zmod}
\end{figure}

\section{\ce{Na4HfP2O9} Attempted Synthesis}

Synthesis attempt of \ce{Na4HfP2O9} with A-Lab. \ce{Na2CO3}, \ce{NH4H2PO4}, and \ce{HfO2} precursors were combined with wet mixing and heated to 1,100$^\circ$ for 8 hours in air. Samples were cooled naturally in the closed furnace. 

No target phase is claimed. XRD pattern analysis, matched with an in-house tree search algorithm, and then refined with with matched phases and the target phase, show a significant amount of unmatched peaks. \ce{HfO2} and \ce{Na4P2O7} are found in relatively large phase fractions. Sample is seen under SEM and EDS to be heterogeneous in composition and morphology. 

\begin{figure}[H]
    \centering
    \includegraphics[width=1.0\linewidth]{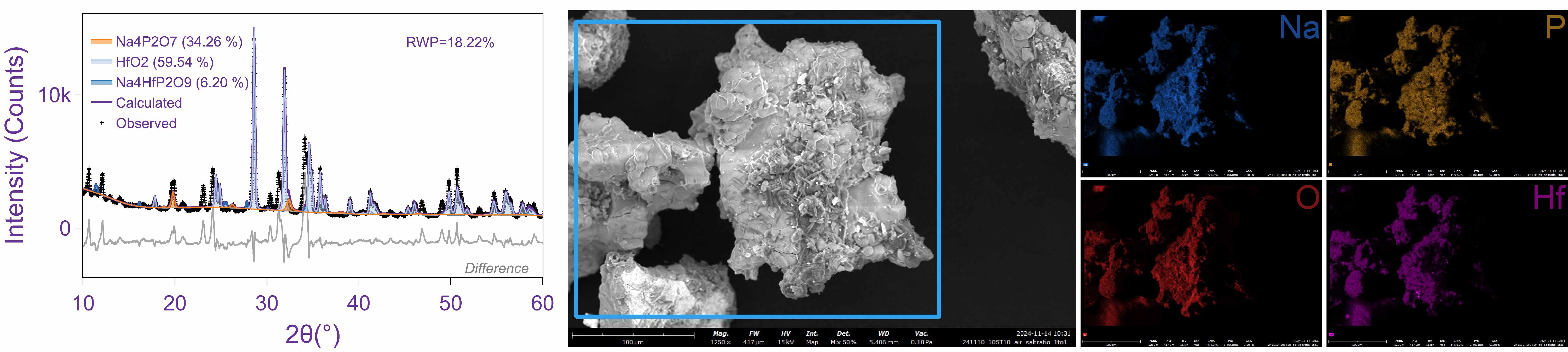}
    \caption{\ce{Na4HfP2O9} trial characterization results depicting powder XRD, imaging with SEM, and elemental mapping with EDS.}
    \label{fig:na4hfp2o9-trial}
\end{figure}

\vfill
\bibliography{AMain}